\documentclass[prl,aps,reprint,amssymb,superscriptaddress]{revtex4-2}
\usepackage{amsmath}
\usepackage{bm}
\usepackage[dvipsnames]{xcolor}
\usepackage{graphicx}
\usepackage{nicematrix}
\usepackage{epstopdf}
\usepackage{epsfig}
\usepackage{amsfonts}
\usepackage[pdfencoding=auto,naturalnames]{hyperref}
\usepackage{hypcap}
\usepackage{mathtools}
\usepackage{verbatim}
\usepackage{tabularx}
\usepackage{bbm}
\usepackage{esvect}
\usepackage{subfigure}
\usepackage{slashed}
\usepackage{physics}
\usepackage{times}
\usepackage{orcidlink}
\usepackage{comment}

\def\bea{\begin{eqnarray}}
\def\eea{\end{eqnarray}}

\def\ba{\begin{array}}
\def\ea{\end{array}}

\hypersetup{colorlinks=true, citecolor=blue, urlcolor=blue, linkcolor=blue}

\begin{document}
\title{Holographic dual of defect CFT with corner contributions}
\author{Xinyu Sun}
\affiliation{Institute for Advanced Study, Tsinghua University, Beijing 100084, China}
\author{Shao-Kai Jian}
\email{sjian@tulane.edu}
\affiliation{Department of Physics and Engineering Physics, Tulane University, New Orleans, Louisiana, 70118, USA}

\date{\today}

\begin{abstract}
We study defect CFT within the framework of holographic duality, emphasizing the impact of corner contributions. 
We model distinct conformal defects using interface branes that differ in tensions and are connected by a corner. 
Employing the relationship between CFT scaling dimensions and Euclidean gravity actions, we outline a general procedure for calculating the anomalous dimensions of defect changing operators at nontrivial cusps. 
Several analytical results are obtained, including the cusp anomalous dimensions at big and small angles. 
While $1/\phi$ universal divergence appears for small cusp angles due to the fusion of two defects, more interestingly, we uncover a bubble phase rendered by a near zero angle cusp, in which the divergence is absent. 
\end{abstract}

\maketitle

\begin{figure}
    \centering
    \subfigure[]{
    \includegraphics[width=0.22\textwidth]{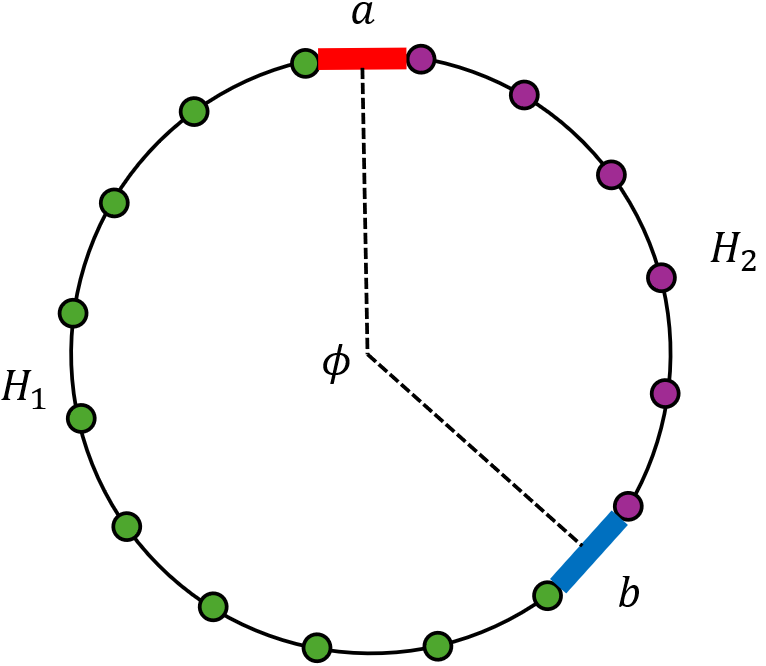}} 
    \subfigure[]{
    \includegraphics[width=0.24\textwidth]{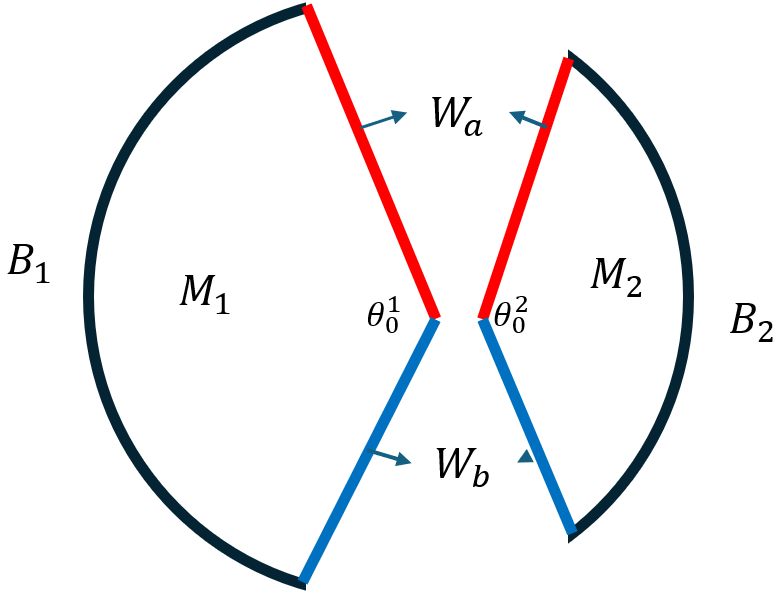}}
    \caption{Illustration of (a) the defect CFT model and (b) the holographic model.
    See the main text for detailed explanations.}
    \label{fig:defect}
\end{figure}

\textcolor{blue}{\it Introduction.---}Defect conformal field theory (CFT) appears across various disciplines in  physics~\cite{andrei2018boundary}, ranging from Kondo effect~\cite{kondo1964resistance,wilson1975the} in solid-state physics to the description of D-branes in the string theory~\cite{recknagel2013boundary}. 
Fathoming defect CFT is not only of great importance but also has broad applications. 
Consider a general defect CFT on a circle $S^1$, hosting two conformal defects, as illustrated in Fig.~\ref{fig:defect} (a). 
Two CFT Hamiltonian, $H_1$ and $H_2$, are defined on lattice sites with green and purple colors, respectively. 
The two defects of type $a$ and $b$ separating the bulk CFT are denoted by the red and blue bonds, respectively. 
The full Hamiltonian reads,
\begin{equation} \label{eq:defect_Hamiltonian}
    \begin{split}
    H_{ab} &= \sum_{r=1}^{L_1} h_1(r, r+1) + \sum_{r=L_1+1}^{L_1+L_2} h_2(r, r+1) \\
    & + \kappa_a h_a(L_1 + L_2, 1) +  \kappa_b h_b(L_1, L_1 + 1)\, .
    \end{split}
\end{equation}
Here, $h_1$ and $h_2$ denote local terms in $H_1$ and $H_2$ defined in $r=1,...,L_1$ and $r=L_1 +1 ,...,L_1 + L_2$.
$h_a$ and $h_b$ with the strength $\kappa_a$ and $\kappa_b$ are the local terms of defect type $a$ and $b$, respectively. 
The total number of sites is $L = L_1 + L_2$, and $r+L = r$ due to the periodic boundary condition. 
The angle spanned by these two defects is $\phi = 2\pi L_1/L$.  

In the low-energy limit, this lattice model under RG will flow to a defect CFT. 
The bulk CFT Hamiltonian $H_1$ and $H_2$ correspond to two CFTs separated by different conformal defects in the most general case. 
The ground state energy has the following form~\cite{christe2008introduction},
\begin{equation} \label{eq:GS_energy}
    E_\text{GS} = \epsilon_\text{bulk} L + \epsilon_\text{defect} + \frac{\Delta_{ab}^{12}(\phi) }{L} + \mathcal O(L^{-2})\,,
\end{equation}
in which $\epsilon_\text{bulk}$ and $\epsilon_\text{defect}$ denote the bulk and defect energy density. 
The third term that is proportional to $1/L$ is universal, and it gives the scaling dimension, $\Delta_{ab}^{12}(\phi)$, of the defect changing operator with an additional cusp $\phi$. 
While a partial list of known examples includes a few minimal models~\cite{henkel1989ising,kane1992transmission,delfino1994scattering,oshikawa1996defect,oshikawa1997boundary,quella2006reflection,frohlich2006duality,bachas2007fusion,kormos2009defect,bachas2012a,gang2008superconformal,makabe2017defects,cogburn2024cft}, the $O(N)$ model~\cite{gliozzi2015boundary,krishnan2023a,trepanier2023surface,giombi2023notes,zhou2023g,cuomo2024impurities}, and supersymmetric Yang-Mills models~\cite{drukker1999wilson,bachas2001permeable,makeenko2006cusped,correa2012the,gromov2015quantum,grozin2015the,kravchuk2024effective}, 
the solution to this defect problem is relatively limited, even in 2D. 
For instance, the product of two minimal models in the folding trick is no longer a minimal model, and the conformal boundary condition~\cite{cardy1984conformal,cardy1989boundary} is not known in general.

In this paper, we propose a bottom-up holographic dual~\cite{maldacena1997the,witten1998anti,gubser1998gauge} to gain a general understanding. 
More explicitly, the CFT is dual to a bulk gravity and each defect is dual to an interface brane~\cite{karch2000locally,karch2000open,takayanagi2011holographic,fujita2011aspect}, as illustrated in Fig.~\ref{fig:defect} (b). 
$B_{1,2}$ ($M_{1,2}$) denotes the asymptotic boundary (the bulk region) of the AdS gravity dual to the CFTs.
Two branes $W_{a,b}$ are dual to the defects $a,b$, and they intersect at a corner with internal angles $\theta_{0}^{1,2}$ in region $M_{1,2}$. 
It should be noted that the corner angle $\theta_0^{1,2}$ is distinct from the cusp angle $\phi$.
Holographic models for defect CFT have been widely explored~\cite{dewolfe2001holography,azeyanagi2007holographic,erdmenger2014bending,bachas2020energy,simidzija2020holoween,karch2021universal,bachas2021phases,miyaji2022holographic,karch2022universal,karch2023universal,tang2023universal}.
Similar setups either in a single CFT~\cite{miyaji2022holographic} or without a corner contribution~\cite{bachas2021phases} were considered, but we obtain new results as summarized in the following.
We outline a general procedure of solving the brane solution and getting $\Delta_{ab}^{12}(\phi)$. 
This gives a universal methodology to obtain arbitrary  $\Delta_{ab}^{12}(\phi)$ numerically in holography.
For some special cases, we can make simplifications to  get analytical results.
Specifically, we focus on (a) the defect changing operator between different conformal defects, $ \Delta_{ab}^{11}(\pi)$, without a cusp, see Eq.~\eqref{eq:defect_changing_operator}, and (b) the cusp anomalous dimension for the same kind of defect between the distinct CFTs $\Delta_{aa}^{12}(\phi)$, see Eq.~\eqref{eq:cusp_dimension_big} [Eq.~\eqref{eq:cusp_dimension_small}] for a big (small) cusp angle, or the same CFTs with a nontrivial corner angle $\Delta_{ab}^{11}(\phi)$, see Eq.~\eqref{eq:cusp_dimension_corner}. 
More interestingly, we uncover a bubble-solution phase at near zero cusp angle, $\phi \approx 0$, where the fusion between defects is unconventional.

\textcolor{blue}{\it Holographic duality and general solutions.---}
We consider a dual model of Eq.~\eqref{eq:defect_Hamiltonian} described by 3D Euclidean gravity action: 
\begin{equation}
\label{meq:total action for ads space}
    I_{\rm tot}=I_{\rm EH}+I_T+I_{\rm surface}+I_{\rm corner}\, .
\end{equation}
Here $I_{\rm EH}=-\frac{1}{2}\sum_{i=1,2}\int_{M_i}\sqrt{g_i}\left(R_i+\frac{2}{l_i^2}\right)$ is the bulk EH action for the two AdS regions $M_i$ with $g_i$, $R_i$, and $l_i$ being the determinant of metric, the Ricci scalar, and the AdS radius, respectively.
The boundaries are $\partial M_i=S_a^i+S_b^i+B_i$, where $S_{a,b}^i$ are the interface branes and $B_i$ is an asymptotic boundary.
$I_T=\sum_{\alpha=a,b}\int_{W_\alpha}\sqrt{\hat{g}_\alpha}\ T_\alpha$, and $I_{\rm surface}=-\sum_{i=1,2}\int_{\partial M_i}\sqrt{\hat{g}_i}\ K_i$ describes the interface brane $W_{a,b}$, $W_{a,b}=S^1_{a,b}=S^2_{a,b}$, with $\hat g$, $K$ and $T$ denoting the induced metric, the external curvature, and the corresponding tension, respectively.
The corner contribution joining two interface branes is described by
\begin{equation} \label{eq:corner_action}
    \begin{split}
    I_{\rm corner}  = 
    -\sum_{i=1,2}\int_{S_a^i\cap S_b^i}\sqrt{\hat{g}}\left(\theta_0^i-\theta^i\right).
    \end{split}
\end{equation}
with $\theta_{0}^i$ being the angle spanning between the tangents of the two branes at the corner.
Note that we omit the corner at the asymptotic boundary and a counter term $I_{\rm c.t.}$, the detail of which can be found in the Supplemental Material.

Consider a gravity dual for CFT in a long strip of width $L_i$, the AdS bulk solution is
\begin{equation}
\label{meq:general AdS metric}
    {\rm d}s_i^2=(r^2-M_i l_i^2){\rm d}\tau^2+\frac{l_i^2{\rm d}r^2}{r^2-M_i l_i^2}+r^2{\rm d}x^2,
\end{equation}
where $l_{1,2}$ is the radius for $AdS_{1,2}$ related to the central charge $c_{1,2}=\frac{3l_{12}}{2G_N}$, where $G_N$ is the Newton constant, and $M_i$ is the parameter which is related to the width $L_i$. 
The Euclidean time is periodic, $\tau = \tau + T_\text{DCFT}^{-1}$, $T_\text{DCFT}^{-1} \gg L$. 
It is worth explaining the physics meaning of $M_i$.
For a pure AdS geometry, $M_i < 0$ determines the perimeter in the compactified coordinate, $x = x + 2\pi / \sqrt{-M_i} $. 
Nevertheless, in our geometry shown in Fig.~\ref{fig:defect} (b), two AdS geometries are joined via interface branes, so that the compactified coordinate is not determined by solely either $M_1$ or $M_2$, but both of them.
We will need to determine them according to the length $L_{1,2}$ in the dual CFTs.
 
The two interface branes dual to the defects are given by $(x(\sigma),r(\sigma))$ at a constant time slice, with $\sigma$ being a parameter. 
Two matching conditions are~\cite{bachas2021phases}
\begin{equation}
\label{meq:matching condition for branes}
\begin{split}
     {\rm d}s^2|_{S_\alpha^1} ={\rm d}s^2|_{S_\alpha^2}\,, \quad 
     (K^1+K^2)_{\mu\nu}  =T_\alpha h_{\mu\nu}\, ,
\end{split}
\end{equation}
where $ \alpha = a, b$. 
The first equation states the metric on the brane is continuous.
The second equation relates the tension of the brane $T_\alpha$ to the extrinsic curvatures $K^{1,2}$ from the two regions, with $h_{\mu\nu}$ being the induced metric.
The second equation gives a constraint on the tension  
$T_\text{min}\le T \le T_\text{max}$ with
$T_\text{min} = |l_1^{-1}-l_2^{-1}|$, $T_\text{max} = l_1^{-1}+l_2^{-1}$.
With the corner term, the two interface branes shall not only satisfy the matching condition~\eqref{meq:matching condition for branes}, but they will join at a tangent angle $\theta^{1,2}_0$. 
Although there are two corners in two AdS spacetime, we can prove that two angles are not independent, and in the following, we require that the angle in $AdS_1$ is $\theta_0 \equiv \theta_0^1$.
The general procedure to proceed is to solve the brane trajectory with given $(l_{1,2},T_{a,b},M_{1,2},\theta_0)$, and then match the interval $L_{1,2}$ for the two CFTs, which will then lead to an equation for $M_{1,2}$ and $L_{1,2}$.

Without loss of generality, we assume $l_1<l_2$.
After solving the equations of motion with matching conditions \eqref{meq:matching condition for branes}, with a proper coordinate $\sigma=r^2-Ml^2$, we have the differential equations $\dot{x}_i^\alpha$ for branes trajectories with tension $T$ at each AdS spacetime, where the explicit form is given in the Supplemental Material.
Here $i=1,2$ label the brane in $AdS_{1,2}$, and $T_\alpha=T_{a,b}$ for different branes.
With the condition that the angle at the corner of two branes in $AdS_1$ is $\theta_0$, we can solve the coordinate of the crossing point of two branes with $\sigma=\sigma_0$~\footnote{Note that $\sigma \rightarrow \infty$ approaches the intersection between the interface brane and the asymptotic boundary}.
Then an equation can be constructed to match $L_{1,2}$. 
To this end, it is convenient to define two dimensionless variables $\gamma=\frac{L_1}{L_2} = \frac{\phi}{2\pi - \phi}$ and $\mu= \frac{M_2}{M_1}$, then, we have
\begin{equation}
\label{meq:equation of gamma mu}
     \gamma=\frac{I_1'+\pi(1-{\rm sgn}(I_1'))}{I_2'+\frac{\pi}{\sqrt{\mu}}(1-{\rm sgn}(I_2'))} \,,
\end{equation}
where $I_{1,2}'$ correspond to two integrals that are related to the brane trajectory, whose expression can be found in the Supplemental Material. 
In the following, we will use $\gamma$ and $\phi$ interchangeably.
In general, Eq.~\eqref{meq:equation of gamma mu} is a complicated integral function about $\mu$ and cannot be solved analytically.
However, in some special cases, we can solve it to get $\mu(\gamma)$, which is shown below.

Now, let's discuss the effect of the corner term Eq.~\eqref{eq:corner_action}. 
In AdS/BCFT, it is known that a corner term is necessary when tensions are different $T_a \ne T_b$, and plays the role of boundary condition changing operator.
In our model, a corner term plays the role of defect changing operator.
Specifically, a nontrivial defect changing operator exists for $\theta_0 \ne \pi$. 
To extract the scaling dimension of defect changing operator, we evaluate the onshell action:
\begin{equation}
\label{meq:saddle point action}
    I_{\rm tot}=\frac{M_1 l_1 L_1+M_2 l_2 L_2}{2T_{\rm DCFT}} \,.
\end{equation}
Although $I_{\rm tot}$ seems to have a simple form, the complexity is hidden in $M_{1,2}$ as we need to express them as $(l_{1,2}, T_{a,b}, L_{1,2}, \theta_0)$ by solving Eq.~\eqref{meq:equation of gamma mu}.
As the defect CFT is dual to the gravitational model with interface branes joined by a corner, this partition function reveals the information of $\Delta_{ab}^{12}(\phi)$.  
In particular, 
\begin{equation} \label{eq:action_anomalous_dimension}
    I_{\rm tot} = \frac{E_{ab}^{12}(\gamma)}{T_{\rm DCFT}}\,,~   \Delta^{12}_{ab}(\phi)= E^{12}_{ab}(\gamma)- \frac12 (E^{12}_{aa}+E^{12}_{bb})\,, 
\end{equation}
in which 
$E_{aa}^{12} \equiv E_{aa}^{12}(1)$ is the eigenvalue at $\gamma=1$. 
With the procedure outlined above, a general $\Delta^{12}_{ab}(\phi)$ can be solved numerically.
Next, we will discuss several special cases with simpler analytical expressions.

\textcolor{blue}{\it Defect changing operator.---} Consider the defect changing operator between different conformal defects without a cusp in the same CFT: $\gamma=1$ and $l_1=l_2=l$, the geometry has a $\mathbb{Z}_2$ symmetry, so we know $\mu=1$ without solving Eq.~\eqref{meq:equation of gamma mu}.
Plugging it into Eq.~\eqref{meq:saddle point action}, we have $I_{\rm tot}=-\frac{2l}{T_{\rm DCFT}L}\left[\pi+{\rm sgn}(\theta_0-\pi)(\pi-\arccos{\beta})\right]^2$ with
\begin{equation}
    \beta=\frac{\cos{\theta_0}+(lT_{a}/2)(lT_{b}/2)}{\sqrt{(1-(lT_{a}/2)^2)(1-(lT_{b}/2)^2)}}\,,
\end{equation}
and consequently, $E_{ab}^{12}(\gamma = 1) = I_\text{tot} T_\text{DCFT}$.
After subtracting $E_{aa}=-2\pi^2 l/L$ for $T_a=T_b$ and $\theta_0=\pi$, we arrive at 
\begin{equation}
\label{eq:defect_changing_operator}
    \Delta_{ab}^{11}(\pi) = \frac{2l}{L}(\pi^2 - \arccos^2{\beta}).
\end{equation} 
in which we require $\theta_0 \le \pi$ to satisfy $\Delta_{ab}>\Delta_{aa}=0$. 

\begin{figure}
    \centering
    \subfigure[]{
    \includegraphics[width=0.23\textwidth]{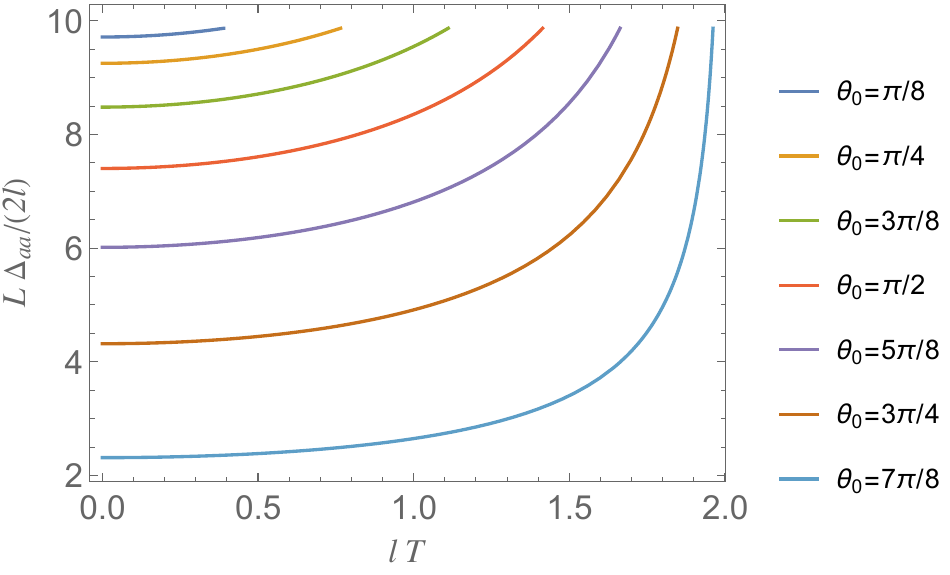}} 
    \subfigure[]{
    \includegraphics[width=0.23\textwidth]{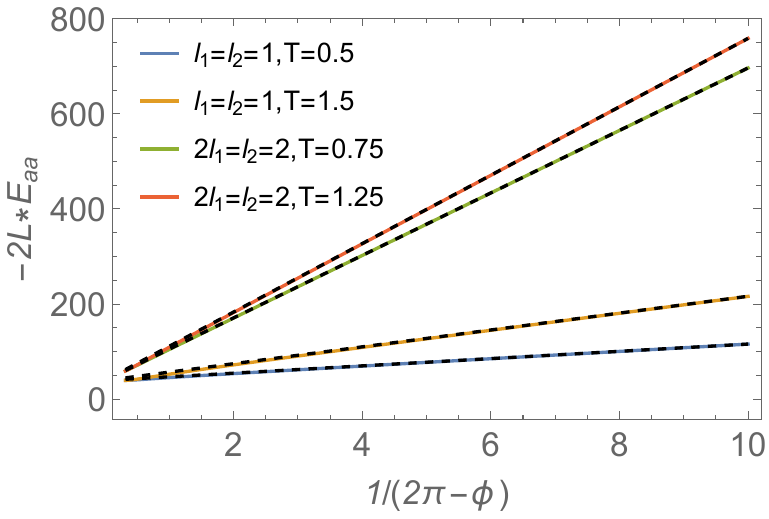}} 
    \subfigure[]{
    \includegraphics[width=0.23\textwidth]{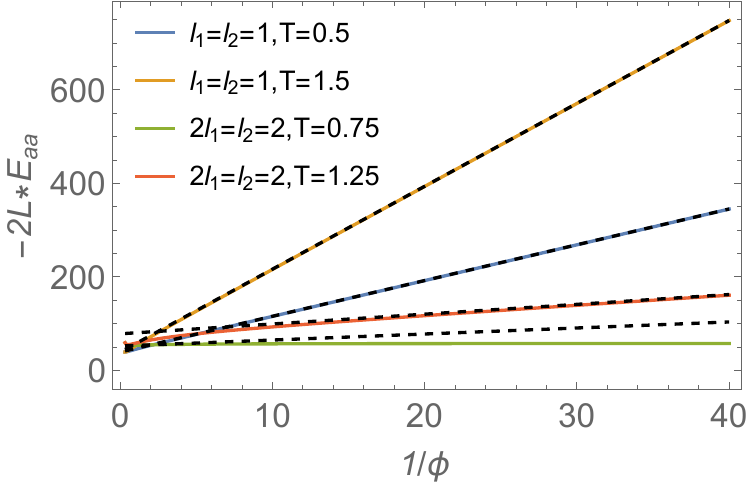}}
    \subfigure[]{
    \includegraphics[width=0.23\textwidth]{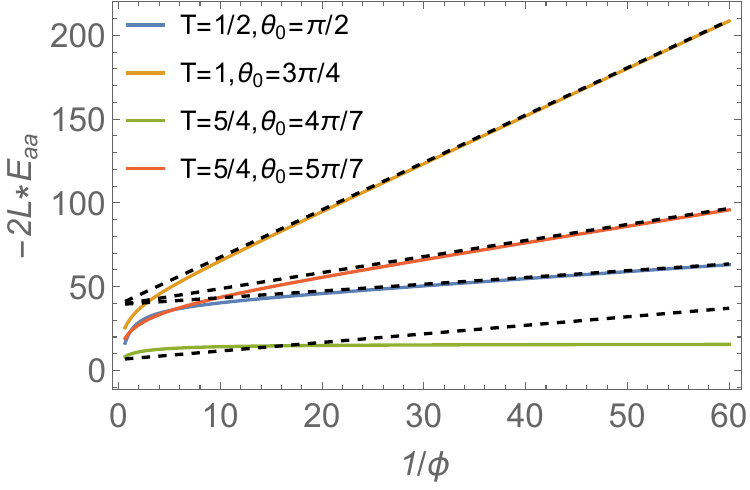}}
    \caption{(a) Scaling dimension of the defect changing operator with $l_1=l_2=l$ and $T_a=T_b=T$ with different $\theta_0$.
    (b,c) On-shell action for (b) $\phi\rightarrow2\pi$ limit and (c) $\phi\rightarrow0$ limit without corners.
    (d) On-shell action for $\phi\rightarrow0$ limit with a corner.
    The solid (dash) curves are numerical (analytical) results.}
    \label{fig:on-shell action for special limits}
\end{figure}

To better understand the effect of the corner contribution, we set $T_a=T_b=T$ which leads to $\Delta_{ab}|_{T_\alpha = T}=\frac{2l}{L}[\pi^2-\arccos^2{(\frac{4\cos{\theta_0}+(Tl)^2}{4-(Tl)^2})}]$ with the constraint $0<T<T_c=2\sin{(\theta_0/2)}$. 
Note that, although $T_a = T_b$, it still corresponds to two different defects due to a nontrivial $\theta_0$. 
The scaling dimension of the defect changing operator is plotted in Fig.~\ref{fig:on-shell action for special limits} (a).
A greater tension $T$ and a larger deviation from $\theta_0 = \pi$ will lead to a greater scaling dimension of defect changing operators.
Specially, we have $\Delta_{ab}|_{T_\alpha=T_c}=\frac{2l}{L}\pi^2$ and $\Delta_{ab}|_{T_\alpha=0}=\frac{2l}{L}(\pi^2-\theta_0^2)$.

\textcolor{blue}{\it Cusp anomalous dimension.---} Apart from the defect changing operator given above, interesting conformal data exists even for the same defect: the cusp anomalous dimension. 
The same types of defects means $T_a=T_b=T$ and $\theta_0=\pi$. 
Then we look at a big angle $\phi \sim \pi$ and small angle $\phi \sim 0, 2\pi$, respectively. 
Consider $\phi \sim \pi$, $\gamma = 1 + \delta_\gamma$ and $\delta_\gamma \ll 1$.
We can check $\mu=1$ is a solution for $\gamma=1$.
Therefore, we can expand  $\mu=1 + \delta_\mu$ with $\delta_\mu \ll1$, to get $\delta_\mu =\frac{2l_1l_2T}{l_1+l_2+l_1l_2T}\delta_\gamma + \mathcal{O}(\delta_\gamma^2)$, where the expression of the second order $\mathcal{O}(\delta_\gamma^2)$ is given in the Supplemental Material.
With this solution and Eq.~\eqref{meq:saddle point action}, the cusp anomalous dimension near $\phi = \pi$ reads
\begin{equation}
\label{eq:cusp_dimension_big}
    \Delta_{aa}^{12}(\phi) =- \frac{l_1}{L}\left[ \pi\left(\frac{l_2}{l_1}-1\right)\delta_\phi+\frac{2(T-T_{\rm min})}{T+T_{\rm max}}\delta_\phi^2\right].
\end{equation}
where $\delta_\phi=\phi-\pi$.
When $l_1=l_2=l$ with a $\mathbb{Z}_2$ symmetry, the linear term in Eq.~\eqref{eq:cusp_dimension_big} vanishes, and the quadratic term is negative, consistent with reflection positivity~\cite{bachas2006comment,diatlyk2024effective,cuomo2024impurities}.

Next, consider the cusp anomalous dimension near $\phi \rightarrow 0$ or $\phi \rightarrow 2\pi$ for the same defect: $T_a= T_b = T$ and $\theta_0=\pi$.
Note that these two limits are different when the two CFTs are distinct $l_1 \ne l_2$. 
As $ \phi \rightarrow 0$ implies $\gamma\rightarrow0$,
we simplify Eq.~\eqref{meq:equation of gamma mu} as $\sqrt{\mu}=\frac{2\pi}{B(T, l_1, l_2)}\gamma+\mathcal{O}(\gamma^2)$, where $B(T,l_1,l_2)=2\pi\Theta\left(l_2T-1\right)-\Xi_1(T,l_1,l_2)$ with
\begin{equation}
\label{meq:xi equation}
    \Xi_{u_0}(T,l_1,l_2)=\sqrt{\frac{4v}{-u}}\left[K\left(u_0,v\right)+\frac{l_1^2l_2^2T^2-l_1^2}{l_2^2}\Pi\left(u_0,u,v\right)\right],
\end{equation}
and $u_0=1$.
We define $v=\frac{(T-T_{\rm max})(T+T_{\rm min})}{(T+T_{\rm max})(T-T_{\rm min})}$, $u=l_1^2(T_{\rm max}-T)(T+T_{\rm min})$,
$\Theta(\cdot)$ is the step function, and $K(\cdot,\cdot)$ ($\Pi(\cdot, \cdot, \cdot)$) is the incomplete elliptic integral of the first (third) kind.
With this solution and Eq.~\eqref{meq:saddle point action}, we can obtain the onshell action explicitly,
\begin{equation}
\label{meq:saddle point action for no bubble l1l2l}
    I_{\rm tot}=-\frac{1}{2T_{\rm DCFT}L}\left(\frac{l\cdot B(T,l_1, l_2)^2}{\gamma}+\mathcal{O}(1)\right).
\end{equation}

The other limit, $\phi \rightarrow 2\pi$, implies instead $\gamma\rightarrow\infty$, which can be solved similarly. 
With Eq.~\eqref{eq:action_anomalous_dimension}, the cusp anomalous dimension is given by
\begin{equation} \label{eq:cusp_dimension_small}
    \Delta_{aa}^{12}(\phi) = \begin{cases} 
    - \frac{l_2 B(T,l_2,l_1)^2}{2L} \frac{2\pi}{2\pi- \phi}  & \phi \rightarrow 2\pi \\
    - \frac{l_1 B(T,l_1,l_2)^2}{2L} \frac{2\pi}{\phi}  & \phi \rightarrow 0
    \end{cases} \,.
\end{equation}

The cusp anomalous dimension diverges as $1/\phi$ or $1/(2\pi - \phi)$, under the limit~\cite{cuomo2024impurities,Estienne_2022}.
On the CFT side, the two defects fuse into a trivial defect.
The divergent behavior can be interpreted the Casimir energy and the prefactor is the fusion coefficient. 
It is positive or negative for repulsive and attractive interaction.

We plot the numerical results obtained by our general procedure and analytical results in Fig.~\ref{fig:on-shell action for special limits} (b) and (c), in which Fig.~\ref{fig:on-shell action for special limits} (b),(c) show the limit $\phi \rightarrow 2\pi, 0 $, respectively.
For $\phi \rightarrow 2\pi$, the two results, i.e., numerical (analytical) results shown by solid (dashed) curves, agree quite well in Fig.~\ref{fig:on-shell action for special limits} (b).
However, for $\phi \rightarrow 0$, some results agree, but some do not, for instance, the green solid curve with $2l_1 =l_2 =2$, $T=0.75$ shows a significant deviation from the analytical solution plotted in a dashed line.
The reason is that there can emerge a bubble solution in the limit $\phi \rightarrow 0$, as is discussed next.

\textcolor{blue}{\it Bubble phase and exotic fusion.---} As mentioned above, for $\phi \rightarrow 0$, two defects into one, and lead to a divergent behavior. 
However, it is not always the case.
In the following, we consider more general solutions, and also discuss the role of the corner contribution with $\theta_0\neq\pi$.

\begin{figure}
    \centering
    \subfigure[]{
    \includegraphics[width=0.3\textwidth]{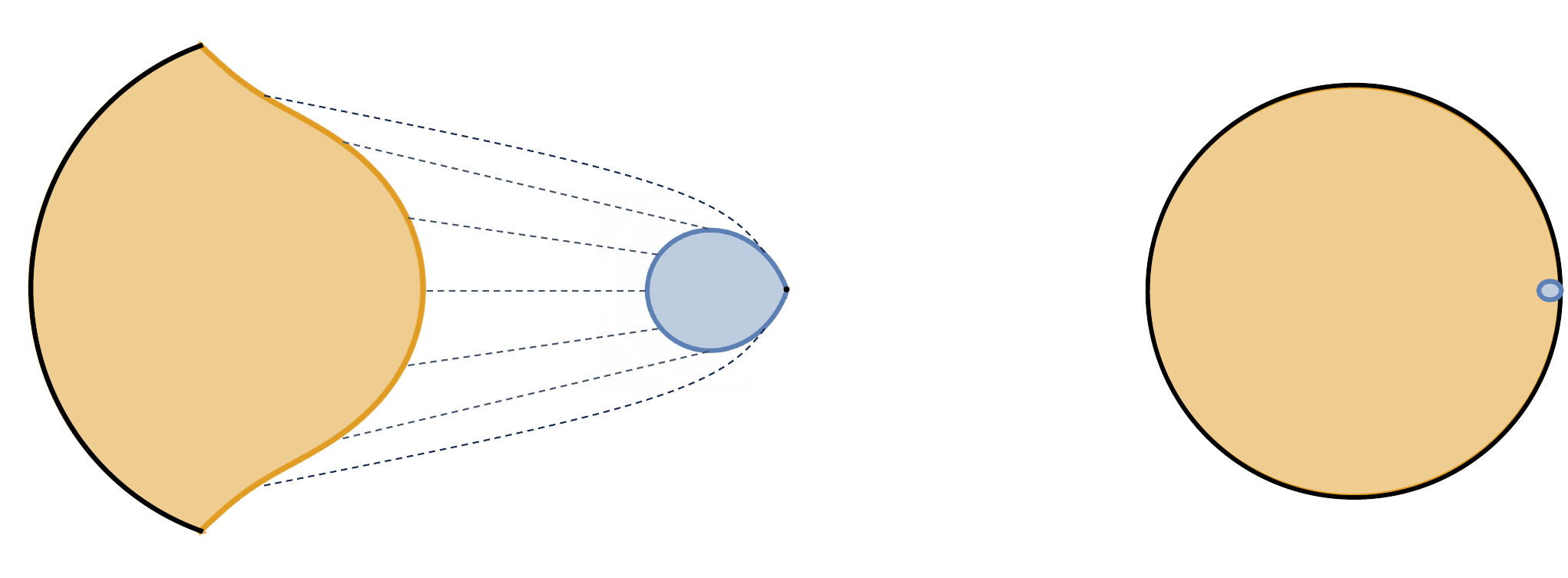}} 
    \subfigure[]{
    \includegraphics[width=0.21\textwidth]{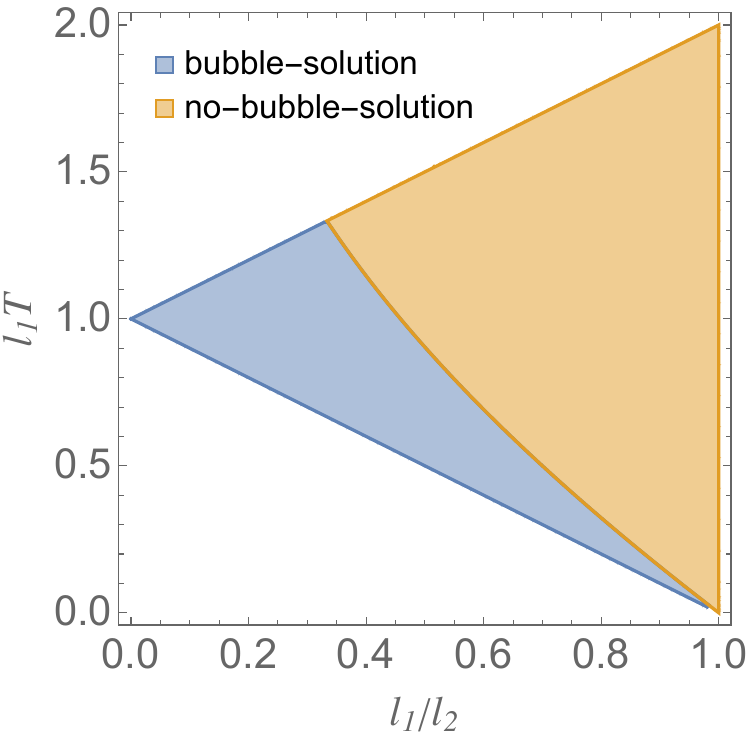}} 
    \subfigure[]{
    \includegraphics[width=0.22\textwidth]{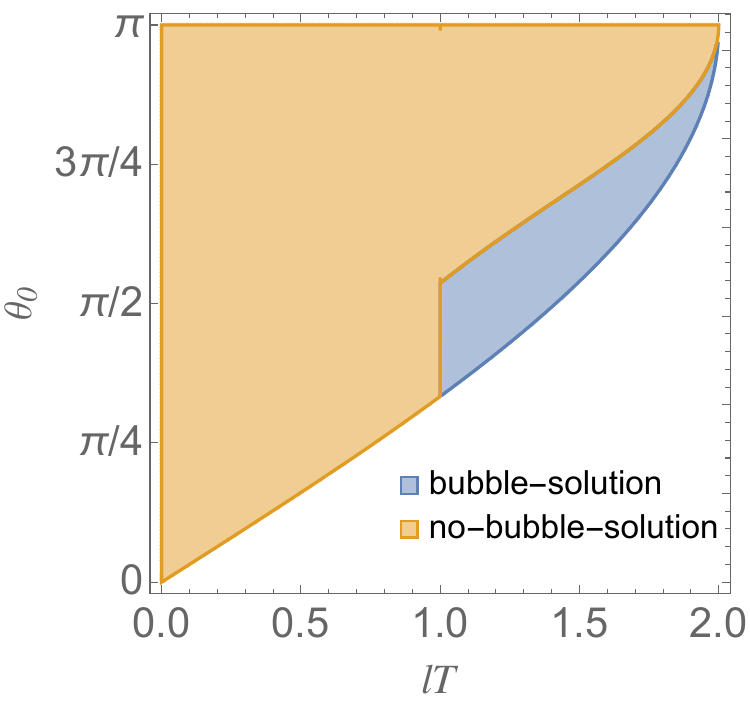}}
    \caption{(a) Configurations of the bubble-solution phase on the left panel and no-bubble-solution phase on the right panel, where the blue (orange) region illustrates $AdS_{1 (2)}$.
    (b) Phase diagram from small cusp angles $\phi \sim 0 $ for $\theta_0=\pi$.
    (c) Phase diagram from small cusp angles $\phi \sim 0 $ with a corner.}
    \label{fig:phase diagram}
\end{figure}

First, we consider the same defect, $T_a=T_b$, $l_1\neq l_2$ and $\theta_0=\pi$. 
Given $B(T,l_1, l_2) > 0$, the solution $\sqrt{\mu}=\frac{2\pi}{B(T, l_1, l_2)}\gamma$ becomes $2\pi L_1/ L_2 = B(T,l_1, l_2) \sqrt{M_2/M_1}$.
It shows that when the length of one CFT shrinks to zero, $L_1 \ll L_2$, its dual AdS also shrinks to zero $M_1 \gg M_2$. 
A schematic plot of the solution is given in Fig.~\ref{fig:phase diagram} (a) right panel. 
However, if $B(T,l_1, l_2) \rightarrow 0$, that the length of one CFT shrinks to zero, $L_1 \ll L_2$, does not imply its dual AdS shrinks to zero. 
Instead, we find that at  $B(T,l_1, l_2) = 0$, the solution of Eq.~\eqref{meq:equation of gamma mu} at $\gamma \rightarrow 0 $ is given by a finite $M_2/M_1 \rightarrow \mu_0 >0$.
A schematic plot of the two geometries given in Fig.~\ref{fig:phase diagram} (a), showing a qualitative difference between the two solutions.
The one with $M_2 /M_1 \rightarrow 0$ is termed as the no-bubble-solution phase, and the one with $M_2 /M_1 \rightarrow \mu_0 > 0 $ is termed as the bubble-solution phase.
Their phase boundary is precisely given by $B(T, l_1, l_2) = 0$. 
We can show that $B(T,l_1,l_2) = 0$ is possible only when $l_1 < l_2$.
It means that the dual of CFTs with a greater central charge can support a bubble of CFTs with a smaller central charge, but the opposite cannot occur. 
With independent parameters $(l_1/l_2,T l_1)$ and $l_1/l_2 \in [0,1], Tl_1\in[1-l_1/l_2,1+l_1/l_2]$, the phase diagram is plotted in Fig.~\ref{fig:phase diagram} (b).

We discuss the implication for the cusp anomalous dimension. 
For $B(T,l_1, l_2) > 0$, namely, in the no-bubble-solution phase, the cusp anomalous dimension given in Eq.~\eqref{eq:cusp_dimension_small} is valid.
However, in the bubble-solution phase, the solution with $\mu = \mu_0$ renders a finite onshell action without the $1/\phi$ divergence, in contrast to Eq.~\eqref{meq:saddle point action for no bubble l1l2l}.  
Therefore, we expect the existence of the bubble-solution phase corresponds to an exotic fusion of two defects.

We are ready to explain Fig.~\ref{fig:on-shell action for special limits} (b) and (c). 
For parameters located in the no-bubble-solution phase, the numerical and analytical results agree, whereas for parameters located in the bubble-solution phase, specially, for $l_2=2l_1=2$, $T=0.75$, the cusp anomalous dimension converges at $\phi \rightarrow 0$ shown by the solid curve. 

Next, we consider a more general case with a corner contribution $\theta_0\neq\pi$ that indicates a nontrivial defect changing operator.
We further assume $l_1=l_2=l$ and $T_a=T_b=T$.
Then, similar to the previous discussion,
In the limit $\phi \rightarrow 0$, or equivalently $\gamma \rightarrow 0$, we obtain $\sqrt{\mu}=\frac{2\pi}{\tilde B( T, l, \theta_0)}\gamma+\mathcal{O}(\gamma^2)$,
where $\tilde B(T, l, \theta_0) $ is defined in the Supplemental Material.
For $\theta_0<\pi$ and $T>1$, it can be simplified to $\Tilde{B}(T,l,\theta_0)=2\pi-2\Xi_1(T,l,l)+ \Xi_{u_0}(T,l,l)$,
where $u_0=[\frac{l^2T^2+2\cos{\theta_0}-2}{(lT-2)(\sqrt{(l^2T^2+l^2T^2\cos{\theta_0}-2)(\cos{\theta_0}-1)}-lT\cos{\theta_0})}]^{1/2}$.
Again, $\tilde B(T, l, \theta_0) = 0$ defines the phase transition between the no-bubble-solution phase and the bubble-solution phase. 
With $lT\in[0,2]$ and $\cos{\theta_0}\in[-1,1-(lT)^2/2]$ which is required to have a solution, the phase diagram is shown in Fig.~\ref{fig:phase diagram} (c). 
From the phase diagram, we can find that there is no bubble-solution phase for $lT<1$.  

The asymptotic behavior of the cusp anomalous dimension for $\phi\rightarrow0$ in the no-bubble-solution phase is, 
\begin{equation} \label{eq:cusp_dimension_corner}
    \Delta_{ab}(\phi) = 
    - \frac{l \tilde B(T,l, \theta_0)^2}{2L} \frac{2\pi}{\phi} \,,
\end{equation}
which has a $1/\phi$ singularity from the Casimir energy. 
Finally, we compare the numerical and analytical results of the anomalous dimension and plot them in Fig.~\ref{fig:on-shell action for special limits} (d).

It is illustrative to consider $lT=1$.
In this case, $\tilde B(1/l, l, \theta_0) = \theta_0-\theta_0^c $  with $\theta_0^c=\frac{2}{\sqrt{3}}\cdot K\left(-\frac{1}{3}\right)$.
Hence, the bubble-solution phase and the no-bubble-solution phase exist for $\theta_0\in(\frac{\pi}{3},\theta_0^c)$ and $\theta_0\in(\theta_0^c,\frac{5\pi}{3})$, respectively.

\textcolor{blue}{\it Concluding remarks.---} To illustrate the connection between our results and concrete CFT models, we consider the defect Ising model $H=-\frac{1}{2}\sum'_r [\sigma^z(r)+\sigma^x(r)\sigma^x(r+1)] - \frac{1}{2}  \kappa_a \sigma^x(L_1+L_2)\sigma^x(1) - \frac12 \kappa_b \sigma^x(L_1)\sigma^x(L_1+1)$, where $\sigma^\mu(r) $ denotes the Pauli matrix at site $r$ and the first summation omits the terms at the two defect bonds.
We calculate the defect changing operator at $\phi = \pi$ (see Supplemental Material)
\begin{equation}
    \Delta_{ab}^{11}(\pi) = \frac{1}{\pi } \left( \arctan \kappa_a - (-1)^Q \arctan \kappa_b \right)^2\,.
\end{equation}
where $Q=0$ ($1$) denotes the symmetric (antisymmetric) sector.
We observe that in the $Q=1$ sector, when $\kappa_a = \kappa_b = \kappa$, $\Delta_{ab}^{11}(\pi) = \frac{4}{\pi } \arctan^2 \kappa  $ still nontrivially depends on $\kappa$.  
Note that this happens in the holographic model when $\theta_0 \ne \pi$.
Hence, the effect of a nontrivial corner term is similar to the different sector $Q=0,1$.
Further investigation on the corner angle $\theta_0$ on the CFT side is important.

We also consider the cusp anomalous dimension of the same types of defects in a weak defect limit (see Supplemental Material)
\begin{equation}
\label{meq:ising_cusp}
    \Delta_{aa}(\phi) =-\frac{1}{2L}\frac{1}{\pi}\left(\frac{\pi-\phi}{\sin{\phi}}-1\right)\delta\kappa^2 \,, \quad \kappa=1 - \delta\kappa\,.
\end{equation} 
It is quadratic near $\phi \sim \pi$ and $1/\phi$ divergent near $\phi \sim 0$ both with a negative prefactor, consistent with our result in the no-bubble-solution phase. 

In the exotic fusion, if two defects are the same with $\theta_0=\pi$, the bubble region will have a smaller AdS radius, and act as a true vacuum~\cite{bachas2021phases}.
It is exotic because interface and anti-interface do not fuse to a trivial identity.
More interestingly, we find that a nontrivial corner term $\theta_0\neq\pi$ can also support a bubble solution even with $l_1=l_2$ and $T_a=T_b$.
The exotic fusion on the CFT side is also an interesting open question.

{\it Acknowledgement.} XS acknowledges the support from the Lavin-Bernick Grant during his visit to Tulane University, where the work was conducted.
The work of SKJ is supported by a startup fund at Tulane University.

\bibliography{reference.bib}

\begin{thebibliography}{58}%
\makeatletter
\providecommand \@ifxundefined [1]{%
 \@ifx{#1\undefined}
}%
\providecommand \@ifnum [1]{%
 \ifnum #1\expandafter \@firstoftwo
 \else \expandafter \@secondoftwo
 \fi
}%
\providecommand \@ifx [1]{%
 \ifx #1\expandafter \@firstoftwo
 \else \expandafter \@secondoftwo
 \fi
}%
\providecommand \natexlab [1]{#1}%
\providecommand \enquote  [1]{``#1''}%
\providecommand \bibnamefont  [1]{#1}%
\providecommand \bibfnamefont [1]{#1}%
\providecommand \citenamefont [1]{#1}%
\providecommand \href@noop [0]{\@secondoftwo}%
\providecommand \href [0]{\begingroup \@sanitize@url \@href}%
\providecommand \@href[1]{\@@startlink{#1}\@@href}%
\providecommand \@@href[1]{\endgroup#1\@@endlink}%
\providecommand \@sanitize@url [0]{\catcode `\\12\catcode `\$12\catcode
  `\&12\catcode `\#12\catcode `\^12\catcode `\_12\catcode `\%12\relax}%
\providecommand \@@startlink[1]{}%
\providecommand \@@endlink[0]{}%
\providecommand \url  [0]{\begingroup\@sanitize@url \@url }%
\providecommand \@url [1]{\endgroup\@href {#1}{\urlprefix }}%
\providecommand \urlprefix  [0]{URL }%
\providecommand \Eprint [0]{\href }%
\providecommand \doibase [0]{https://doi.org/}%
\providecommand \selectlanguage [0]{\@gobble}%
\providecommand \bibinfo  [0]{\@secondoftwo}%
\providecommand \bibfield  [0]{\@secondoftwo}%
\providecommand \translation [1]{[#1]}%
\providecommand \BibitemOpen [0]{}%
\providecommand \bibitemStop [0]{}%
\providecommand \bibitemNoStop [0]{.\EOS\space}%
\providecommand \EOS [0]{\spacefactor3000\relax}%
\providecommand \BibitemShut  [1]{\csname bibitem#1\endcsname}%
\let\auto@bib@innerbib\@empty
\bibitem [{\citenamefont {Andrei}\ \emph {et~al.}(2020)\citenamefont {Andrei}
  \emph {et~al.}}]{andrei2018boundary}%
  \BibitemOpen
  \bibfield  {author} {\bibinfo {author} {\bibfnamefont {N.}~\bibnamefont
  {Andrei}} \emph {et~al.},\ }\bibfield  {title} {\bibinfo {title} {{Boundary
  and Defect CFT: Open Problems and Applications}},\ }\href
  {https://doi.org/10.1088/1751-8121/abb0fe} {\bibfield  {journal} {\bibinfo
  {journal} {J. Phys. A}\ }\textbf {\bibinfo {volume} {53}},\ \bibinfo {pages}
  {453002} (\bibinfo {year} {2020})},\ \Eprint
  {https://arxiv.org/abs/1810.05697} {arXiv:1810.05697 [hep-th]} \BibitemShut
  {NoStop}%
\bibitem [{\citenamefont {Kondo}(1964)}]{kondo1964resistance}%
  \BibitemOpen
  \bibfield  {author} {\bibinfo {author} {\bibfnamefont {J.}~\bibnamefont
  {Kondo}},\ }\bibfield  {title} {\bibinfo {title} {{Resistance Minimum in
  Dilute Magnetic Alloys}},\ }\href {https://doi.org/10.1143/PTP.32.37}
  {\bibfield  {journal} {\bibinfo  {journal} {Progress of Theoretical Physics}\
  }\textbf {\bibinfo {volume} {32}},\ \bibinfo {pages} {37} (\bibinfo {year}
  {1964})},\ \Eprint
  {https://arxiv.org/abs/https://academic.oup.com/ptp/article-pdf/32/1/37/5193092/32-1-37.pdf}
  {https://academic.oup.com/ptp/article-pdf/32/1/37/5193092/32-1-37.pdf}
  \BibitemShut {NoStop}%
\bibitem [{\citenamefont {Wilson}(1975)}]{wilson1975the}%
  \BibitemOpen
  \bibfield  {author} {\bibinfo {author} {\bibfnamefont {K.~G.}\ \bibnamefont
  {Wilson}},\ }\bibfield  {title} {\bibinfo {title} {The renormalization group:
  Critical phenomena and the kondo problem},\ }\href
  {https://doi.org/10.1103/RevModPhys.47.773} {\bibfield  {journal} {\bibinfo
  {journal} {Rev. Mod. Phys.}\ }\textbf {\bibinfo {volume} {47}},\ \bibinfo
  {pages} {773} (\bibinfo {year} {1975})}\BibitemShut {NoStop}%
\bibitem [{\citenamefont {Recknagel}\ and\ \citenamefont
  {Schomerus}(2013)}]{recknagel2013boundary}%
  \BibitemOpen
  \bibfield  {author} {\bibinfo {author} {\bibfnamefont {A.}~\bibnamefont
  {Recknagel}}\ and\ \bibinfo {author} {\bibfnamefont {V.}~\bibnamefont
  {Schomerus}},\ }\href {https://doi.org/10.1017/CBO9780511806476} {\emph
  {\bibinfo {title} {{Boundary Conformal Field Theory and the Worldsheet
  Approach to D-Branes}}}},\ Cambridge Monographs on Mathematical Physics\
  (\bibinfo  {publisher} {Cambridge University Press},\ \bibinfo {year}
  {2013})\BibitemShut {NoStop}%
\bibitem [{\citenamefont {Christe}\ and\ \citenamefont
  {Henkel}(2008)}]{christe2008introduction}%
  \BibitemOpen
  \bibfield  {author} {\bibinfo {author} {\bibfnamefont {P.}~\bibnamefont
  {Christe}}\ and\ \bibinfo {author} {\bibfnamefont {M.}~\bibnamefont
  {Henkel}},\ }\href@noop {} {\emph {\bibinfo {title} {Introduction to
  conformal invariance and its applications to critical phenomena}}},\
  Vol.~\bibinfo {volume} {16}\ (\bibinfo  {publisher} {Springer Science \&
  Business Media},\ \bibinfo {year} {2008})\BibitemShut {NoStop}%
\bibitem [{\citenamefont {Henkel}\ \emph {et~al.}(1989)\citenamefont {Henkel},
  \citenamefont {Patkos},\ and\ \citenamefont {Schlottmann}}]{henkel1989ising}%
  \BibitemOpen
  \bibfield  {author} {\bibinfo {author} {\bibfnamefont {M.}~\bibnamefont
  {Henkel}}, \bibinfo {author} {\bibfnamefont {A.}~\bibnamefont {Patkos}},\
  and\ \bibinfo {author} {\bibfnamefont {M.}~\bibnamefont {Schlottmann}},\
  }\bibfield  {title} {\bibinfo {title} {{The Ising Quantum Chain With Defects.
  1. The Exact Solution}},\ }\href
  {https://doi.org/10.1016/0550-3213(89)90410-0} {\bibfield  {journal}
  {\bibinfo  {journal} {Nucl. Phys. B}\ }\textbf {\bibinfo {volume} {314}},\
  \bibinfo {pages} {609} (\bibinfo {year} {1989})}\BibitemShut {NoStop}%
\bibitem [{\citenamefont {Kane}\ and\ \citenamefont
  {Fisher}(1992)}]{kane1992transmission}%
  \BibitemOpen
  \bibfield  {author} {\bibinfo {author} {\bibfnamefont {C.~L.}\ \bibnamefont
  {Kane}}\ and\ \bibinfo {author} {\bibfnamefont {M.~P.~A.}\ \bibnamefont
  {Fisher}},\ }\bibfield  {title} {\bibinfo {title} {Transmission through
  barriers and resonant tunneling in an interacting one-dimensional electron
  gas},\ }\href {https://doi.org/10.1103/PhysRevB.46.15233} {\bibfield
  {journal} {\bibinfo  {journal} {Phys. Rev. B}\ }\textbf {\bibinfo {volume}
  {46}},\ \bibinfo {pages} {15233} (\bibinfo {year} {1992})}\BibitemShut
  {NoStop}%
\bibitem [{\citenamefont {Delfino}\ \emph {et~al.}(1994)\citenamefont
  {Delfino}, \citenamefont {Mussardo},\ and\ \citenamefont
  {Simonetti}}]{delfino1994scattering}%
  \BibitemOpen
  \bibfield  {author} {\bibinfo {author} {\bibfnamefont {G.}~\bibnamefont
  {Delfino}}, \bibinfo {author} {\bibfnamefont {G.}~\bibnamefont {Mussardo}},\
  and\ \bibinfo {author} {\bibfnamefont {P.}~\bibnamefont {Simonetti}},\
  }\bibfield  {title} {\bibinfo {title} {{Scattering theory and correlation
  functions in statistical models with a line of defect}},\ }\href
  {https://doi.org/10.1016/0550-3213(94)90032-9} {\bibfield  {journal}
  {\bibinfo  {journal} {Nucl. Phys. B}\ }\textbf {\bibinfo {volume} {432}},\
  \bibinfo {pages} {518} (\bibinfo {year} {1994})},\ \Eprint
  {https://arxiv.org/abs/hep-th/9409076} {arXiv:hep-th/9409076} \BibitemShut
  {NoStop}%
\bibitem [{\citenamefont {Oshikawa}\ and\ \citenamefont
  {Affleck}(1996)}]{oshikawa1996defect}%
  \BibitemOpen
  \bibfield  {author} {\bibinfo {author} {\bibfnamefont {M.}~\bibnamefont
  {Oshikawa}}\ and\ \bibinfo {author} {\bibfnamefont {I.}~\bibnamefont
  {Affleck}},\ }\bibfield  {title} {\bibinfo {title} {{Defect lines in the
  Ising model and boundary states on orbifolds}},\ }\href
  {https://doi.org/10.1103/PhysRevLett.77.2604} {\bibfield  {journal} {\bibinfo
   {journal} {Phys. Rev. Lett.}\ }\textbf {\bibinfo {volume} {77}},\ \bibinfo
  {pages} {2604} (\bibinfo {year} {1996})},\ \Eprint
  {https://arxiv.org/abs/hep-th/9606177} {arXiv:hep-th/9606177} \BibitemShut
  {NoStop}%
\bibitem [{\citenamefont {Oshikawa}\ and\ \citenamefont
  {Affleck}(1997)}]{oshikawa1997boundary}%
  \BibitemOpen
  \bibfield  {author} {\bibinfo {author} {\bibfnamefont {M.}~\bibnamefont
  {Oshikawa}}\ and\ \bibinfo {author} {\bibfnamefont {I.}~\bibnamefont
  {Affleck}},\ }\bibfield  {title} {\bibinfo {title} {{Boundary conformal field
  theory approach to the critical two-dimensional Ising model with a defect
  line}},\ }\href {https://doi.org/10.1016/S0550-3213(97)00219-8} {\bibfield
  {journal} {\bibinfo  {journal} {Nucl. Phys. B}\ }\textbf {\bibinfo {volume}
  {495}},\ \bibinfo {pages} {533} (\bibinfo {year} {1997})},\ \Eprint
  {https://arxiv.org/abs/cond-mat/9612187} {arXiv:cond-mat/9612187}
  \BibitemShut {NoStop}%
\bibitem [{\citenamefont {Quella}\ \emph {et~al.}(2007)\citenamefont {Quella},
  \citenamefont {Runkel},\ and\ \citenamefont {Watts}}]{quella2006reflection}%
  \BibitemOpen
  \bibfield  {author} {\bibinfo {author} {\bibfnamefont {T.}~\bibnamefont
  {Quella}}, \bibinfo {author} {\bibfnamefont {I.}~\bibnamefont {Runkel}},\
  and\ \bibinfo {author} {\bibfnamefont {G.~M.~T.}\ \bibnamefont {Watts}},\
  }\bibfield  {title} {\bibinfo {title} {{Reflection and transmission for
  conformal defects}},\ }\href {https://doi.org/10.1088/1126-6708/2007/04/095}
  {\bibfield  {journal} {\bibinfo  {journal} {JHEP}\ }\textbf {\bibinfo
  {volume} {04}},\ \bibinfo {pages} {095}},\ \Eprint
  {https://arxiv.org/abs/hep-th/0611296} {arXiv:hep-th/0611296} \BibitemShut
  {NoStop}%
\bibitem [{\citenamefont {Frohlich}\ \emph {et~al.}(2007)\citenamefont
  {Frohlich}, \citenamefont {Fuchs}, \citenamefont {Runkel},\ and\
  \citenamefont {Schweigert}}]{frohlich2006duality}%
  \BibitemOpen
  \bibfield  {author} {\bibinfo {author} {\bibfnamefont {J.}~\bibnamefont
  {Frohlich}}, \bibinfo {author} {\bibfnamefont {J.}~\bibnamefont {Fuchs}},
  \bibinfo {author} {\bibfnamefont {I.}~\bibnamefont {Runkel}},\ and\ \bibinfo
  {author} {\bibfnamefont {C.}~\bibnamefont {Schweigert}},\ }\bibfield  {title}
  {\bibinfo {title} {{Duality and defects in rational conformal field
  theory}},\ }\href {https://doi.org/10.1016/j.nuclphysb.2006.11.017}
  {\bibfield  {journal} {\bibinfo  {journal} {Nucl. Phys. B}\ }\textbf
  {\bibinfo {volume} {763}},\ \bibinfo {pages} {354} (\bibinfo {year}
  {2007})},\ \Eprint {https://arxiv.org/abs/hep-th/0607247}
  {arXiv:hep-th/0607247} \BibitemShut {NoStop}%
\bibitem [{\citenamefont {Bachas}\ and\ \citenamefont
  {Brunner}(2008)}]{bachas2007fusion}%
  \BibitemOpen
  \bibfield  {author} {\bibinfo {author} {\bibfnamefont {C.}~\bibnamefont
  {Bachas}}\ and\ \bibinfo {author} {\bibfnamefont {I.}~\bibnamefont
  {Brunner}},\ }\bibfield  {title} {\bibinfo {title} {{Fusion of conformal
  interfaces}},\ }\href {https://doi.org/10.1088/1126-6708/2008/02/085}
  {\bibfield  {journal} {\bibinfo  {journal} {JHEP}\ }\textbf {\bibinfo
  {volume} {02}},\ \bibinfo {pages} {085}},\ \Eprint
  {https://arxiv.org/abs/0712.0076} {arXiv:0712.0076 [hep-th]} \BibitemShut
  {NoStop}%
\bibitem [{\citenamefont {Kormos}\ \emph {et~al.}(2009)\citenamefont {Kormos},
  \citenamefont {Runkel},\ and\ \citenamefont {Watts}}]{kormos2009defect}%
  \BibitemOpen
  \bibfield  {author} {\bibinfo {author} {\bibfnamefont {M.}~\bibnamefont
  {Kormos}}, \bibinfo {author} {\bibfnamefont {I.}~\bibnamefont {Runkel}},\
  and\ \bibinfo {author} {\bibfnamefont {G.~M.~T.}\ \bibnamefont {Watts}},\
  }\bibfield  {title} {\bibinfo {title} {{Defect flows in minimal models}},\
  }\href {https://doi.org/10.1088/1126-6708/2009/11/057} {\bibfield  {journal}
  {\bibinfo  {journal} {JHEP}\ }\textbf {\bibinfo {volume} {11}},\ \bibinfo
  {pages} {057}},\ \Eprint {https://arxiv.org/abs/0907.1497} {arXiv:0907.1497
  [hep-th]} \BibitemShut {NoStop}%
\bibitem [{\citenamefont {Bachas}\ \emph {et~al.}(2012)\citenamefont {Bachas},
  \citenamefont {Brunner},\ and\ \citenamefont {Roggenkamp}}]{bachas2012a}%
  \BibitemOpen
  \bibfield  {author} {\bibinfo {author} {\bibfnamefont {C.}~\bibnamefont
  {Bachas}}, \bibinfo {author} {\bibfnamefont {I.}~\bibnamefont {Brunner}},\
  and\ \bibinfo {author} {\bibfnamefont {D.}~\bibnamefont {Roggenkamp}},\
  }\bibfield  {title} {\bibinfo {title} {{A worldsheet extension of
  O(d,d:Z)}},\ }\href {https://doi.org/10.1007/JHEP10(2012)039} {\bibfield
  {journal} {\bibinfo  {journal} {JHEP}\ }\textbf {\bibinfo {volume} {10}},\
  \bibinfo {pages} {039}},\ \Eprint {https://arxiv.org/abs/1205.4647}
  {arXiv:1205.4647 [hep-th]} \BibitemShut {NoStop}%
\bibitem [{\citenamefont {Gang}\ and\ \citenamefont
  {Yamaguchi}(2008)}]{gang2008superconformal}%
  \BibitemOpen
  \bibfield  {author} {\bibinfo {author} {\bibfnamefont {D.}~\bibnamefont
  {Gang}}\ and\ \bibinfo {author} {\bibfnamefont {S.}~\bibnamefont
  {Yamaguchi}},\ }\bibfield  {title} {\bibinfo {title} {{Superconformal defects
  in the tricritical Ising model}},\ }\href
  {https://doi.org/10.1088/1126-6708/2008/12/076} {\bibfield  {journal}
  {\bibinfo  {journal} {JHEP}\ }\textbf {\bibinfo {volume} {12}},\ \bibinfo
  {pages} {076}},\ \Eprint {https://arxiv.org/abs/0809.0175} {arXiv:0809.0175
  [hep-th]} \BibitemShut {NoStop}%
\bibitem [{\citenamefont {Makabe}\ and\ \citenamefont
  {Watts}(2017)}]{makabe2017defects}%
  \BibitemOpen
  \bibfield  {author} {\bibinfo {author} {\bibfnamefont {I.}~\bibnamefont
  {Makabe}}\ and\ \bibinfo {author} {\bibfnamefont {G.~M.~T.}\ \bibnamefont
  {Watts}},\ }\bibfield  {title} {\bibinfo {title} {{Defects in the
  Tri-critical Ising model}},\ }\href {https://doi.org/10.1007/JHEP09(2017)013}
  {\bibfield  {journal} {\bibinfo  {journal} {JHEP}\ }\textbf {\bibinfo
  {volume} {09}},\ \bibinfo {pages} {013}},\ \Eprint
  {https://arxiv.org/abs/1703.09148} {arXiv:1703.09148 [hep-th]} \BibitemShut
  {NoStop}%
\bibitem [{\citenamefont {Cogburn}\ \emph {et~al.}(2024)\citenamefont
  {Cogburn}, \citenamefont {Fitzpatrick},\ and\ \citenamefont
  {Geng}}]{cogburn2024cft}%
  \BibitemOpen
  \bibfield  {author} {\bibinfo {author} {\bibfnamefont {C.~V.}\ \bibnamefont
  {Cogburn}}, \bibinfo {author} {\bibfnamefont {A.~L.}\ \bibnamefont
  {Fitzpatrick}},\ and\ \bibinfo {author} {\bibfnamefont {H.}~\bibnamefont
  {Geng}},\ }\bibfield  {title} {\bibinfo {title} {Cft and lattice correlators
  near an rg domain wall between minimal models},\ }\href@noop {} {\bibfield
  {journal} {\bibinfo  {journal} {SciPost Physics Core}\ }\textbf {\bibinfo
  {volume} {7}},\ \bibinfo {pages} {021} (\bibinfo {year} {2024})}\BibitemShut
  {NoStop}%
\bibitem [{\citenamefont {Gliozzi}\ \emph {et~al.}(2015)\citenamefont
  {Gliozzi}, \citenamefont {Liendo}, \citenamefont {Meineri},\ and\
  \citenamefont {Rago}}]{gliozzi2015boundary}%
  \BibitemOpen
  \bibfield  {author} {\bibinfo {author} {\bibfnamefont {F.}~\bibnamefont
  {Gliozzi}}, \bibinfo {author} {\bibfnamefont {P.}~\bibnamefont {Liendo}},
  \bibinfo {author} {\bibfnamefont {M.}~\bibnamefont {Meineri}},\ and\ \bibinfo
  {author} {\bibfnamefont {A.}~\bibnamefont {Rago}},\ }\bibfield  {title}
  {\bibinfo {title} {{Boundary and Interface CFTs from the Conformal
  Bootstrap}},\ }\href {https://doi.org/10.1007/JHEP05(2015)036} {\bibfield
  {journal} {\bibinfo  {journal} {JHEP}\ }\textbf {\bibinfo {volume} {05}},\
  \bibinfo {pages} {036}},\ \Eprint {https://arxiv.org/abs/1502.07217}
  {arXiv:1502.07217 [hep-th]} \BibitemShut {NoStop}%
\bibitem [{\citenamefont {Krishnan}\ and\ \citenamefont
  {Metlitski}(2023)}]{krishnan2023a}%
  \BibitemOpen
  \bibfield  {author} {\bibinfo {author} {\bibfnamefont {A.}~\bibnamefont
  {Krishnan}}\ and\ \bibinfo {author} {\bibfnamefont {M.~A.}\ \bibnamefont
  {Metlitski}},\ }\bibfield  {title} {\bibinfo {title} {{A plane defect in the
  3d O(N) model}},\ }\href {https://doi.org/10.21468/SciPostPhys.15.3.090}
  {\bibfield  {journal} {\bibinfo  {journal} {SciPost Phys.}\ }\textbf
  {\bibinfo {volume} {15}},\ \bibinfo {pages} {090} (\bibinfo {year} {2023})},\
  \Eprint {https://arxiv.org/abs/2301.05728} {arXiv:2301.05728
  [cond-mat.str-el]} \BibitemShut {NoStop}%
\bibitem [{\citenamefont {Tr\'epanier}(2023)}]{trepanier2023surface}%
  \BibitemOpen
  \bibfield  {author} {\bibinfo {author} {\bibfnamefont {M.}~\bibnamefont
  {Tr\'epanier}},\ }\bibfield  {title} {\bibinfo {title} {{Surface defects in
  the O(N) model}},\ }\href {https://doi.org/10.1007/JHEP09(2023)074}
  {\bibfield  {journal} {\bibinfo  {journal} {JHEP}\ }\textbf {\bibinfo
  {volume} {09}},\ \bibinfo {pages} {074}},\ \Eprint
  {https://arxiv.org/abs/2305.10486} {arXiv:2305.10486 [hep-th]} \BibitemShut
  {NoStop}%
\bibitem [{\citenamefont {Giombi}\ and\ \citenamefont
  {Liu}(2023)}]{giombi2023notes}%
  \BibitemOpen
  \bibfield  {author} {\bibinfo {author} {\bibfnamefont {S.}~\bibnamefont
  {Giombi}}\ and\ \bibinfo {author} {\bibfnamefont {B.}~\bibnamefont {Liu}},\
  }\bibfield  {title} {\bibinfo {title} {{Notes on a surface defect in the O(N)
  model}},\ }\href {https://doi.org/10.1007/JHEP12(2023)004} {\bibfield
  {journal} {\bibinfo  {journal} {JHEP}\ }\textbf {\bibinfo {volume} {12}},\
  \bibinfo {pages} {004}},\ \Eprint {https://arxiv.org/abs/2305.11402}
  {arXiv:2305.11402 [hep-th]} \BibitemShut {NoStop}%
\bibitem [{\citenamefont {Zhou}\ \emph {et~al.}(2023)\citenamefont {Zhou},
  \citenamefont {Gaiotto}, \citenamefont {He},\ and\ \citenamefont
  {Zou}}]{zhou2023g}%
  \BibitemOpen
  \bibfield  {author} {\bibinfo {author} {\bibfnamefont {Z.}~\bibnamefont
  {Zhou}}, \bibinfo {author} {\bibfnamefont {D.}~\bibnamefont {Gaiotto}},
  \bibinfo {author} {\bibfnamefont {Y.-C.}\ \bibnamefont {He}},\ and\ \bibinfo
  {author} {\bibfnamefont {Y.}~\bibnamefont {Zou}},\ }\bibfield  {title}
  {\bibinfo {title} {The $ g $-function and defect changing operators from
  wavefunction overlap on a fuzzy sphere},\ }\href@noop {} {\bibfield
  {journal} {\bibinfo  {journal} {arXiv preprint arXiv:2401.00039}\ } (\bibinfo
  {year} {2023})}\BibitemShut {NoStop}%
\bibitem [{\citenamefont {Cuomo}\ \emph {et~al.}(2024)\citenamefont {Cuomo},
  \citenamefont {He},\ and\ \citenamefont {Komargodski}}]{cuomo2024impurities}%
  \BibitemOpen
  \bibfield  {author} {\bibinfo {author} {\bibfnamefont {G.}~\bibnamefont
  {Cuomo}}, \bibinfo {author} {\bibfnamefont {Y.-C.}\ \bibnamefont {He}},\ and\
  \bibinfo {author} {\bibfnamefont {Z.}~\bibnamefont {Komargodski}},\ }\href
  {https://arxiv.org/abs/2406.10186} {\bibinfo {title} {Impurities with a cusp:
  general theory and 3d ising}} (\bibinfo {year} {2024}),\ \Eprint
  {https://arxiv.org/abs/2406.10186} {arXiv:2406.10186 [hep-th]} \BibitemShut
  {NoStop}%
\bibitem [{\citenamefont {Drukker}\ \emph {et~al.}(1999)\citenamefont
  {Drukker}, \citenamefont {Gross},\ and\ \citenamefont
  {Ooguri}}]{drukker1999wilson}%
  \BibitemOpen
  \bibfield  {author} {\bibinfo {author} {\bibfnamefont {N.}~\bibnamefont
  {Drukker}}, \bibinfo {author} {\bibfnamefont {D.~J.}\ \bibnamefont {Gross}},\
  and\ \bibinfo {author} {\bibfnamefont {H.}~\bibnamefont {Ooguri}},\
  }\bibfield  {title} {\bibinfo {title} {{Wilson loops and minimal surfaces}},\
  }\href {https://doi.org/10.1103/PhysRevD.60.125006} {\bibfield  {journal}
  {\bibinfo  {journal} {Phys. Rev. D}\ }\textbf {\bibinfo {volume} {60}},\
  \bibinfo {pages} {125006} (\bibinfo {year} {1999})},\ \Eprint
  {https://arxiv.org/abs/hep-th/9904191} {arXiv:hep-th/9904191} \BibitemShut
  {NoStop}%
\bibitem [{\citenamefont {Bachas}\ \emph {et~al.}(2002)\citenamefont {Bachas},
  \citenamefont {de~Boer}, \citenamefont {Dijkgraaf},\ and\ \citenamefont
  {Ooguri}}]{bachas2001permeable}%
  \BibitemOpen
  \bibfield  {author} {\bibinfo {author} {\bibfnamefont {C.}~\bibnamefont
  {Bachas}}, \bibinfo {author} {\bibfnamefont {J.}~\bibnamefont {de~Boer}},
  \bibinfo {author} {\bibfnamefont {R.}~\bibnamefont {Dijkgraaf}},\ and\
  \bibinfo {author} {\bibfnamefont {H.}~\bibnamefont {Ooguri}},\ }\bibfield
  {title} {\bibinfo {title} {{Permeable conformal walls and holography}},\
  }\href {https://doi.org/10.1088/1126-6708/2002/06/027} {\bibfield  {journal}
  {\bibinfo  {journal} {JHEP}\ }\textbf {\bibinfo {volume} {06}},\ \bibinfo
  {pages} {027}},\ \Eprint {https://arxiv.org/abs/hep-th/0111210}
  {arXiv:hep-th/0111210} \BibitemShut {NoStop}%
\bibitem [{\citenamefont {Makeenko}\ \emph {et~al.}(2006)\citenamefont
  {Makeenko}, \citenamefont {Olesen},\ and\ \citenamefont
  {Semenoff}}]{makeenko2006cusped}%
  \BibitemOpen
  \bibfield  {author} {\bibinfo {author} {\bibfnamefont {Y.}~\bibnamefont
  {Makeenko}}, \bibinfo {author} {\bibfnamefont {P.}~\bibnamefont {Olesen}},\
  and\ \bibinfo {author} {\bibfnamefont {G.~W.}\ \bibnamefont {Semenoff}},\
  }\bibfield  {title} {\bibinfo {title} {{Cusped SYM Wilson loop at two loops
  and beyond}},\ }\href {https://doi.org/10.1016/j.nuclphysb.2006.05.002}
  {\bibfield  {journal} {\bibinfo  {journal} {Nucl. Phys. B}\ }\textbf
  {\bibinfo {volume} {748}},\ \bibinfo {pages} {170} (\bibinfo {year}
  {2006})},\ \Eprint {https://arxiv.org/abs/hep-th/0602100}
  {arXiv:hep-th/0602100} \BibitemShut {NoStop}%
\bibitem [{\citenamefont {Correa}\ \emph {et~al.}(2012)\citenamefont {Correa},
  \citenamefont {Maldacena},\ and\ \citenamefont {Sever}}]{correa2012the}%
  \BibitemOpen
  \bibfield  {author} {\bibinfo {author} {\bibfnamefont {D.}~\bibnamefont
  {Correa}}, \bibinfo {author} {\bibfnamefont {J.}~\bibnamefont {Maldacena}},\
  and\ \bibinfo {author} {\bibfnamefont {A.}~\bibnamefont {Sever}},\ }\bibfield
   {title} {\bibinfo {title} {{The quark anti-quark potential and the cusp
  anomalous dimension from a TBA equation}},\ }\href
  {https://doi.org/10.1007/JHEP08(2012)134} {\bibfield  {journal} {\bibinfo
  {journal} {JHEP}\ }\textbf {\bibinfo {volume} {08}},\ \bibinfo {pages}
  {134}},\ \Eprint {https://arxiv.org/abs/1203.1913} {arXiv:1203.1913 [hep-th]}
  \BibitemShut {NoStop}%
\bibitem [{\citenamefont {Gromov}\ and\ \citenamefont
  {Levkovich-Maslyuk}(2016)}]{gromov2015quantum}%
  \BibitemOpen
  \bibfield  {author} {\bibinfo {author} {\bibfnamefont {N.}~\bibnamefont
  {Gromov}}\ and\ \bibinfo {author} {\bibfnamefont {F.}~\bibnamefont
  {Levkovich-Maslyuk}},\ }\bibfield  {title} {\bibinfo {title} {{Quantum
  Spectral Curve for a cusped Wilson line in $ \mathcal{N}=4 $ SYM}},\ }\href
  {https://doi.org/10.1007/JHEP04(2016)134} {\bibfield  {journal} {\bibinfo
  {journal} {JHEP}\ }\textbf {\bibinfo {volume} {04}},\ \bibinfo {pages}
  {134}},\ \Eprint {https://arxiv.org/abs/1510.02098} {arXiv:1510.02098
  [hep-th]} \BibitemShut {NoStop}%
\bibitem [{\citenamefont {Grozin}\ \emph {et~al.}(2016)\citenamefont {Grozin},
  \citenamefont {Henn}, \citenamefont {Korchemsky},\ and\ \citenamefont
  {Marquard}}]{grozin2015the}%
  \BibitemOpen
  \bibfield  {author} {\bibinfo {author} {\bibfnamefont {A.}~\bibnamefont
  {Grozin}}, \bibinfo {author} {\bibfnamefont {J.~M.}\ \bibnamefont {Henn}},
  \bibinfo {author} {\bibfnamefont {G.~P.}\ \bibnamefont {Korchemsky}},\ and\
  \bibinfo {author} {\bibfnamefont {P.}~\bibnamefont {Marquard}},\ }\bibfield
  {title} {\bibinfo {title} {{The three-loop cusp anomalous dimension in QCD
  and its supersymmetric extensions}},\ }\href
  {https://doi.org/10.1007/JHEP01(2016)140} {\bibfield  {journal} {\bibinfo
  {journal} {JHEP}\ }\textbf {\bibinfo {volume} {01}},\ \bibinfo {pages}
  {140}},\ \Eprint {https://arxiv.org/abs/1510.07803} {arXiv:1510.07803
  [hep-ph]} \BibitemShut {NoStop}%
\bibitem [{\citenamefont {Kravchuk}\ \emph {et~al.}(2024)\citenamefont
  {Kravchuk}, \citenamefont {Radcliffe},\ and\ \citenamefont
  {Sinha}}]{kravchuk2024effective}%
  \BibitemOpen
  \bibfield  {author} {\bibinfo {author} {\bibfnamefont {P.}~\bibnamefont
  {Kravchuk}}, \bibinfo {author} {\bibfnamefont {A.}~\bibnamefont
  {Radcliffe}},\ and\ \bibinfo {author} {\bibfnamefont {R.}~\bibnamefont
  {Sinha}},\ }\bibfield  {title} {\bibinfo {title} {Effective theory for fusion
  of conformal defects},\ }\href@noop {} {\bibfield  {journal} {\bibinfo
  {journal} {arXiv preprint arXiv:2406.04561}\ } (\bibinfo {year}
  {2024})}\BibitemShut {NoStop}%
\bibitem [{\citenamefont {Cardy}(1984)}]{cardy1984conformal}%
  \BibitemOpen
  \bibfield  {author} {\bibinfo {author} {\bibfnamefont {J.~L.}\ \bibnamefont
  {Cardy}},\ }\bibfield  {title} {\bibinfo {title} {{Conformal Invariance and
  Surface Critical Behavior}},\ }\href
  {https://doi.org/10.1016/0550-3213(84)90241-4} {\bibfield  {journal}
  {\bibinfo  {journal} {Nucl. Phys. B}\ }\textbf {\bibinfo {volume} {240}},\
  \bibinfo {pages} {514} (\bibinfo {year} {1984})}\BibitemShut {NoStop}%
\bibitem [{\citenamefont {Cardy}(1989)}]{cardy1989boundary}%
  \BibitemOpen
  \bibfield  {author} {\bibinfo {author} {\bibfnamefont {J.~L.}\ \bibnamefont
  {Cardy}},\ }\bibfield  {title} {\bibinfo {title} {{Boundary Conditions,
  Fusion Rules and the Verlinde Formula}},\ }\href
  {https://doi.org/10.1016/0550-3213(89)90521-X} {\bibfield  {journal}
  {\bibinfo  {journal} {Nucl. Phys. B}\ }\textbf {\bibinfo {volume} {324}},\
  \bibinfo {pages} {581} (\bibinfo {year} {1989})}\BibitemShut {NoStop}%
\bibitem [{\citenamefont {Maldacena}(1998)}]{maldacena1997the}%
  \BibitemOpen
  \bibfield  {author} {\bibinfo {author} {\bibfnamefont {J.~M.}\ \bibnamefont
  {Maldacena}},\ }\bibfield  {title} {\bibinfo {title} {{The Large N limit of
  superconformal field theories and supergravity}},\ }\href
  {https://doi.org/10.4310/ATMP.1998.v2.n2.a1} {\bibfield  {journal} {\bibinfo
  {journal} {Adv. Theor. Math. Phys.}\ }\textbf {\bibinfo {volume} {2}},\
  \bibinfo {pages} {231} (\bibinfo {year} {1998})},\ \Eprint
  {https://arxiv.org/abs/hep-th/9711200} {arXiv:hep-th/9711200} \BibitemShut
  {NoStop}%
\bibitem [{\citenamefont {Witten}(1998)}]{witten1998anti}%
  \BibitemOpen
  \bibfield  {author} {\bibinfo {author} {\bibfnamefont {E.}~\bibnamefont
  {Witten}},\ }\bibfield  {title} {\bibinfo {title} {{Anti-de Sitter space and
  holography}},\ }\href {https://doi.org/10.4310/ATMP.1998.v2.n2.a2} {\bibfield
   {journal} {\bibinfo  {journal} {Adv. Theor. Math. Phys.}\ }\textbf {\bibinfo
  {volume} {2}},\ \bibinfo {pages} {253} (\bibinfo {year} {1998})},\ \Eprint
  {https://arxiv.org/abs/hep-th/9802150} {arXiv:hep-th/9802150} \BibitemShut
  {NoStop}%
\bibitem [{\citenamefont {Gubser}\ \emph {et~al.}(1998)\citenamefont {Gubser},
  \citenamefont {Klebanov},\ and\ \citenamefont {Polyakov}}]{gubser1998gauge}%
  \BibitemOpen
  \bibfield  {author} {\bibinfo {author} {\bibfnamefont {S.~S.}\ \bibnamefont
  {Gubser}}, \bibinfo {author} {\bibfnamefont {I.~R.}\ \bibnamefont
  {Klebanov}},\ and\ \bibinfo {author} {\bibfnamefont {A.~M.}\ \bibnamefont
  {Polyakov}},\ }\bibfield  {title} {\bibinfo {title} {{Gauge theory
  correlators from noncritical string theory}},\ }\href
  {https://doi.org/10.1016/S0370-2693(98)00377-3} {\bibfield  {journal}
  {\bibinfo  {journal} {Phys. Lett. B}\ }\textbf {\bibinfo {volume} {428}},\
  \bibinfo {pages} {105} (\bibinfo {year} {1998})},\ \Eprint
  {https://arxiv.org/abs/hep-th/9802109} {arXiv:hep-th/9802109} \BibitemShut
  {NoStop}%
\bibitem [{\citenamefont {Karch}\ and\ \citenamefont
  {Randall}(2001{\natexlab{a}})}]{karch2000locally}%
  \BibitemOpen
  \bibfield  {author} {\bibinfo {author} {\bibfnamefont {A.}~\bibnamefont
  {Karch}}\ and\ \bibinfo {author} {\bibfnamefont {L.}~\bibnamefont
  {Randall}},\ }\bibfield  {title} {\bibinfo {title} {{Locally localized
  gravity}},\ }\href {https://doi.org/10.1088/1126-6708/2001/05/008} {\bibfield
   {journal} {\bibinfo  {journal} {JHEP}\ }\textbf {\bibinfo {volume} {05}},\
  \bibinfo {pages} {008}},\ \Eprint {https://arxiv.org/abs/hep-th/0011156}
  {arXiv:hep-th/0011156} \BibitemShut {NoStop}%
\bibitem [{\citenamefont {Karch}\ and\ \citenamefont
  {Randall}(2001{\natexlab{b}})}]{karch2000open}%
  \BibitemOpen
  \bibfield  {author} {\bibinfo {author} {\bibfnamefont {A.}~\bibnamefont
  {Karch}}\ and\ \bibinfo {author} {\bibfnamefont {L.}~\bibnamefont
  {Randall}},\ }\bibfield  {title} {\bibinfo {title} {{Open and closed string
  interpretation of SUSY CFT's on branes with boundaries}},\ }\href
  {https://doi.org/10.1088/1126-6708/2001/06/063} {\bibfield  {journal}
  {\bibinfo  {journal} {JHEP}\ }\textbf {\bibinfo {volume} {06}},\ \bibinfo
  {pages} {063}},\ \Eprint {https://arxiv.org/abs/hep-th/0105132}
  {arXiv:hep-th/0105132} \BibitemShut {NoStop}%
\bibitem [{\citenamefont {Takayanagi}(2011)}]{takayanagi2011holographic}%
  \BibitemOpen
  \bibfield  {author} {\bibinfo {author} {\bibfnamefont {T.}~\bibnamefont
  {Takayanagi}},\ }\bibfield  {title} {\bibinfo {title} {{Holographic Dual of
  BCFT}},\ }\href {https://doi.org/10.1103/PhysRevLett.107.101602} {\bibfield
  {journal} {\bibinfo  {journal} {Phys. Rev. Lett.}\ }\textbf {\bibinfo
  {volume} {107}},\ \bibinfo {pages} {101602} (\bibinfo {year} {2011})},\
  \Eprint {https://arxiv.org/abs/1105.5165} {arXiv:1105.5165 [hep-th]}
  \BibitemShut {NoStop}%
\bibitem [{\citenamefont {Fujita}\ \emph {et~al.}(2011)\citenamefont {Fujita},
  \citenamefont {Takayanagi},\ and\ \citenamefont {Tonni}}]{fujita2011aspect}%
  \BibitemOpen
  \bibfield  {author} {\bibinfo {author} {\bibfnamefont {M.}~\bibnamefont
  {Fujita}}, \bibinfo {author} {\bibfnamefont {T.}~\bibnamefont {Takayanagi}},\
  and\ \bibinfo {author} {\bibfnamefont {E.}~\bibnamefont {Tonni}},\ }\bibfield
   {title} {\bibinfo {title} {{Aspects of AdS/BCFT}},\ }\href
  {https://doi.org/10.1007/JHEP11(2011)043} {\bibfield  {journal} {\bibinfo
  {journal} {JHEP}\ }\textbf {\bibinfo {volume} {11}},\ \bibinfo {pages}
  {043}},\ \Eprint {https://arxiv.org/abs/1108.5152} {arXiv:1108.5152 [hep-th]}
  \BibitemShut {NoStop}%
\bibitem [{\citenamefont {DeWolfe}\ \emph {et~al.}(2002)\citenamefont
  {DeWolfe}, \citenamefont {Freedman},\ and\ \citenamefont
  {Ooguri}}]{dewolfe2001holography}%
  \BibitemOpen
  \bibfield  {author} {\bibinfo {author} {\bibfnamefont {O.}~\bibnamefont
  {DeWolfe}}, \bibinfo {author} {\bibfnamefont {D.~Z.}\ \bibnamefont
  {Freedman}},\ and\ \bibinfo {author} {\bibfnamefont {H.}~\bibnamefont
  {Ooguri}},\ }\bibfield  {title} {\bibinfo {title} {{Holography and defect
  conformal field theories}},\ }\href
  {https://doi.org/10.1103/PhysRevD.66.025009} {\bibfield  {journal} {\bibinfo
  {journal} {Phys. Rev. D}\ }\textbf {\bibinfo {volume} {66}},\ \bibinfo
  {pages} {025009} (\bibinfo {year} {2002})},\ \Eprint
  {https://arxiv.org/abs/hep-th/0111135} {arXiv:hep-th/0111135} \BibitemShut
  {NoStop}%
\bibitem [{\citenamefont {Azeyanagi}\ \emph {et~al.}(2008)\citenamefont
  {Azeyanagi}, \citenamefont {Karch}, \citenamefont {Takayanagi},\ and\
  \citenamefont {Thompson}}]{azeyanagi2007holographic}%
  \BibitemOpen
  \bibfield  {author} {\bibinfo {author} {\bibfnamefont {T.}~\bibnamefont
  {Azeyanagi}}, \bibinfo {author} {\bibfnamefont {A.}~\bibnamefont {Karch}},
  \bibinfo {author} {\bibfnamefont {T.}~\bibnamefont {Takayanagi}},\ and\
  \bibinfo {author} {\bibfnamefont {E.~G.}\ \bibnamefont {Thompson}},\
  }\bibfield  {title} {\bibinfo {title} {{Holographic calculation of boundary
  entropy}},\ }\href {https://doi.org/10.1088/1126-6708/2008/03/054} {\bibfield
   {journal} {\bibinfo  {journal} {JHEP}\ }\textbf {\bibinfo {volume} {03}},\
  \bibinfo {pages} {054}},\ \Eprint {https://arxiv.org/abs/0712.1850}
  {arXiv:0712.1850 [hep-th]} \BibitemShut {NoStop}%
\bibitem [{\citenamefont {Erdmenger}\ \emph {et~al.}(2015)\citenamefont
  {Erdmenger}, \citenamefont {Flory},\ and\ \citenamefont
  {Newrzella}}]{erdmenger2014bending}%
  \BibitemOpen
  \bibfield  {author} {\bibinfo {author} {\bibfnamefont {J.}~\bibnamefont
  {Erdmenger}}, \bibinfo {author} {\bibfnamefont {M.}~\bibnamefont {Flory}},\
  and\ \bibinfo {author} {\bibfnamefont {M.-N.}\ \bibnamefont {Newrzella}},\
  }\bibfield  {title} {\bibinfo {title} {{Bending branes for DCFT in two
  dimensions}},\ }\href {https://doi.org/10.1007/JHEP01(2015)058} {\bibfield
  {journal} {\bibinfo  {journal} {JHEP}\ }\textbf {\bibinfo {volume} {01}},\
  \bibinfo {pages} {058}},\ \Eprint {https://arxiv.org/abs/1410.7811}
  {arXiv:1410.7811 [hep-th]} \BibitemShut {NoStop}%
\bibitem [{\citenamefont {Bachas}\ \emph {et~al.}(2020)\citenamefont {Bachas},
  \citenamefont {Chapman}, \citenamefont {Ge},\ and\ \citenamefont
  {Policastro}}]{bachas2020energy}%
  \BibitemOpen
  \bibfield  {author} {\bibinfo {author} {\bibfnamefont {C.}~\bibnamefont
  {Bachas}}, \bibinfo {author} {\bibfnamefont {S.}~\bibnamefont {Chapman}},
  \bibinfo {author} {\bibfnamefont {D.}~\bibnamefont {Ge}},\ and\ \bibinfo
  {author} {\bibfnamefont {G.}~\bibnamefont {Policastro}},\ }\bibfield  {title}
  {\bibinfo {title} {Energy reflection and transmission at 2d holographic
  interfaces},\ }\href {https://doi.org/10.1103/PhysRevLett.125.231602}
  {\bibfield  {journal} {\bibinfo  {journal} {Phys. Rev. Lett.}\ }\textbf
  {\bibinfo {volume} {125}},\ \bibinfo {pages} {231602} (\bibinfo {year}
  {2020})}\BibitemShut {NoStop}%
\bibitem [{\citenamefont {Simidzija}\ and\ \citenamefont
  {Van~Raamsdonk}(2020)}]{simidzija2020holoween}%
  \BibitemOpen
  \bibfield  {author} {\bibinfo {author} {\bibfnamefont {P.}~\bibnamefont
  {Simidzija}}\ and\ \bibinfo {author} {\bibfnamefont {M.}~\bibnamefont
  {Van~Raamsdonk}},\ }\bibfield  {title} {\bibinfo {title} {{Holo-ween}},\
  }\href {https://doi.org/10.1007/JHEP12(2020)028} {\bibfield  {journal}
  {\bibinfo  {journal} {JHEP}\ }\textbf {\bibinfo {volume} {12}},\ \bibinfo
  {pages} {028}},\ \Eprint {https://arxiv.org/abs/2006.13943} {arXiv:2006.13943
  [hep-th]} \BibitemShut {NoStop}%
\bibitem [{\citenamefont {Karch}\ \emph {et~al.}(2021)\citenamefont {Karch},
  \citenamefont {Luo},\ and\ \citenamefont {Sun}}]{karch2021universal}%
  \BibitemOpen
  \bibfield  {author} {\bibinfo {author} {\bibfnamefont {A.}~\bibnamefont
  {Karch}}, \bibinfo {author} {\bibfnamefont {Z.-X.}\ \bibnamefont {Luo}},\
  and\ \bibinfo {author} {\bibfnamefont {H.-Y.}\ \bibnamefont {Sun}},\
  }\bibfield  {title} {\bibinfo {title} {{Universal relations for holographic
  interfaces}},\ }\href {https://doi.org/10.1007/JHEP09(2021)172} {\bibfield
  {journal} {\bibinfo  {journal} {JHEP}\ }\textbf {\bibinfo {volume} {09}},\
  \bibinfo {pages} {172}},\ \Eprint {https://arxiv.org/abs/2107.02165}
  {arXiv:2107.02165 [hep-th]} \BibitemShut {NoStop}%
\bibitem [{\citenamefont {Bachas}\ and\ \citenamefont
  {Papadopoulos}(2021)}]{bachas2021phases}%
  \BibitemOpen
  \bibfield  {author} {\bibinfo {author} {\bibfnamefont {C.}~\bibnamefont
  {Bachas}}\ and\ \bibinfo {author} {\bibfnamefont {V.}~\bibnamefont
  {Papadopoulos}},\ }\bibfield  {title} {\bibinfo {title} {{Phases of
  Holographic Interfaces}},\ }\href {https://doi.org/10.1007/JHEP04(2021)262}
  {\bibfield  {journal} {\bibinfo  {journal} {JHEP}\ }\textbf {\bibinfo
  {volume} {04}},\ \bibinfo {pages} {262}},\ \Eprint
  {https://arxiv.org/abs/2101.12529} {arXiv:2101.12529 [hep-th]} \BibitemShut
  {NoStop}%
\bibitem [{\citenamefont {Miyaji}\ and\ \citenamefont
  {Murdia}(2022)}]{miyaji2022holographic}%
  \BibitemOpen
  \bibfield  {author} {\bibinfo {author} {\bibfnamefont {M.}~\bibnamefont
  {Miyaji}}\ and\ \bibinfo {author} {\bibfnamefont {C.}~\bibnamefont
  {Murdia}},\ }\bibfield  {title} {\bibinfo {title} {{Holographic BCFT with a
  Defect on the End-of-the-World brane}},\ }\href
  {https://doi.org/10.1007/JHEP11(2022)123} {\bibfield  {journal} {\bibinfo
  {journal} {JHEP}\ }\textbf {\bibinfo {volume} {11}},\ \bibinfo {pages}
  {123}},\ \Eprint {https://arxiv.org/abs/2208.13783} {arXiv:2208.13783
  [hep-th]} \BibitemShut {NoStop}%
\bibitem [{\citenamefont {Karch}\ and\ \citenamefont
  {Wang}(2023)}]{karch2022universal}%
  \BibitemOpen
  \bibfield  {author} {\bibinfo {author} {\bibfnamefont {A.}~\bibnamefont
  {Karch}}\ and\ \bibinfo {author} {\bibfnamefont {M.}~\bibnamefont {Wang}},\
  }\bibfield  {title} {\bibinfo {title} {{Universal behavior of entanglement
  entropies in interface CFTs from general holographic spacetimes}},\ }\href
  {https://doi.org/10.1007/JHEP06(2023)145} {\bibfield  {journal} {\bibinfo
  {journal} {JHEP}\ }\textbf {\bibinfo {volume} {06}},\ \bibinfo {pages}
  {145}},\ \Eprint {https://arxiv.org/abs/2211.09148} {arXiv:2211.09148
  [hep-th]} \BibitemShut {NoStop}%
\bibitem [{\citenamefont {Karch}\ \emph {et~al.}(2023)\citenamefont {Karch},
  \citenamefont {Kusuki}, \citenamefont {Ooguri}, \citenamefont {Sun},\ and\
  \citenamefont {Wang}}]{karch2023universal}%
  \BibitemOpen
  \bibfield  {author} {\bibinfo {author} {\bibfnamefont {A.}~\bibnamefont
  {Karch}}, \bibinfo {author} {\bibfnamefont {Y.}~\bibnamefont {Kusuki}},
  \bibinfo {author} {\bibfnamefont {H.}~\bibnamefont {Ooguri}}, \bibinfo
  {author} {\bibfnamefont {H.-Y.}\ \bibnamefont {Sun}},\ and\ \bibinfo {author}
  {\bibfnamefont {M.}~\bibnamefont {Wang}},\ }\bibfield  {title} {\bibinfo
  {title} {{Universality of effective central charge in interface CFTs}},\
  }\href {https://doi.org/10.1007/JHEP11(2023)126} {\bibfield  {journal}
  {\bibinfo  {journal} {JHEP}\ }\textbf {\bibinfo {volume} {11}},\ \bibinfo
  {pages} {126}},\ \Eprint {https://arxiv.org/abs/2308.05436} {arXiv:2308.05436
  [hep-th]} \BibitemShut {NoStop}%
\bibitem [{\citenamefont {Tang}\ \emph {et~al.}(2024)\citenamefont {Tang},
  \citenamefont {Wei}, \citenamefont {Tang}, \citenamefont {Wen},\ and\
  \citenamefont {Zhu}}]{tang2023universal}%
  \BibitemOpen
  \bibfield  {author} {\bibinfo {author} {\bibfnamefont {Q.}~\bibnamefont
  {Tang}}, \bibinfo {author} {\bibfnamefont {Z.}~\bibnamefont {Wei}}, \bibinfo
  {author} {\bibfnamefont {Y.}~\bibnamefont {Tang}}, \bibinfo {author}
  {\bibfnamefont {X.}~\bibnamefont {Wen}},\ and\ \bibinfo {author}
  {\bibfnamefont {W.}~\bibnamefont {Zhu}},\ }\bibfield  {title} {\bibinfo
  {title} {{Universal entanglement signatures of interface conformal field
  theories}},\ }\href {https://doi.org/10.1103/PhysRevB.109.L041104} {\bibfield
   {journal} {\bibinfo  {journal} {Phys. Rev. B}\ }\textbf {\bibinfo {volume}
  {109}},\ \bibinfo {pages} {L041104} (\bibinfo {year} {2024})},\ \Eprint
  {https://arxiv.org/abs/2308.03646} {arXiv:2308.03646 [cond-mat.stat-mech]}
  \BibitemShut {NoStop}%
\bibitem [{Note1()}]{Note1}%
  \BibitemOpen
  \bibinfo {note} {Note that $\sigma \rightarrow \infty $ approaches the
  intersection between the interface brane and the asymptotic
  boundary}\BibitemShut {NoStop}%
\bibitem [{\citenamefont {Bachas}(2007)}]{bachas2006comment}%
  \BibitemOpen
  \bibfield  {author} {\bibinfo {author} {\bibfnamefont {C.~P.}\ \bibnamefont
  {Bachas}},\ }\bibfield  {title} {\bibinfo {title} {{Comment on the sign of
  the Casimir force}},\ }\href {https://doi.org/10.1088/1751-8113/40/30/028}
  {\bibfield  {journal} {\bibinfo  {journal} {J. Phys. A}\ }\textbf {\bibinfo
  {volume} {40}},\ \bibinfo {pages} {9089} (\bibinfo {year} {2007})},\ \Eprint
  {https://arxiv.org/abs/quant-ph/0611082} {arXiv:quant-ph/0611082}
  \BibitemShut {NoStop}%
\bibitem [{\citenamefont {Diatlyk}\ \emph {et~al.}(2024)\citenamefont
  {Diatlyk}, \citenamefont {Khanchandani}, \citenamefont {Popov},\ and\
  \citenamefont {Wang}}]{diatlyk2024effective}%
  \BibitemOpen
  \bibfield  {author} {\bibinfo {author} {\bibfnamefont {O.}~\bibnamefont
  {Diatlyk}}, \bibinfo {author} {\bibfnamefont {H.}~\bibnamefont
  {Khanchandani}}, \bibinfo {author} {\bibfnamefont {F.~K.}\ \bibnamefont
  {Popov}},\ and\ \bibinfo {author} {\bibfnamefont {Y.}~\bibnamefont {Wang}},\
  }\bibfield  {title} {\bibinfo {title} {Effective field theory of conformal
  boundaries},\ }\href@noop {} {\bibfield  {journal} {\bibinfo  {journal}
  {arXiv preprint arXiv:2406.01550}\ } (\bibinfo {year} {2024})}\BibitemShut
  {NoStop}%
\bibitem [{\citenamefont {Estienne}\ \emph {et~al.}(2022)\citenamefont
  {Estienne}, \citenamefont {Stéphan},\ and\ \citenamefont
  {Witczak-Krempa}}]{Estienne_2022}%
  \BibitemOpen
  \bibfield  {author} {\bibinfo {author} {\bibfnamefont {B.}~\bibnamefont
  {Estienne}}, \bibinfo {author} {\bibfnamefont {J.-M.}\ \bibnamefont
  {Stéphan}},\ and\ \bibinfo {author} {\bibfnamefont {W.}~\bibnamefont
  {Witczak-Krempa}},\ }\bibfield  {title} {\bibinfo {title} {Cornering the
  universal shape of fluctuations},\ }\bibfield  {journal} {\bibinfo  {journal}
  {Nature Communications}\ }\textbf {\bibinfo {volume} {13}},\ \href
  {https://doi.org/10.1038/s41467-021-27727-1} {10.1038/s41467-021-27727-1}
  (\bibinfo {year} {2022})\BibitemShut {NoStop}%
\bibitem [{\citenamefont {Anous}\ \emph {et~al.}(2022)\citenamefont {Anous},
  \citenamefont {Meineri}, \citenamefont {Pelliconi},\ and\ \citenamefont
  {Sonner}}]{Anous_2022}%
  \BibitemOpen
  \bibfield  {author} {\bibinfo {author} {\bibfnamefont {T.}~\bibnamefont
  {Anous}}, \bibinfo {author} {\bibfnamefont {M.}~\bibnamefont {Meineri}},
  \bibinfo {author} {\bibfnamefont {P.}~\bibnamefont {Pelliconi}},\ and\
  \bibinfo {author} {\bibfnamefont {J.}~\bibnamefont {Sonner}},\ }\bibfield
  {title} {\bibinfo {title} {Sailing past the end of the world and discovering
  the island},\ }\bibfield  {journal} {\bibinfo  {journal} {SciPost Physics}\
  }\textbf {\bibinfo {volume} {13}},\ \href
  {https://doi.org/10.21468/scipostphys.13.3.075}
  {10.21468/scipostphys.13.3.075} (\bibinfo {year} {2022})\BibitemShut
  {NoStop}%
\bibitem [{\citenamefont {Sun}\ and\ \citenamefont {Jian}(2023)}]{Sun_2023}%
  \BibitemOpen
  \bibfield  {author} {\bibinfo {author} {\bibfnamefont {X.}~\bibnamefont
  {Sun}}\ and\ \bibinfo {author} {\bibfnamefont {S.-K.}\ \bibnamefont {Jian}},\
  }\bibfield  {title} {\bibinfo {title} {Holographic weak measurement},\
  }\bibfield  {journal} {\bibinfo  {journal} {Journal of High Energy Physics}\
  }\textbf {\bibinfo {volume} {2023}},\ \href
  {https://doi.org/10.1007/jhep12(2023)157} {10.1007/jhep12(2023)157} (\bibinfo
  {year} {2023})\BibitemShut {NoStop}%
\bibitem [{\citenamefont {Kusuki}(2022)}]{PhysRevD.106.066020}%
  \BibitemOpen
  \bibfield  {author} {\bibinfo {author} {\bibfnamefont {Y.}~\bibnamefont
  {Kusuki}},\ }\bibfield  {title} {\bibinfo {title} {Semiclassical gravity from
  averaged boundaries in two-dimensional boundary conformal field theories},\
  }\href {https://doi.org/10.1103/PhysRevD.106.066020} {\bibfield  {journal}
  {\bibinfo  {journal} {Phys. Rev. D}\ }\textbf {\bibinfo {volume} {106}},\
  \bibinfo {pages} {066020} (\bibinfo {year} {2022})}\BibitemShut {NoStop}%
\end{thebibliography}%

\onecolumngrid

\setcounter{secnumdepth}{3}
\setcounter{equation}{0}
\setcounter{figure}{0}
\renewcommand{\theequation}{S\arabic{equation}}
\renewcommand{\thefigure}{S\arabic{figure}}
\renewcommand\figurename{Supplementary Figure}
\renewcommand\tablename{Supplementary Table}
\newcommand\Scite[1]{[S\citealp{#1}]}
\makeatletter \renewcommand\@biblabel[1]{[S#1]} \makeatother

\newpage
\section*{Supplemental Material}

\subsection{General AdS space with defects}
\label{sec:General AdS space with defects}

In this section, we construct the holographic dual to the defect CFT described in the main text, and outline a general procedure for solving the geometry.
To have two distinct interface branes, a corner term at their intersection is necessary.
Collecting all terms~\cite{bachas2021phases,miyaji2022holographic} with $8\pi G_N=1$, the Euclidean action reads 
\begin{subequations}
\label{eq:total action}
\begin{equation}
    I_{\rm tot}=I_{\rm EH}+I_T+I_{\rm surface}+I_{\rm corner}+I_{\rm c.t.},
\end{equation}
where
\begin{equation}
    I_{\rm EH}=-\frac{1}{2}\sum_{i=1,2}\int_{M_i}\sqrt{g_i}\left(R_i+\frac{2}{l_i^2}\right),
\end{equation}
\begin{equation}
    I_T=\sum_{\alpha=a,b}\int_{W_\alpha}\sqrt{\hat{g}_\alpha}\ T_\alpha,
\end{equation}
\begin{equation}
    I_{\rm surface}=-\sum_{i=1,2}\int_{\partial M_i}\sqrt{\hat{g}_i}\ K_i,
\end{equation}
\begin{equation}
    I_{\rm corner}=\sum_{\alpha=a,b}\int_{B_1\cap B_2}\sqrt{\hat{g}}\left(\theta_\alpha-\pi\right)-\sum_{i=1,2}\int_{S_a^i\cap S_b^i}\sqrt{\hat{g}}\left(\theta_0^i-\theta^i\right),
\end{equation}
\begin{equation}
    I_{\rm c.t.}=\sum_{i=1,2}\frac{1}{l_i}\int_{B_i}\sqrt{\hat{g}_i}-\int_{B_1\cap B_2}\sqrt{\hat{g}}\left(\theta_1+\theta_2\right),
\end{equation}
\end{subequations}
where $\alpha=a,b$ labels different interface branes between two AdS space labeled by $i=1,2$.
Here the boundary of $M_i$ is $\partial M_i=S_a^i+S_b^i+B_i$, where $S_{a,b}^i$ are two branes of AdS bulk $M_i$, and $B_i$ is its asymptotic boundary.
We also denote the interface brane as $W_{a,b}$ with corresponding tension $T_{a,b}$. 
Besides, we specify the direction of each surface that the direction of extrinsic curvatures, $K_{1,2}$, points outside the AdS bulk. 
There are also codimension-2 terms in the action.
$B_1\cap B_2$ corresponds to the intersection of asymptotic boundaries, with $\theta_\alpha$ being the corresponding angle.
$S_a^i\cap S_b^i$ corresponds to the corner with an angle $\theta^i$ in the intersection of the two branes labeled by $i=1,2$.
The counter terms are included for completeness~\cite{bachas2021phases}.

With the saddle point approximation, we have two equations of motion (e.o.m.) for the branes and the corner on them. 
\begin{subequations}
\label{eq:eom for brane and defect}
\begin{equation}
\label{eq:eom for brane and defect a}
    \left(K^1+K^2\right)_{\mu\nu}=T_\alpha h_{\mu\nu},
\end{equation}
\begin{equation}
\label{eq:eom for brane and defect b}
    \theta^i=\theta_{0}^i,
\end{equation}
\end{subequations}
where $\alpha=a,b$ for two branes which are dual to different defects and have different tension $T_\alpha$, and the e.o.m. from the corner term requires two branes to form an angle $\theta^i_0$, which may deviate from $\pi$ corresponding to a nontrivial corner on branes. 
We will use the first equation to solve the brane trajectory with undetermined parameters of the AdS spacetime, and then the second equation \eqref{eq:eom for brane and defect b} will become a constraint for the parameters.

In the following, we use a similar convention of Ref.~\cite{bachas2021phases}.
The metrics of the two AdS spacetime are
\begin{equation}
\label{eq:metric for two AdS}
    {\rm d}s^2=(r_i^2-M_i l_i^2){\rm d}\tau^2+\frac{l_i^2{\rm d}r_i^2}{r_i^2-M_i l_i^2}+r_i^2{\rm d}x_i^2,
\end{equation}
where $(\tau,r_i,x_i)$ are coordinates for two AdS spacetime labeled by $i=1,2$, and $M_i<0$ corresponds to the cold phase in Ref.~\cite{bachas2021phases}. 
We denote $\tau_1=\tau_2=\tau$ by identifying the time for two AdS spaces. 
With the parameter $\sigma$ defined on the brane, we have brane coordinates $(r_i=r_i(\sigma),x_i=x_i(\sigma))$ on a constant time slice.
Because of the continuity of metrics on the brane, ${\rm d}s^2=f(\sigma){\rm d}\tau^2+g(\sigma){\rm d}\sigma^2$, we have
\begin{subequations}
\label{eq:continuity condition for metrics}
\begin{equation}
\label{eq:continuity condition for metrics a}
    r_1^2-M_1 l_1^2=r_2^2-M_2 l_2^2\equiv f(\sigma),
\end{equation}
\begin{equation}
\label{eq:continuity condition for metrics b}
    \frac{l_1^2 \dot{r}_1^2}{r_1^2-M_1 l_1^2}+r_1^2\dot{x}_1^2=\frac{l_2^2 \dot{r}_2^2}{r_2^2-M_2 l_2^2}+r_2^2\dot{x}_2^2\equiv g(\sigma),
\end{equation}
\end{subequations}
Then, simplifying the junction condition \eqref{eq:eom for brane and defect a} leads to
\begin{equation}
\label{eq:simplified eom b}
    \frac{\dot{x}_1 r_1^2}{l_1}+\frac{\dot{x}_2 r_2^2}{l_2}=-T_\alpha\sqrt{fg}.
\end{equation}
We exploit the freedom to choose a proper coordinate $\sigma$: $\sigma=f(\sigma)$,  which leads to
\begin{equation}
\label{eq:relation od sigma and coordinate}
    r_i=\sqrt{\sigma+M_i l_i^2}.
\end{equation}
With \eqref{eq:continuity condition for metrics}, \eqref{eq:simplified eom b} and \eqref{eq:relation od sigma and coordinate}, we can solve the brane trajectory $x_i=x_i(\sigma)$.
In the following, we drop the label $\alpha$ for simplicity, and discuss the general property of the brane trajectory.
Later, we will use the constraint from \eqref{eq:eom for brane and defect b} to solve the full geometry in which the label $\alpha$ will be restored. 
We simplify the function~\cite{bachas2021phases} $g(\sigma)$ 
\begin{equation}
\label{eq:simplify g}
    g(\sigma)=T^2\left[4\frac{r_1^2 r_2^2}{l_1^2 l_2^2}-\left(T^2\sigma-\frac{r_1^2}{l_1^2}-\frac{r_2^2}{l_2^2}\right)^2\right]^{-1}\equiv\frac{T^2}{A\sigma^2+2B\sigma+C},
\end{equation}
where
\begin{subequations}
\label{eq:definition of ABC}
\begin{equation}
    A=\frac{4}{l_1^2 l_2^2}-\left(T^2-\frac{1}{l_1^2}-\frac{1}{l_2^2}\right)^2=(T^2-T_{\rm min}^2)(T_{\rm max}^2-T^2),
\end{equation}
\begin{equation}
    B=\frac{2}{l_1^2 l_2^2}\left(M_1 l_1^2+M_2 l_2^2\right)+\left(M_1+M_2\right)\left(T^2-\frac{1}{l_1^2}-\frac{1}{l_2^2}\right)=\left(M_1+M_2\right)T^2-\left(M_1-M_2\right)T_0^2,
\end{equation}
\begin{equation}
    C=-\left(M_1-M_2\right)^2.
\end{equation}
\end{subequations}
Here and in the following, we assume AdS radius $l_1\leq l_2$. 
Therefore, the range of the tension is $T_{\rm min}<T<T_{\rm max}$ with $T_{\rm min}=\frac{1}{l_1}-\frac{1}{l_2}$ and $T_{\rm max}=\frac{1}{l_1}+\frac{1}{l_2}$. 
Besides, we define $T_0=\sqrt{\frac{1}{l_1^2}-\frac{1}{l_2^2}}$ for later convenience.
The denominator in \eqref{eq:simplify g} has two zeros, which we denote as $\sigma_{\pm}$ and $\sigma_+>\sigma_-$.
Because $A>0$ and $C<0$, we have $\sigma_+>0$.

Plugging \eqref{eq:simplify g} into \eqref{eq:continuity condition for metrics b}, we reach the differential equations of the branes
\begin{subequations}
\label{eq:brane equation x-sigma}
\begin{equation}
\label{eq:brane equation x1-sigma}
    (\dot{x}_1)^2=l_1^2\frac{\left[M_1-M_2+\sigma\left(T^2+T_0^2\right)\right]^2}{4(l_1^2 M_1+\sigma)^2\sigma(A\sigma^2+2B\sigma+C)},
\end{equation}
\begin{equation}
\label{eq:brane equation x2-sigma}
    (\dot{x}_2)^2=l_2^2\frac{\left[M_2-M_1+\sigma\left(T^2-T_0^2\right)\right]^2}{4(l_2^2 M_2+\sigma)^2\sigma(A\sigma^2+2B\sigma+C)}.
\end{equation}
\end{subequations}
There is a sign ambiguity that can be fixed by choosing one half the brane.
\begin{figure}
    \centering
    \subfigure[]{\includegraphics[width=0.4\textwidth]{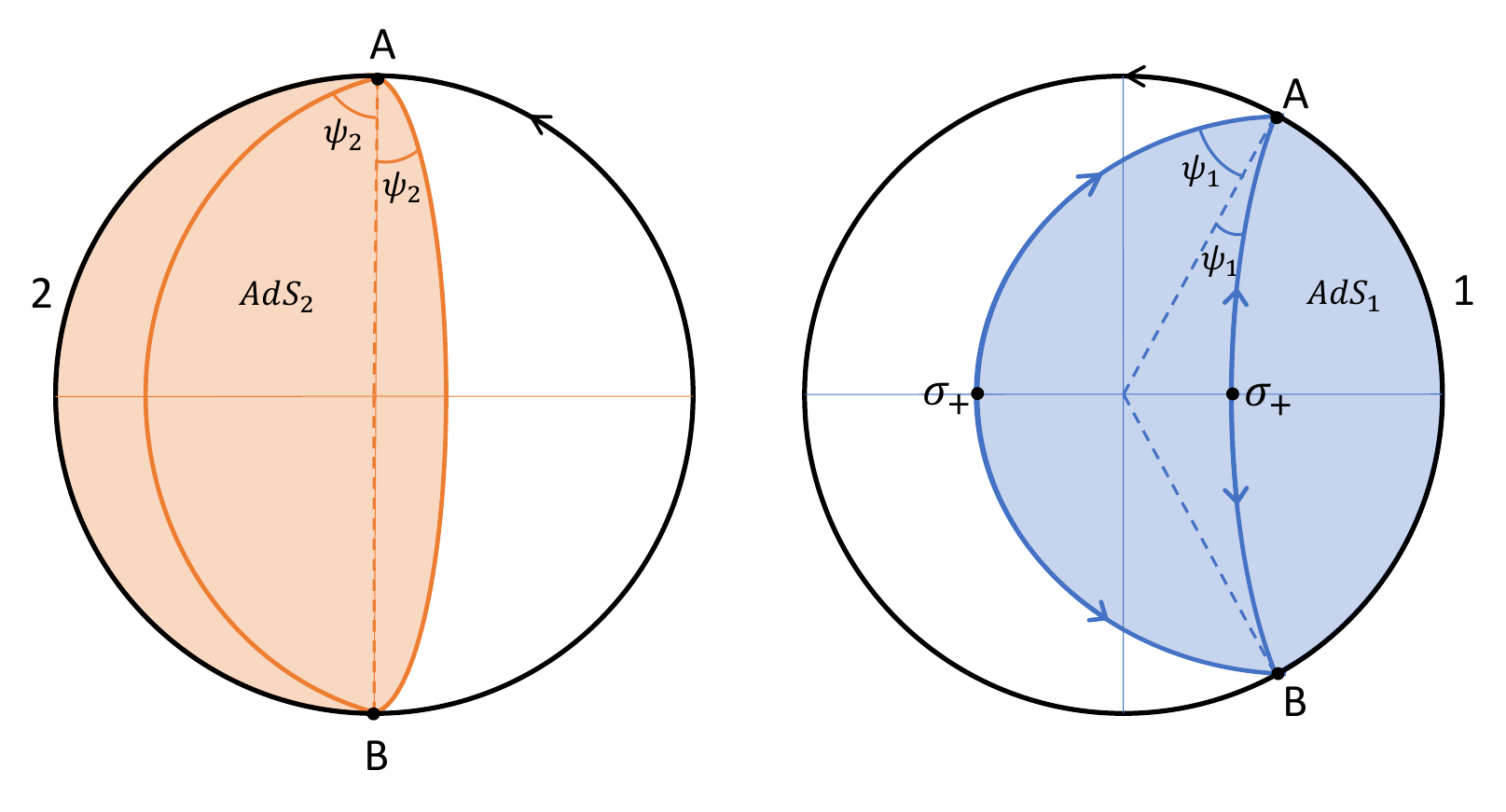}} \qquad \qquad
    \subfigure[]{\includegraphics[width=0.37\textwidth]{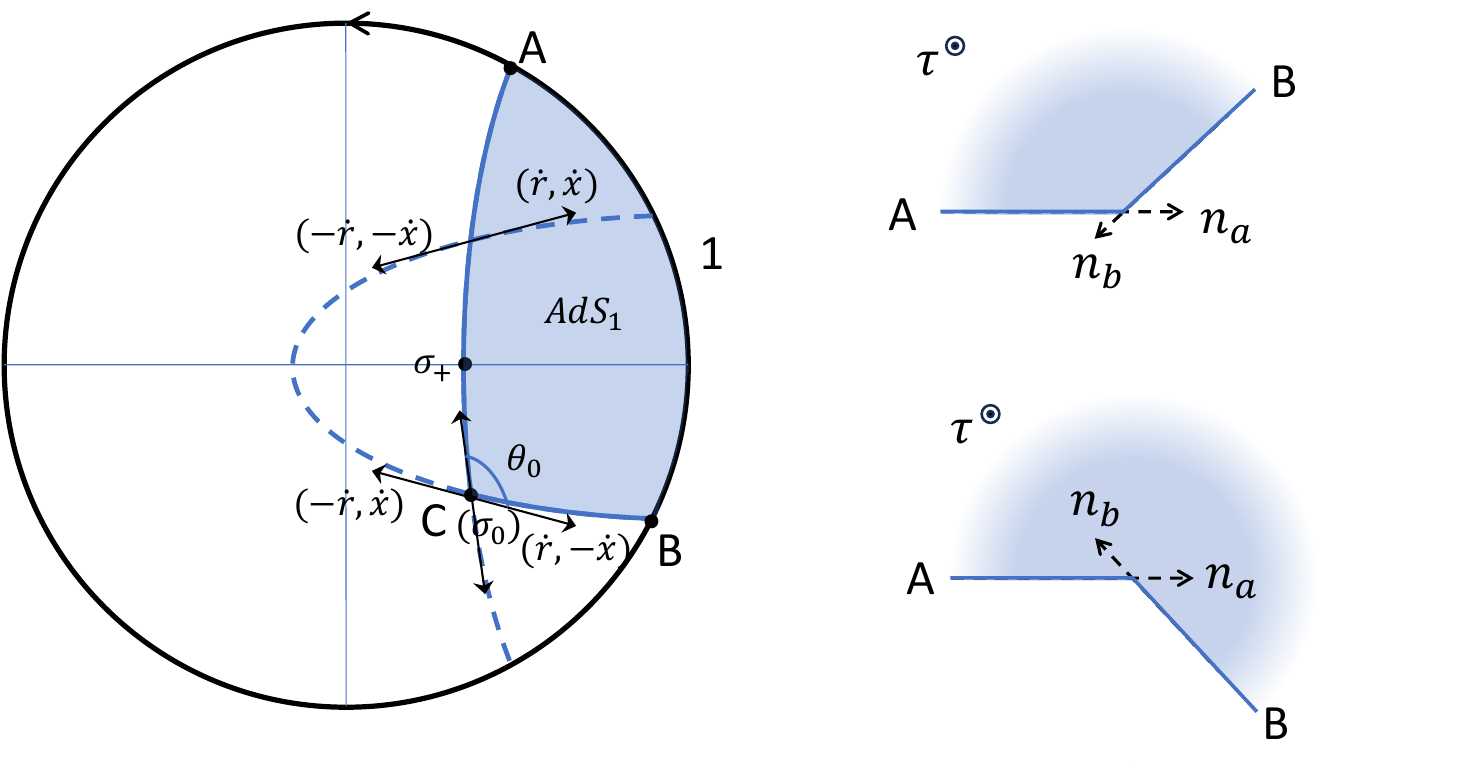}}
    \caption{Notations and the conventions of brane. 
    (a) The geometry of AdS space without the corner on branes. 
    The left and right panels correspond to $AdS_{1,2}$, and $\psi_{1,2}$ are related to the tension of branes in \eqref{eq:ange of brane equation with Poincare metric}. $1$ and $2$ label the two asymptotic boundaries dual to the two CFTs, and the positive direction of the coordinate $x$ is clockwise. 
    A and B are two defects that connect two different CFTs. 
    Two branes in each AdS space correspond to the phases E1 that includes the center $r=0$ and E2 that excludes the center. 
    In $AdS_1$, the blue arrows indicate the positive direction of $\sigma\in[\sigma_+,\infty)$. 
    (b) The geometry of AdS space with a corner on branes. 
    In $AdS_1$, the two branes starting from defects A and B intersect at C with the angle $\theta_0$. 
    The tangent vectors at the intersecting point and its symmetric point are shown with black arrows, which are consistent with \eqref{eq:direction of branes}, where $\iota_A=1,-1 (\iota_B=-1,1)$ corresponds to the half brane close to or further from the defect A (B).
    For $\theta_0<\pi(>\pi)$, we zoom in the intersecting point with different signs of $\boldsymbol{n}_a\times\boldsymbol{n}_b\cdot\boldsymbol{n}_\tau$ on the right upper (lower) panel.}
    \label{fig:Notations and the conventions of brane}
\end{figure}
Namely, as shown in Fig.~\ref{fig:Notations and the conventions of brane} (a), there are two halves of the brane that are symmetric under reflection in each AdS space.
It is easy to see from \eqref{eq:simplify g} that $\sigma_+$ is the symmetric point, which has the smallest $r_i$ on the brane trajectory. 
Different sign choices in \eqref{eq:brane equation x-sigma} correspond to different halves of brane.
We take the minus sign for both $\dot{x}_1$ and $\dot{x}_2$, so they correspond to the half brane starting from A for the $AdS_1$ and the half brane starting from B for the $AdS_2$.
We set the counterclockwise direction as the positive direction of $x$, and from the Fig.~\ref{fig:Notations and the conventions of brane} (a), we know that $\dot{x}(\sigma_+)<0$ $(>0)$ corresponds to an AdS region with (without) center $r=0$, which is denoted as E1 (E2) phase~\cite{bachas2021phases}.

With an explicit brane trajectory, we can construct equations for the AdS spacetime parameters $M_{1,2}$.
We will restore the label $\alpha=a,b$ in the following discussion.
In Fig.~\ref{fig:Notations and the conventions of brane} (b), we consider the branes in $AdS_1$, where the black arrows label the directions of the branes on the right panel.
Let's denote the intersecting point coordinate as $x_C(\sigma_0)$.
Then we have 
\begin{equation}  
    x_C=x_A+\int_{+\infty}^{\sigma_+}\dot{x}_a{\rm d}\sigma+\iota_A\int_{\sigma_+}^{\sigma_0}\dot{x}_a{\rm d}\sigma, 
\end{equation}
where $\iota_A=+1$ ($-1$) corresponds to the crossing point locating on the negative (positive) half of the brane for the brane starting from point A (here the negative and positive half correspond to the signs of $\dot{x}$). 
Remember that we take the minus sign for $\dot{x}$ in \eqref{eq:brane equation x-sigma}.
Similarly, for the point B we have 
\begin{equation}
    x_C'=x_B+\int_{+\infty}^{\sigma_+}(-\dot{x}_b){\rm d}\sigma-\iota_B\int_{\sigma_+}^{\sigma_0}(-\dot{x}_b){\rm d}\sigma,
\end{equation}
where $\iota_B=+1$ ($-1$) corresponds to the crossing point on the negative (positive) half of the brane for the brane starting from point B.
Because of the periodic boundary condition, $x_C$ and $x_C'$ are related by 
\begin{equation}
    x_C=x_C'+2\pi \frac{n}{\sqrt{-M}}.
\end{equation}
where $n=1$ ($n=0$) for the E1 (E2) phase with (without) the center.
Relating this to the interval in the dual CFT, we arrive at
\begin{equation}
\label{eq:constraint for M with L}
    L = x_A-x_B =\int_{\sigma_+}^{+\infty}(\dot{x}^a+\dot{x}^b){\rm d}\sigma-\int_{\sigma_+}^{\sigma_0}(\iota_A\ \dot{x}^a-\iota_B\ \dot{x}^b){\rm d}\sigma+\frac{2\pi n}{\sqrt{-M}},
\end{equation}
where $a,b$ corresponds to the branes for the $AdS_i$ with different tensions $T_{a,b}$.
To summarize, we have for $AdS_i$ spacetime
\begin{subequations}
\label{eq:equations for M}
\begin{equation}
\label{eq:constraint for Mi with Li}
    L_i=I_i+\frac{2\pi}{\sqrt{-M_i}}\cdot\frac{1}{2}(1-{\rm sgn}(I_i)),
\end{equation}
\begin{equation}
\label{eq:equation of Ii}
    I_i=\int_{\sigma_+}^{+\infty}(\dot{x}_i^a+\dot{x}_i^b){\rm d}\sigma-\int_{\sigma_+}^{\sigma_0}(\iota_A\ \dot{x}_i^a-\iota_B\ \dot{x}_i^b){\rm d}\sigma,
\end{equation}
\end{subequations}
where we use ${\rm sgn}(I_i)$ to express $n=0,1$, and $\dot{x}_1^\alpha$ and $\dot{x}_2^\alpha$ are
\begin{subequations}
\label{eq:explicit expression of x}
\begin{equation}
\label{eq:explicit expression of x1}
    \dot{x}_1^\alpha=-l_1\frac{M_1-M_2+\sigma\left(T_\alpha^2+T_0^2\right)}{2(l_1^2 M_1+\sigma)\sqrt{\sigma(A\sigma^2+2B\sigma+C)}},
\end{equation}
\begin{equation}
\label{eq:explicit expression of x2}
    \dot{x}^\alpha_2=-l_2\frac{M_2-M_1+\sigma\left(T_\alpha^2-T_0^2\right)}{2(l_2^2 M_2+\sigma)\sqrt{\sigma(A\sigma^2+2B\sigma+C)}}.
\end{equation}
\end{subequations}
Notice $A,B$ and $C$ are now functions of $T=T_\alpha$ from Eq.~\eqref{eq:definition of ABC}.

It is straightforward to check that ${\rm sgn}(I_i) = -1$ ($+1$) corresponds to E1 (E2) phase, i.e.,$n=0$ $(1)$.  
Therefor, for later convenience, we also introduce the following notation,
\begin{equation} 
    2\delta_{i,{\rm E1}} = 1-{\rm sgn}(I_i').
\end{equation}  
We will use these two notations interchangeably.

Here is a remark: we should be careful about different coordinates for different AdS spacetime.
But in Eq. \eqref{eq:equation of Ii} we only introduce one parameter $\sigma_0$.
It is because the coordinates of the intersecting point (corner) on different branes are supposed to be the same due to the junction condition.

In the discussion above, we have introduced parameters $\iota_A,\iota_B$ and $\sigma_0$.
Now we construct equations to solve their values.
As shown in Fig.~\ref{fig:Notations and the conventions of brane} (b), the directions of the brane $a$ and $b$ which start from boundary points A and B are 
\begin{equation}
\label{eq:direction of branes}
    \boldsymbol{n}_a=(-\iota_A\ \dot{r}_a,-\dot{x}_a),\quad\quad \boldsymbol{n}_b=(\iota_B\ \dot{r}_b,\dot{x}_b).
\end{equation}
We can get an equation from the relation $\cos{\theta_0^1}=\boldsymbol{n}_a\cdot\boldsymbol{n}_b/|\boldsymbol{n}_a| |\boldsymbol{n}_b|$.
Additionally, we need another equation to distinguish between $\theta_0^1 < \pi $ and $\theta_0^1 > \pi $ as illustrated in Fig.~\ref{fig:Notations and the conventions of brane} (b).
We use $\boldsymbol{n}_a\times(-\boldsymbol{n}_b)\cdot\boldsymbol{n}_\tau>0$ ($<0$) for $\theta_0^1<\pi$ ($>\pi$), where we regard  $\boldsymbol{n}_a=(0,-\iota_A\ \dot{r}_a,-\dot{x}_a)$, $\boldsymbol{n}_b=(0,\iota_B\ \dot{r}_b,\dot{x}_b)$, and $\boldsymbol{n}_\tau=(1,0,0)$ as three-dimensional vectors.
These considerations give
\begin{equation}
\label{eq:constraint for theta0 with 1-1 map}
    \theta_0^1=\pi+{\rm sgn}[\boldsymbol{n}_a\times\boldsymbol{n}_b\cdot\boldsymbol{n}_\tau]\left(\pi-\arccos{\frac{\boldsymbol{n}_a\cdot\boldsymbol{n}_b}{|\boldsymbol{n}_a| |\boldsymbol{n}_b|}}\right).
\end{equation}
The equation above is equivalent to 
\begin{subequations}
\label{eq:equations for theta}
\begin{equation}
\label{eq:equations for theta a}
    \cos{\theta_0^1}=\frac{-\left[\iota_A\iota_B+\frac{M_1-M_2+\sigma\left(T_a^2+T_0^2\right)}{\sqrt{A_a(\sigma-\sigma_+^a)(\sigma-\sigma_-^a)}}\frac{M_1-M_2+\sigma\left(T_b^2+T_0^2\right)}{\sqrt{A_b(\sigma-\sigma_+^b)(\sigma-\sigma_-^b)}}\right]}{\sqrt{\left[1+\frac{\left[M_1-M_2+\sigma\left(T_a^2+T_0^2\right)\right]^2}{A_a(\sigma-\sigma_+^a)(\sigma-\sigma_-^a)}\right]\left[1+\frac{\left[M_1-M_2+\sigma\left(T_b^2+T_0^2\right)\right]}{A_b(\sigma-\sigma_+^b)(\sigma-\sigma_-^b)}\right]}}.
\end{equation}
\begin{equation}
\label{eq:equations for theta b}
    {\rm sgn}{[\theta_0^1-\pi]}=\iota_A\ {\rm sgn}\left[\frac{M_1-M_2+\sigma\left(T_b^2+T_0^2\right)}{\sqrt{A_b(\sigma-\sigma_+^b)(\sigma-\sigma_-^b)}}-\iota_A\iota_B\frac{M_1-M_2+\sigma\left(T_a^2+T_0^2\right)}{\sqrt{A_a(\sigma-\sigma_+^a)(\sigma-\sigma_-^a)}}\right].
\end{equation}
\end{subequations}
It is easy to show that $(\iota_A,\iota_B)$ is unique for a given $\sigma$. 
Then, \eqref{eq:equations for theta a} can be simplified and reduced to a quadratic equation for $\sigma$ that $A'\sigma^2 +2B'\sigma+C'=0$, with
\begin{equation}
\label{eq:coefficient of equation}
\begin{split}
    A'=&-(T_a^2+T_b^2)(T_0^4+T_a^2 T_b^2)+4T_a^2 T_b^2\left({l_2^{-2}}-{l_1^{-2}}\cos^2{\theta_0^1}\right)-2T_a T_b\cos{\theta_0^1}(T_a^2+T_0^2)(T_b^2+T_0^2),\\
    B'=&-T_0^2(T_a^2+T_b^2)(M_1-M_2)-T_a T_b(2T_0^2+T_a^2+T_b^2)(M_1-M_2)\cos{\theta_0^1}+2T_a^2 T_b^2(M_2-\cos^2{\theta_0^1}M_1),\\
    C'=&-(M_1-M_2)^2(T_a^2+T_b^2+2T_a T_b\cos{\theta_0^1}).
\end{split}
\end{equation}
Its solution is nothing but the intersecting point $\sigma_0$. 
To summarize, we can solve $\sigma_0$ according to \eqref{eq:coefficient of equation}, and then get $\iota_A$ and $\iota_B$ from \eqref{eq:equations for theta}.
Finally, solving \eqref{eq:equations for M} we get $M_1$ and $M_2$, i.e., the full geometry is obtained.

For simplicity, we introduce dimensionless parameters~\cite{bachas2021phases}: $\gamma=L_1/L_2$ and $\mu=M_2/M_1$.
Then \eqref{eq:coefficient of equation} and \eqref{eq:equations for theta} can be simplified with a new variable $s=\sigma/|M_1|=-\sigma/M_1$.
The equation of $\sigma_0$ leads to $A''s^2+2B''s+C''=0$ with
\begin{equation}
\label{eq:coefficient of equation with s}
\begin{split}
    A''=&A',\\
    B''=&-T_0^2(T_a^2+T_b^2)(\mu-1)-T_a T_b(2T_0^2+T_a^2+T_b^2)(\mu-1)\cos{\theta_0^1}+2T_a^2 T_b^2(\cos^2{\theta_0^1}-\mu),\\
    C''=&-(\mu-1)^2(T_a^2+T_b^2+2T_a T_b\cos{\theta_0^1}).
\end{split}
\end{equation}
With $s_{\pm}=-\sigma_\pm/M_1$ and $s_0=-\sigma_0/M_1$, $\iota_A,\iota_B$ are given by
\begin{subequations}
\label{eq:equations for theta with s}
\begin{equation}
\label{eq:equations for theta a with s}
\begin{split}
    \iota_A\cdot\iota_B={\rm sgn}\left[-\cos{\theta_0^1}\prod_{\alpha=a,b}\sqrt{A_\alpha(s-s_+^\alpha)(s-s_-^\alpha)+\left[\mu-1+s\left(T_\alpha^2+T_0^2\right)\right]^2}\left.-\prod_{\alpha=a,b}\left[\mu-1+s\left(T_\alpha^2+T_0^2\right)\right]\right]\right|_{s=s_0},
\end{split}
\end{equation}
\begin{equation}
\label{eq:equations for theta b with s}
    \iota_A={\rm sgn}{[\theta_0^1-\pi]}\left.\cdot{\rm sgn}\left[\frac{\mu-1+s\left(T_b^2+T_0^2\right)}{\sqrt{A_b(s-s_+^b)(s-s_-^b)}}-\iota_A\iota_B\frac{\mu-1+s\left(T_a^2+T_0^2\right)}{\sqrt{A_a(s-s_+^a)(s-s_-^a)}}\right]\right|_{s=s_0}.
\end{equation}
\end{subequations}
Finally, the equation for $\mu$ becomes
\begin{equation}
\label{eq:equation for mu}
    \gamma = \gamma(\mu) \equiv \frac{I_1'+\pi(1-{\rm sgn}(I_1'))}{I_2'+\frac{\pi}{\sqrt{\mu}}(1-{\rm sgn}(I_2'))},
\end{equation}
where $I_{1,2}'=\sqrt{-M_1}\cdot I_{1,2}$.
With a tedious calculation, $I_{1,2}'$ can be simplified to
\begin{subequations}
\label{eq:equation Ip with elliptic integrals}
\begin{equation}
\label{eq:Ip1 with elliptic integrals}
\begin{split}
    I_1'=&(1-\iota_A)\frac{l_1}{\sqrt{A s_+}}\left.\left[\frac{\mu-1}{l_1^2}\cdot K\left(\frac{s_-}{s_+}\right)-\left(T_a^2+T_0^2+\frac{\mu-1}{l_1^2}\right)\cdot\Pi\left(\frac{l_1^2}{s_+},\frac{s_-}{s_+}\right)\right]\right|_a\\
    +&(1+\iota_B)\frac{l_1}{\sqrt{A s_+}}\left.\left[\frac{\mu-1}{l_1^2}\cdot K\left(\frac{s_-}{s_+}\right)-\left(T_b^2+T_0^2+\frac{\mu-1}{l_1^2}\right)\cdot\Pi\left(\frac{l_1^2}{s_+},\frac{s_-}{s_+}\right)\right]\right|_b\\
    +&\iota_A\frac{l_1}{\sqrt{A s_+}}\left.\left[\frac{\mu-1}{l_1^2}\cdot K\left(\sqrt{\frac{s_+}{s_0}},\frac{s_-}{s_+}\right)-\left(T_a^2+T_0^2+\frac{\mu-1}{l_1^2}\right)\cdot\Pi\left(\sqrt{\frac{s_+}{s_0}},\frac{l_1^2}{s_+},\frac{s_-}{s_+}\right)\right]\right|_a\\
    -&\iota_B\frac{l_1}{\sqrt{A s_+}}\left.\left[\frac{\mu-1}{l_1^2}\cdot K\left(\sqrt{\frac{s_+}{s_0}},\frac{s_-}{s_+}\right)-\left(T_b^2+T_0^2+\frac{\mu-1}{l_1^2}\right)\cdot\Pi\left(\sqrt{\frac{s_+}{s_0}},\frac{l_1^2}{s_+},\frac{s_-}{s_+}\right)\right]\right|_b,
\end{split}
\end{equation}
\begin{equation}
\label{eq:Ip2 with elliptic integrals}
\begin{split}
    I_2'=&(1-\iota_A)\frac{l_2}{\sqrt{A s_+}}\left.\left[\frac{1-\mu}{l_2^2\mu}\cdot K\left(\frac{s_-}{s_+}\right)-\left(T_a^2-T_0^2+\frac{1-\mu}{l_2^2\mu}\right)\cdot\Pi\left(\frac{l_2^2\mu}{s_+},\frac{s_-}{s_+}\right)\right]\right|_a\\
    +&(1+\iota_B)\frac{l_2}{\sqrt{A s_+}}\left.\left[\frac{1-\mu}{l_2^2\mu}\cdot K\left(\frac{s_-}{s_+}\right)-\left(T_b^2-T_0^2+\frac{1-\mu}{l_2^2\mu}\right)\cdot\Pi\left(\frac{l_2^2\mu}{s_+},\frac{s_-}{s_+}\right)\right]\right|_b\\
    +&\iota_A\frac{l_2}{\sqrt{A s_+}}\left.\left[\frac{1-\mu}{l_2^2\mu}\cdot K\left(\sqrt{\frac{s_+}{s_0}},\frac{s_-}{s_+}\right)-\left(T_a^2-T_0^2+\frac{1-\mu}{l_2^2\mu}\right)\cdot\Pi\left(\sqrt{\frac{s_+}{s_0}},\frac{l_2^2\mu}{s_+},\frac{s_-}{s_+}\right)\right]\right|_a\\
    -&\iota_B\frac{l_2}{\sqrt{A s_+}}\left.\left[\frac{1-\mu}{l_2^2\mu}\cdot K\left(\sqrt{\frac{s_+}{s_0}},\frac{s_-}{s_+}\right)-\left(T_b^2-T_0^2+\frac{1-\mu}{l_2^2\mu}\right)\cdot\Pi\left(\sqrt{\frac{s_+}{s_0}},\frac{l_2^2\mu}{s_+},\frac{s_-}{s_+}\right)\right]\right|_b,
\end{split}
\end{equation}
\end{subequations}
where we use $(\cdot)|_{a,b}$ to indicate that $s_\pm$ and $s_0$ are the functions of $T=T_{a,b}$.
$K(y_0,v)$ and $\Pi(y_0,u,v)$ are elliptic integrals and $K(v)=K(1,v)$ and $\Pi(u,v)=\Pi(1,u,v)$.
The details of elliptic integrals and the derivation of equation \eqref{eq:equation Ip with elliptic integrals} are shown in Appendix~\ref{sec:Arc equations and special functions}.

We summarize the general procedure to get the solution
\begin{enumerate}
    \item with coefficients given in \eqref{eq:coefficient of equation with s}, solve $s_0$;
    
    \item plug $s_0$ into \eqref{eq:equations for theta with s}, to solve $\iota_A$ and $\iota_B$, which determine the geometry of branes and the location of the intersection point;
    \item use $\iota_A$ and $\iota_B$ to solve \eqref{eq:equation for mu} to get $\mu$, which then will give $M_{1,2}$;
    \item use $M_{1,2}$ to evaluate the onshell action given in~\eqref{eq:total action final results}.
\end{enumerate}

From the discussion above, we find that with the constraint $\theta^1=\theta_0^1$, we can already solve the AdS geometry and the brane trajectories.
It means $\theta_0^1$ and $\theta_0^2$ in the action \eqref{eq:total action} are not independent. 
Hence, we define $\theta_0 = \theta_0^1$ for simplicity.

\subsection{Defect changing operator without a cusp}

\label{Symmetric case with a defect}

In this section, we derive the scaling dimension of the defect changing operator given in Eq.~\eqref{eq:defect_changing_operator} in the main text.
To this end, we consider the geometry with $\mathbb{Z}_2$ symmetry, $L_1=L_2$ and $l_1=l_2$. 
Because of the $\mathbb{Z}_2$ symmetry, we have $M_1=M_2=M$ and $\mu=1$ without solving the complicated equations outlined in the last section. 
Using $l_1=l_2=l$ and $T_0^2=l_1^{-2}-l_2^{-2}=0$, the coefficients in \eqref{eq:coefficient of equation} can be simplified as 
\begin{equation}
\label{eq:coefficient of equation with l1eql2}
\begin{split}
    A'=-T_a^2 T_b^2(T_a^2+T_b^2+2T_a T_b\cos{\theta_0^1}-4l^{-2}\sin^2{\theta_0^1}), \qquad
    B'=2MT_a^2 T_b^2\sin^2{\theta_0^1}, \qquad C'=0,
\end{split}
\end{equation}
and $\sigma_0=-2B'/A'$.

Next, we need to solve \eqref{eq:equations for theta with s}.
Firstly, since $r=\sqrt{\sigma+Ml^2} \ge 0$, we must have $\sigma_0 +Ml^2 \ge 0$, and it
leads to the constraint for $\cos{\theta_0^1}$,
\begin{equation}
\label{eq:constraint of theta01}
    -\mathcal{T}_a\mathcal{T}_b-\sqrt{(1-\mathcal{T}_a^2)(1-\mathcal{T}_b^2)}<\cos{\theta_0^1}<-\mathcal{T}_a\mathcal{T}_b+\sqrt{(1-\mathcal{T}_a^2)(1-\mathcal{T}_b^2)}, 
\end{equation}
where we introduced $\mathcal{T}_{a,b}=l\cdot T_{a,b}/2$ for simplicity.
We assume $T_a>T_b$ without loss of generality. 
Using $l_1=l_2=l$ and $\mu=1$, the coefficient $A_\alpha$ in \eqref{eq:definition of ABC} can be simplified as $A_\alpha=T_\alpha^2(4/l^2-T_\alpha^2)$, and 
\begin{equation}
\label{eq:Aspm equation}
    A_\alpha s_\pm=-\frac{A_\alpha}{M}\frac{-B_\alpha\pm\sqrt{B_\alpha^2-A_\alpha C_\alpha}}{A}=2T_\alpha^2\pm 2T_\alpha^2.
\end{equation}
Therefore, \eqref{eq:equations for theta a with s} can be simplified to $\iota_A\iota_B={\rm sgn}\left[4\cos{\theta_0^1}-s_0\left(\frac{4}{l^2}\cos{\theta_0^1}+T_a T_b\right)\right]$.  
Using $s_0=(-\frac{2B'}{A'})/(-M)$, after a tedious calculation, we have
\begin{equation}
\label{eq:simplified sgnAsgnB final result}
    \iota_A\iota_B=
    \begin{cases}
        1,&-\mathcal{T}_a\mathcal{T}_b-\sqrt{(1-\mathcal{T}_a^2)(1-\mathcal{T}_b^2)}<\cos{\theta_0^1}<-\frac{\mathcal{T}_a}{\mathcal{T}_b},\\
        -1,&-\frac{\mathcal{T}_a}{\mathcal{T}_b}<\cos{\theta_0^1}<-\mathcal{T}_a\mathcal{T}_b+\sqrt{(1-\mathcal{T}_a^2)(1-\mathcal{T}_b^2)},
    \end{cases}
\end{equation}
If $\iota_A\iota_B=-1$, \eqref{eq:equations for theta b with s} leads $\iota_A={\rm sgn}(\theta_0^1-\pi)$.
If $\iota_A\iota_B=1$, with $s_+^\alpha=4/(4l^{-2}-T_\alpha^2)$ and $s_0=4l^2\sin^2{\theta_0^1}/[-l^2(T_a^2+T_b^2+2T_a T_b\cos{\theta_0^1})+4\sin^2{\theta_0^1}]$, we have 
\begin{equation}
\label{eq:sgnA with Z2 symm final result}
\begin{split}
    \iota_A=&{\rm sgn}{[\theta_0^1-\pi]}\cdot{\rm sgn}\left[\left(1-\frac{s_+^a}{s_0}\right)\left(\frac{4}{l^2T_a^2}-1\right)-\left(1-\frac{s_+^b}{s_0}\right)\left(\frac{4}{l^2T_b^2}-1\right)\right]\\
    =&{\rm sgn}{[\theta_0^1-\pi]}\cdot{\rm sgn}\left[\frac{(T_a^2-T_b^2)(T_a^2+T_b^2+2T_a T_b\cos{\theta_0^1})}{-\sin^2{\theta_0^1}T_a^2 T_b^2}\right].
\end{split}
\end{equation}
which leads to
\begin{equation}
\label{eq:simplified sgnA and sgnB final result}
    \begin{cases}
        \iota_A=\iota_B=-{\rm sgn}(\theta_0^1-\pi),&-\mathcal{T}_a\mathcal{T}_b-\sqrt{(1-\mathcal{T}_a^2)(1-\mathcal{T}_b^2)}<\cos{\theta_0^1}<-\frac{\mathcal{T}_a}{\mathcal{T}_b},\\
        \iota_A=-\iota_B={\rm sgn}(\theta_0^1-\pi),&-\frac{\mathcal{T}_a}{\mathcal{T}_b}<\cos{\theta_0^1}<-\mathcal{T}_a\mathcal{T}_b+\sqrt{(1-\mathcal{T}_a^2)(1-\mathcal{T}_b^2)}.
    \end{cases}
\end{equation}

Finally, we are ready to solve \eqref{eq:equations for M} to get the parameter $M$.
Since $s_-=0$ and $T_0=0$, we have $K(y_0,0)=\arcsin{y_0}$, $\Pi(y_0,u,0)=\arctan{\left(y_0\sqrt{\frac{1-u}{1-y_0^2}}\right)}/\sqrt{1-u}$. 
Then plugging $A_\alpha s_+^\alpha=4T_\alpha^2$, $\frac{l_1^2}{s_+^\alpha} = \frac{4-l_1^2T_\alpha^2}{4}$ in \eqref{eq:Ip1 with elliptic integrals}, we obtain
\begin{equation}
\label{eq:Ip1 with elliptic integrals with Z2 symmetry concretely}
\begin{split}
    I_1'=(1-\iota_A)\frac{-\pi}{2}+(1+\iota_B)\frac{-\pi}{2}+\iota_A\left[-\arctan{\left(\sqrt{\frac{s_+^a-l_1^2}{s_0-s_+^a}}\right)}\right]
    -\iota_B\left[-\arctan{\left(\sqrt{\frac{s_+^b-l_1^2}{s_0-s_+^b}}\right)}\right].
\end{split}
\end{equation}

We discuss two possibilities in \eqref{eq:simplified sgnA and sgnB final result}.
For $-\mathcal{T}_a\mathcal{T}_b-\sqrt{(1-\mathcal{T}_a^2)(1-\mathcal{T}_b^2)}<\cos{\theta_0^1}<-\frac{\mathcal{T}_a}{\mathcal{T}_b}$, we can get 
\begin{equation}
\label{eq:Ip1 with elliptic integrals with Z2 symmetry case 1 final}
    I_1'=-\pi+{\rm sgn}(\theta_0^1-\pi)\arccos{\frac{-(\mathcal{T}_a \mathcal{T}_b+\cos{\theta_0^1})}{\sqrt{(1-\mathcal{T}_a^2)(1-\mathcal{T}_b^2)}}}\leq0.
\end{equation}
Then, \eqref{eq:constraint for Mi with Li} gives
\begin{equation}
\label{eq:constraint for L}
    L_1\sqrt{-M_1}=I_i'+2\pi\cdot\frac{1}{2}(1-{\rm sgn}(I_i'))=\pi+{\rm sgn}(\theta_0^1-\pi)\arccos{\frac{-(\mathcal{T}_a \mathcal{T}_b+\cos{\theta_0^1})}{\sqrt{(1-\mathcal{T}_a^2)(1-\mathcal{T}_b^2)}}}=\begin{cases}
        2\pi-\arccos{\beta},&\theta_0^1>\pi,\\
        \arccos{\beta},&\theta_0^1<\pi.
    \end{cases},
\end{equation}
where $\beta=(\mathcal{T}_a \mathcal{T}_b+\cos{\theta_0^1})/\sqrt{(1-\mathcal{T}_a^2)(1-\mathcal{T}_b^2)}$. 
Similarly, for $-\frac{\mathcal{T}_a}{\mathcal{T}_b}<\cos{\theta_0^1}<-\mathcal{T}_a\mathcal{T}_b+\sqrt{(1-\mathcal{T}_a^2)(1-\mathcal{T}_b^2)}$, we can get
\begin{equation}
\label{eq:Ip1 with elliptic integrals with Z2 symmetry case 2 final}
    I_1'=-\pi+{\rm sgn}(\theta_0^1-\pi)(\pi-\arccos{\beta})\leq0,
\end{equation}
which leads to the same result as \eqref{eq:constraint for L}.

Then, the onshell action~\eqref{eq:total action final results} reads
\begin{equation}
\label{eq:total action for theta0}
\begin{split}
    I_{\rm tot} = \frac{M_1 l_1 L_1+M_2 l_2 L_2}{2T_{\rm DCFT}}=-\frac{2l}{T_{\rm DCFT}L}[I_1'+\pi(1-{\rm sgn}(I_1'))]^2 =-\frac{2l}{T_{\rm DCFT}L}[\pi+{\rm sgn}(\theta_0^1-\pi)(\pi-\arccos{\beta})]^2.
\end{split}
\end{equation}
This leads to Eq.~\eqref{eq:defect_changing_operator} in the main text.

\subsection{AdS geometry without a corner on the branes}
\label{sec:AdS geometry without point defect}

In this section, we consider the branes without corners, i.e., $\theta_0^1=\theta_0^2=\pi$ and $T_a=T_b=T$.
Note that a similar case has been considered in Ref.~\cite{bachas2021phases} with a different motivation. 
We will derive new results for the cusp anomalous dimension. 

\subsubsection{Continuous brane with the same tension}
\label{Continuous brane with the same tension}

Plugging $\theta_0^1=\theta_0^2=\pi$ into \eqref{eq:coefficient of equation}, we find that $A',B'$ and $C'$ are all proportional to $(T_a-T_b)^2$.
Hence, $T_a=T_b$ means the equation for $\sigma$ is trivial.
Plugging $T_a=T_b$ into \eqref{eq:equations for theta a with s}, we get $\iota_A\iota_B=1$. 
While \eqref{eq:equations for theta b with s} is not applicable.
With the results above, we know that $\iota_A=\iota_B$ and any $\sigma_0$ is a solution, which means two branes starting from A and B will connect smoothly and form a single brane, as one would expect without corners.
Besides, to simplify the problem, we can take $\iota_A=\iota_B=1$ and $\sigma_0=\sigma_+$ ($s_0=s_+$).
Then \eqref{eq:equation Ip with elliptic integrals} can be simplified to
\begin{subequations}
\label{eq:equation Ip with elliptic integrals without defect}
\begin{equation}
\label{eq:Ip1 with elliptic integrals without defect}
\begin{split}
    I_1'=\frac{2l_1}{\sqrt{A s_+}}\left[\frac{\mu-1}{l_1^2}\cdot K\left(\frac{s_-}{s_+}\right)-\left(T^2+T_0^2+\frac{\mu-1}{l_1^2}\right)\cdot\Pi\left(\frac{l_1^2}{s_+},\frac{s_-}{s_+}\right)\right],
\end{split}
\end{equation}
\begin{equation}
\label{eq:Ip2 with elliptic integrals without defect}
\begin{split}
    I_2'=\frac{2l_2}{\sqrt{A s_+}}\left[\frac{1-\mu}{l_2^2\mu}\cdot K\left(\frac{s_-}{s_+}\right)-\left(T^2-T_0^2+\frac{1-\mu}{l_2^2\mu}\right)\cdot\Pi\left(\frac{l_2^2\mu}{s_+},\frac{s_-}{s_+}\right)\right].
\end{split}
\end{equation}
\end{subequations}

\subsubsection{Bubble-solution phase}
\label{sec:Bubble phase condition}

As we discussed in the main text, the existence of the bubble-solution phase is that the solution of \eqref{eq:equation for mu} at $\gamma \rightarrow 0$ is $\mu_0 > 0 $. 
We derive the condition for the bubble-solution phase in this subsection. 
With the help of \eqref{eq:equation Ip with elliptic integrals without defect}, we define the numerator and denominator in \eqref{eq:equation for mu}, respectively, as
\begin{subequations}
\label{eq:numerator and denominator of gamma equation}

\begin{equation}
\label{eq:numerator of gamma equation}
    F_{\rm num} = I_1'+\pi(1-{\rm sgn}(I_1')) 
    =2\pi\delta_{1,{\rm E1}}-\frac{2l_1}{\sqrt{A s_+}}\left[\frac{1-\mu}{l_1^2}\cdot K\left(\frac{s_-}{s_+}\right)+\left(T^2+T_0^2+\frac{\mu-1}{l_1^2}\right)\cdot\Pi\left(\frac{l_1^2}{s_+},\frac{s_-}{s_+}\right)\right],
\end{equation}
\begin{equation}
\label{eq:denominator of gamma equation}
    F_{\rm den}=I_2'+\frac{\pi}{\sqrt{\mu}}(1-{\rm sgn}(I_2'))= \frac{2\pi}{\sqrt{\mu}}\delta_{2,{\rm E1}}-\frac{2l_2}{\sqrt{A s_+}}\left[\frac{\mu-1}{l_2^2\mu}\cdot K\left(\frac{s_-}{s_+}\right)+\left(T^2-T_0^2+\frac{1-\mu}{l_2^2\mu}\right)\cdot\Pi\left(\frac{l_2^2\mu}{s_+},\frac{s_-}{s_+}\right)\right].
\end{equation}
\end{subequations}
We show in Appendix~\ref{sec:Sweeping transition and non-self-intersecting condition} that $\delta_{i,{\rm E1}}$ can be expressed as a function of $\mu$ explicitly. 

\begin{figure}
    \centering
    \subfigure[]{\includegraphics[width=0.4\linewidth]{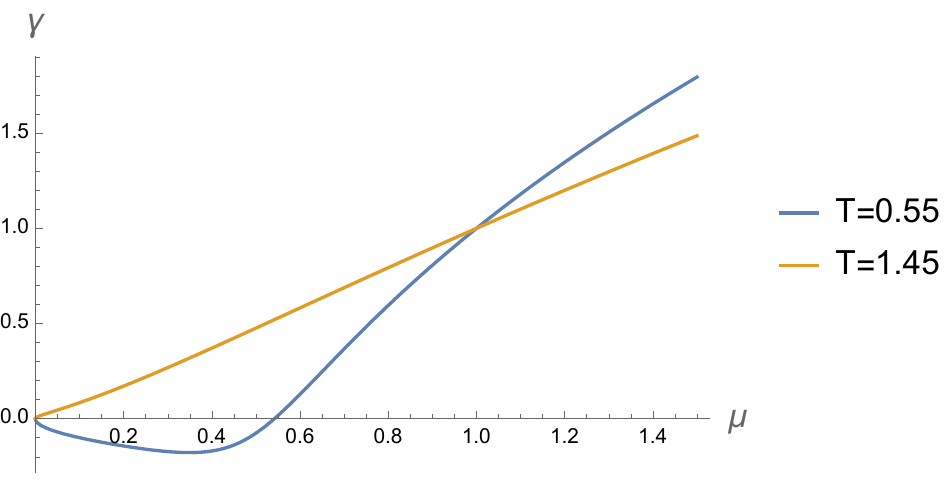}}\qquad\qquad
    \subfigure[]{\includegraphics[width=0.4\linewidth]{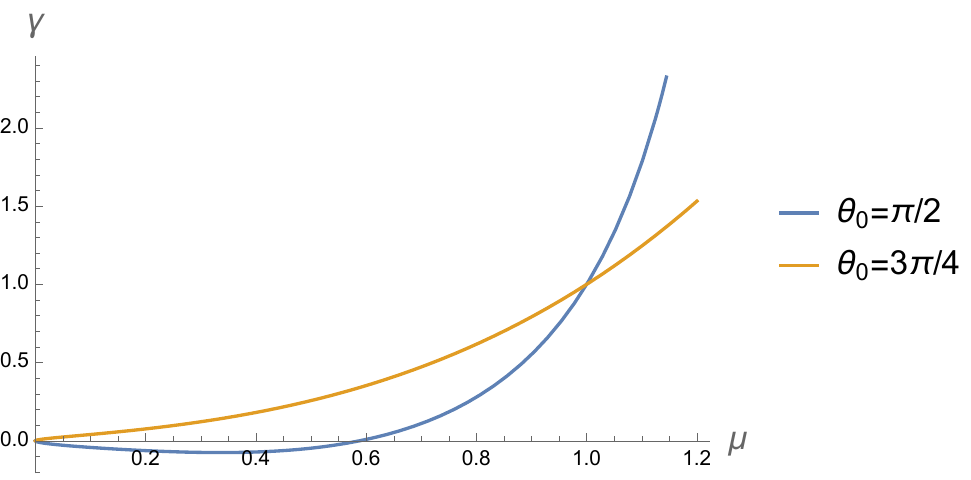}}
    \caption{Examples of $\gamma=\gamma(\mu)$ for different phases without and with a corner on branes. (a) Without a corner, i.e., $\theta_0=\pi$, we plot $l_2=2l_1=2$, so $T=0.55$ ($T=1.45$) corresponds to the (no-)bubble-solution phase, as shown in main text Fig.~\ref{fig:phase diagram} (b). 
    (b) With a nontrivial corner contribution, we plot $l_2=2l_1=2, T=1.25$, so $\theta_0=\pi/2$ ($\theta_0=3\pi/4$) corresponds to the (no-)bubble-solution phase, as shown in main text Fig.~\ref{fig:phase diagram} (c).}
    \label{fig:two example of gamma mu}
\end{figure}

Instead of directly solve $\gamma(\mu) = 0 $, our strategy is to examine the behavior of 
\begin{equation}
    \gamma(\mu) =  \frac{I_1'+\pi(1-{\rm sgn}(I_1')) }{I_2'+\frac{\pi}{\sqrt{\mu}}(1-{\rm sgn}(I_2'))}
    = \frac{F_{\rm den}}{F_{\rm num}}, 
\end{equation}
as a function of $\mu$, from which we can get the phase boundary of the bubble solution. 
The results are summarized in the discussion of~\eqref{eq:gamma_mu_infty} and~\eqref{eq:gamma_mu_0}, with an illustration of $\gamma(\mu)$ shown in Fig.~\ref{fig:two example of gamma mu} (a).

Now. we start our analysis by examining the asymptotic behavior of $F_{\rm num}$ and $F_{\rm den}$, respectively. 
We first consider the limit $\mu\rightarrow\infty$ for $F_{\rm num}$.
We have $\delta_{1,{\rm E1}}=1$ from \eqref{eq:explicit expression of deltaiE1}.
Then by expanding $s_-/s_+$ and $l_1^2/s_+$ in the order of $\mu^{-1}$,
\begin{subequations}
\label{eq:expansion of parameter in special function with large mu}
\begin{equation}
    \frac{s_-}{s_+}=\frac{(T-T_{\rm max})(T-T_{\rm min})}{(T+T_{\rm max})(T+T_{\rm min})}+\mathcal{O}(\mu^{-1}),
\end{equation}
\begin{equation}
    \frac{l_1^2}{s_+}=\frac{1}{\mu}\frac{(T_{\rm max}-T)(T-T_{\rm min})}{l_1^{-2}}+\mathcal{O}(\mu^{-2}),
\end{equation}
\begin{equation}
    \sqrt{As_+}=\mu\cdot(T+T_{\rm max})(T+T_{\rm min})+\mathcal{O}(1),
\end{equation}
\end{subequations}
and substituting them to the special functions in \eqref{eq:numerator of gamma equation}, we can obtain after a simplification
\begin{equation}
\label{eq:simplified numerator with large mu}
\begin{split}
    F_{\rm num}|_{\mu \rightarrow \infty}=&2\pi+\mathcal{O}(1/\sqrt{\mu})\rightarrow2\pi>0.
\end{split}
\end{equation}

Consider the limit $\mu\rightarrow0$ for $F_{\rm num}$.
We have $\delta_{1,{\rm E1}}=\Theta(T l_2-1)$ from \eqref{eq:explicit expression of deltaiE1}.
Hence,
\begin{equation} \label{seq:F_num_0}
    \begin{split}
     F_{\rm num}|_{\mu\rightarrow0} &= 2\pi\Theta(T l_2-1) + I_1'|_{\mu\rightarrow0}, \\
    I_1'|_{\mu\rightarrow0} &= -\frac{2l_1}{\sqrt{A s_+}}\left[\frac{1-\mu}{l_1^2}\cdot K\left(\frac{s_-}{s_+}\right)+\left(T^2+T_0^2+\frac{\mu-1}{l_1^2}\right)\cdot\Pi\left(\frac{l_1^2}{s_+},\frac{s_-}{s_+}\right)\right].
    \end{split}
\end{equation}

Expanding $s_-/s_+$ and $\mu l_2^2/s_+$ in the order of $\mu$ gives 
\begin{subequations}
\label{eq:expansion of parameter in special function with small mu}
\begin{equation}
    \frac{s_-}{s_+}=\frac{(T-T_{\rm max})(T+T_{\rm min})}{(T+T_{\rm max})(T-T_{\rm min})}+\mathcal{O}(\mu),
\end{equation}
\begin{equation}
    \frac{\mu l_2^2}{s_+}=\mu\cdot\frac{(T_{\rm max}-T)(T+T_{\rm min})}{l_2^{-2}}+\mathcal{O}(\mu^{2}),
\end{equation}
\begin{equation}
    \sqrt{As_+}=(T+T_{\rm max})(T-T_{\rm min})+\mathcal{O}(\mu).
\end{equation}
\end{subequations}
Then \eqref{eq:Ip1 with elliptic integrals without defect} can be simplified to 
\begin{equation}
\label{eq:simplified I1 with small mu}
\begin{split}
    {I_1'}|_{\mu\rightarrow0} = -2[(T-T_{\rm min})(T+T_{\rm max})]^{-\frac{1}{2}}&\left[l_1^{-1}\cdot K\left(\frac{(T-T_{\rm max})(T+T_{\rm min})}{(T+T_{\rm max})(T-T_{\rm min})}\right)+l_1\left(T^2-\frac{1}{l_2^2}\right)\right.\\
    \times&\left.\Pi\left(l_1^2(T_{\rm max}-T)(T+T_{\rm min}),\frac{(T-T_{\rm max})(T+T_{\rm min})}{(T+T_{\rm max})(T-T_{\rm min})}\right)\right].
\end{split}
\end{equation}
To discuss its sign, we consider
\begin{equation}
\label{eq:simplified dI1dT with small mu}
    \frac{{\rm d}}{{\rm d} T}{I_1'}|_{\mu\rightarrow0} = \frac{4T(E(v)-K(v))}{l_1(T+T_{\rm min})(T_{\rm max}-T)\sqrt{(T+T_{\rm max})(T-T_{\rm min})}}.
\end{equation}
Note that, for later convenience, we introduce two dimensionless variables,
\begin{equation}
\label{eq:expansion of parameters in the elliptic integrals in I1p with small mu}
    v=\frac{(T-T_{\rm max})(T+T_{\rm min})}{(T+T_{\rm max})(T-T_{\rm min})},\quad
    u=l_1^2(T_{\rm max}-T)(T+T_{\rm min}).
\end{equation}
Because $E(v)-K(v)$ has the property
\begin{equation}
\label{eq:Ev-Kv function}
    E(v)-K(v)=\int_0^1\frac{\sqrt{1-vy^2}}{\sqrt{1-y^2}}{\rm d}y-\int_0^1\frac{{\rm d}y}{\sqrt{(1-y^2)(1-vy^2)}}=\int_0^1\frac{-vy^2{\rm d}y}{\sqrt{(1-y^2)(1-vy^2)}}>0,
\end{equation}
we conclude that $\frac{{\rm d}}{{\rm d} T}{I_1'}|_{\mu\rightarrow0}>0$,

For $T_{\rm min} > l_2^{-1}$, the step function is a constant $\theta(Tl_2 - 1) = 1$, so $\frac{{\rm d}}{{\rm d} T} F_{\rm num}|_{\mu\rightarrow0} =   \frac{{\rm d}}{{\rm d} T}{I_1'}|_{\mu\rightarrow0} > 0$. 
It is more subtle for $T_{\rm min} < l_2^{-1}$.
While the step function has a nontrivial contribution to $F_{\rm num}|_{\mu\rightarrow0}$, it can be shown that $F_{\rm num}|_{\mu\rightarrow0}$ is actually continuous. 
Also, there is a caveat in $\frac{{\rm d}}{{\rm d} T}{I_1'}|_{\mu\rightarrow0}$ at $T = l_2^{-1}$:  
${I_1'}|_{\mu\rightarrow0}$ has a jump at in $T = l_2^{-1}$, but $\lim_{T \rightarrow l_2^{-1}+0^+}\frac{{\rm d}}{{\rm d} T}{I_1'}|_{\mu\rightarrow0} = \lim_{T \rightarrow l_2^{-1}+0^-}\frac{{\rm d}}{{\rm d} T}{I_1'}|_{\mu\rightarrow0}$. 
Hence, we can conclude that $ F_{\rm num}|_{\mu\rightarrow0}$ is a monotonically increasing function of $T$. 
Then we need to discuss the behavior of $ F_{\rm num}|_{\mu\rightarrow0}$ at $T_{\rm min}$ and $T_{\rm max}$. For $T\rightarrow T_{\rm min}^+$, we assume $T=T_{\rm min}\cdot(1+\delta)$ with $\delta\rightarrow0^+$.
Then we can expand the parameters in the elliptic integrals in \eqref{eq:expansion of parameters in the elliptic integrals in I1p with small mu} that $v\approx-\frac{2l_1}{l_2\delta}$, $u\approx\frac{4l_1^2}{l_2}\left(\frac{1}{l_1}-\frac{1}{l_2}\right)$.
Using $K\left(-\frac{a}{\delta}\right)\approx\Pi\left(u,-\frac{a}{\delta}\right)\approx-\ln{\delta}\cdot\sqrt{\delta}/(2\sqrt{a})$ from Appendix~\ref{sec:Special functions}, we arrive at
\begin{equation}
\label{eq:simplified I1 with small mu at small T}
\begin{split}
    I_1'|_{\mu \rightarrow 0, T\rightarrow T_{\rm min}}
    \approx\sqrt{\frac{l_2}{l_1}-1}\cdot\ln{\delta}\rightarrow-\infty.
\end{split}
\end{equation}
While, for $T=T_{\rm max}$ in \eqref{eq:simplified I1 with small mu}, it is straightforward to get
\begin{equation}
\label{eq:simplified I1 with small mu at large T}
\begin{split}
    I_1'|_{\mu \rightarrow 0, T\rightarrow T_{\rm max}} =-\sqrt{\frac{l_2}{l_1}+1}\cdot\pi.
\end{split}
\end{equation}

Therefore, we find that, $F_{\rm num}|_{\mu\rightarrow0}$ will increase as a function of $T$ monotonically, and
\begin{equation} \label{eq:F_num_0}
    F_{\rm num}|_{\mu\rightarrow0,T\rightarrow T_{\rm min}} \approx \sqrt{\frac{l_2}{l_1}-1}\ln(T-T_{\rm min}) \rightarrow -\infty, \qquad F_{\rm num}|_{\mu\rightarrow0,T\rightarrow T_{\rm max}} \rightarrow \left(2-\sqrt{\frac{l_2}{l_1}+1} \right)\cdot\pi.
\end{equation}
It means that, for $l_2>3l_1$, $F_{\rm num}|_{\mu\rightarrow0}$ is less than zero for any $T$, while, for $l_2<3l_1$, there is a zero for $ F_{\rm num}|_{\mu\rightarrow0}$.

Now we discuss the asymptotic behaviors of \eqref{eq:denominator of gamma equation}.
First consider the limit $\mu\rightarrow\infty$, we have $\delta_{2,{\rm E1}}=\theta(Tl_1 - 1)$ from \eqref{eq:explicit expression of deltaiE1}. 
With expansions in \eqref{eq:expansion of parameter in special function with large mu}, \eqref{eq:Ip2 with elliptic integrals without defect} can be simplified to be
\begin{equation}
\label{eq:simplified I2 with large mu}
\begin{split}
    F_{\rm den}|_{\mu \rightarrow \infty}= &\frac{2\pi}{\sqrt{\mu}}\Theta(Tl_1 - 1) + I_2'|_{\mu \rightarrow \infty} , \\
    I_2'|_{\mu \rightarrow \infty}\approx &\frac{-2}{\sqrt{\mu}}[(T+T_{\rm min})(T+T_{\rm max})]^{-\frac{1}{2}}\left[l_2^{-1}\cdot K\left(\frac{(T-T_{\rm max})(T-T_{\rm min})}{(T+T_{\rm max})(T+T_{\rm min})}\right)+l_2\left(T^2-\frac{1}{l_1^2}\right)\right.\\
    &\times\left.\Pi\left(l_2^2(T_{\rm max}-T)(T-T_{\rm min}),\frac{(T-T_{\rm max})(T-T_{\rm min})}{(T+T_{\rm max})(T+T_{\rm min})}\right)\right].
\end{split}
\end{equation}
To proceed, we consider 
\begin{equation}
\label{eq:simplified dI2dT with large mu}
    \frac{{\rm d}}{{\rm d}T} (\sqrt{\mu}I_2'|_{\mu \rightarrow \infty} )\approx\frac{4T(E(v')-K(v'))}{l_2(T-T_{\rm min})(T_{\rm max}-T)\sqrt{(T+T_{\rm max})(T+T_{\rm min})}} > 0,
\end{equation}
where we use that $v'=\frac{(T-T_{\rm max})(T-T_{\rm min})}{(T+T_{\rm max})(T+T_{\rm min})}<0$ and $E(v')-K(v')$ has a similar form to \eqref{eq:Ev-Kv function}, leading to $E(v')-K(v')>0$.
 
Again, there is a subtlety: while the step function has a nontrivial contribution to $F_{\rm den}|_{\mu\rightarrow\infty}$, it can be shown that $F_{\rm den}|_{\mu\rightarrow\infty}$ is actually continuous. 
Also, there is a caveat in $\frac{{\rm d}}{{\rm d}T} (\sqrt{\mu}I_2'|_{\mu \rightarrow \infty} )$ at $T = l_1^{-1}$:  
$\sqrt{\mu}I_2'|_{\mu \rightarrow \infty}$ has a jump at in $T = l_1^{-1}$, but $\lim_{T \rightarrow l_1^{-1}+0^+}\frac{{\rm d}}{{\rm d}T} (\sqrt{\mu}I_2'|_{\mu \rightarrow \infty} ) = \lim_{T \rightarrow l_1^{-1}+0^-}\frac{{\rm d}}{{\rm d}T} (\sqrt{\mu}I_2'|_{\mu \rightarrow \infty} )$. 
Hence, we can conclude that $ F_{\rm den}|_{\mu\rightarrow\infty}$ is a monotonically increasing function of $T$.

Further, it is straightforward to show that
\begin{equation} \label{eq:F_den_infty}
     F_{\rm den}|_{\mu\rightarrow\infty, T = T_{\rm min}} \approx \sqrt{1-\frac{l_1}{l_2}}\cdot\frac{\pi}{\sqrt{\mu}}, \qquad F_{\rm den}|_{\mu\rightarrow\infty, T = T_{\rm max}} \approx \left(2-\sqrt{1+\frac{l_1}{l_2}} \right)\frac{\pi}{\sqrt{\mu}}
\end{equation}
Since we assume $l_2 > l_1$, we have $F_{\rm den}|_{\mu\rightarrow\infty} > 0 $ for all $T$.

Consider the limit $\mu\rightarrow0$, using the expansion \eqref{eq:expansion of parameter in special function with small mu} and $\delta_{2,{\rm E1}}=1$ from \eqref{eq:explicit expression of deltaiE1}, the denominator can be simplified as 
\begin{equation}
\label{eq:simplified denominator with small mu}
\begin{split}
    F_{\rm den}|_{\mu\rightarrow 0} 
    =& \frac{2\pi}{\sqrt{\mu}}+\mathcal{O}(1)\rightarrow+\infty.
\end{split}
\end{equation}

To summarize, at $\mu\rightarrow\infty$, our analysis in \eqref{eq:simplified numerator with large mu} and \eqref{eq:F_den_infty} implies 
\begin{equation} \label{eq:gamma_mu_infty}
      \lim_{\mu \rightarrow \infty }\frac{F_{\rm num}}{F_{\rm den}} = + \infty.
\end{equation} 
On the other hand, at $\mu\rightarrow0^+$, our analysis in \eqref{eq:F_num_0} and \eqref{eq:simplified denominator with small mu} indicates
\begin{equation}
\frac{F_{\rm num}}{F_{\rm den}} \approx \begin{cases} \label{eq:gamma_mu_0}
\sqrt{\frac{l_2}{l_1}-1}\ln(T-T_{\rm min})\cdot \frac{\sqrt{\mu}}{2\pi}, \quad & T \approx T_{\rm min} \\ 
\left(2-\sqrt{\frac{l_2}{l_1}+1} \right)\pi \cdot \frac{\sqrt{\mu}}{2\pi}, \quad & T = T_{\rm max}
\end{cases}.
\end{equation}
This implies that for $l_2<3l_1$ the right-hand side of \eqref{eq:equation for mu} goes to $0^-$ for a small tension $T$, and goes to $0^+$ for a large tension $T$.
While, for $l_2>3l_1$, the right-hand side of \eqref{eq:equation for mu} goes to $0^-$ for $\mu\rightarrow0^+$ with any tension $T$.
We plot two examples with $l_2=2l_1=2$ and $T=0.55,1.55$ in Fig.~\ref{fig:two example of gamma mu} (a).
It shows that for a small tension $T=0.55$, there is a finite solution $\mu_0>0$ for $\gamma\rightarrow0$, which corresponds to a bubble solution.
While, with a large tension $T=1.45$, we have $\mu\rightarrow0$ for $\gamma\rightarrow0$, which corresponds to the no-bubble-solution phase.

Therefore, for $\gamma\rightarrow\infty$, we expect $\mu\rightarrow\infty$ from \eqref{eq:equation for mu}, which means there is no bubble-solution phase.
For $\gamma\rightarrow0^+$, there are two possibilities.
For a large tension and $l_2<3l_1$, we have $\mu\rightarrow0^+$ for $\gamma\rightarrow0^+$, which corresponds to the no-bubble-solution phase.
While, for a small tension with $l_2<3l_1$ or any tension with $l_2>3l_1$, we will have a finite solution $\mu_0>0$ with $\gamma\rightarrow0$ because $\mu\rightarrow0^-$ for $\gamma\rightarrow0^+$.
It is the bubble-solution phase.

Based on the discussion above, the condition for the bubble phase is that $B(T,l_1,l_2) \equiv \lim_{\mu\rightarrow0} F_{\rm num} < 0$. 
Namely, the phase boundary is determined by
\begin{equation}
\label{eq:criterion of bubble phase B}
    \begin{split}
    B(T,l_1,l_2) = & 2\pi\Theta\left(T^2l_2^2 - 1 \right)-2[(T-T_{\rm min})(T+T_{\rm max})]^{-\frac{1}{2}}\left[l_1^{-1}\cdot K\left(\frac{(T-T_{\rm max})(T+T_{\rm min})}{(T+T_{\rm max})(T-T_{\rm min})}\right)\right.\\
    &\left.+l_1\left(T^2-\frac{1}{l_2^2}\right)\Pi\left(l_1^2(T_{\rm max}-T)(T+T_{\rm min}),\frac{(T-T_{\rm max})(T+T_{\rm min})}{(T+T_{\rm max})(T-T_{\rm min})}\right)\right].
    \end{split}
\end{equation}
It is interesting to note that this function is the same as the condition of phase transition in Ref.~\cite{simidzija2020holoween}, though the transition is very different from ours.
Defining
\begin{equation}
\label{eq:Xi special function}
    \Xi_{u_0}(T,l_1,l_2)=\sqrt{\frac{4v}{-u}}\left[K\left(u_0,v\right)+(l_1^2T^2-l_1^2/l_2^2)\cdot\Pi\left(u_0,u,v\right)\right],
\end{equation}
with $u,v$ in \eqref{eq:expansion of parameters in the elliptic integrals in I1p with small mu}, we have $B(T,l_1,l_2) = 2\pi\Theta\left(l_2T-1\right)-\Xi_{1}(T,l_1,l_2)$.
Also, we can observe that
\begin{equation} \label{eq:F_den_infty_B21}
     F_{\rm den}|_{\mu \rightarrow \infty} = \frac{B(T,l_2,l_1) }{\sqrt \mu}.
\end{equation}
Since we have shown that $F_{\rm den}|_{\mu \rightarrow \infty} > 0 $ for any $T$, it means  $B(T,l_2,l_1)>0$.
Note that we have assumed $l_2 > l_1$, so $B(T,l_2,l_1)$ is distinct from $B(T,l_1,l_2)$, the latter can be either positive and negative.

\subsubsection{On-shell action}
\label{sec:Equipartition limit}

We derive the onshell action near big and small cusp angles in this section, which is used to get the cusp anomalous dimension in the main text.
In general, the derivation of onshell action is given by \eqref{eq:total action final results} in Appendix~\ref{sec:Review of the reference bachas2021phases}.

We first discuss big cusp angles. 
At $\phi = \pi$, $\gamma=\frac{L_1}{L_2}=1$ 
we have $\mu=1$, i.e., $M_1=M_2=M$.
From \eqref{eq:explicit expression of deltaiE1}, we have $\delta_{1,{\rm E1}}=1$ and $\delta_{2,{\rm E1}}=\Theta(Tl_1-1)+\Theta(1-Tl_1)\Theta(T^2-T_0^2)=\Theta(T-T_0)$.
Therefore, with \eqref{eq:constraint for Mi with Li}, \eqref{eq:explicit brane equation for x1} and \eqref{eq:explicit brane equation for x2}, we have
\begin{equation}
\label{eq:simplified L12 with L1L2}
    L_1=\frac{2\pi}{\sqrt{-M}}\delta_{1,{\rm E1}}+\left(-\frac{\pi}{\sqrt{-M}}\right)=\frac{\pi}{\sqrt{-M}},\qquad
    L_2=\frac{2\pi}{\sqrt{-M}}\delta_{2,{\rm E1}}+\left(-{\rm sgn}(T^2-T_0^2)\frac{\pi}{\sqrt{-M}}\right)=\frac{\pi}{\sqrt{-M}},
\end{equation}
which give the consistent result that $L_1=L_2$.
Then, the onshell action reads
\begin{equation}
\label{eq:total action final results with L1L2}
\begin{split}
    I_{\rm tot}&=\frac{M L}{2T_{\rm DCFT}}\cdot\frac{l_1+l_2}{2}=\frac{l_1+l_2}{4T_{\rm DCFT}}\cdot\left(-\frac{4\pi^2}{L}\right)=-\frac{\pi^2(l_1+l_2)}{T_{\rm DCFT}}\cdot\frac{1}{L}.
\end{split}
\end{equation}

Now we can expand the result around $\gamma=1$.
Because $\gamma(\mu=1)=1$ for any $(T,l_1,l_2)$, we can consider $\mu=1+\delta_\mu$ with $|\delta_\mu|\ll1$.
From \eqref{eq:explicit expression of deltaiE1}, for $l_1<l_2$ and a small enough $\delta_\mu$, we always have $\delta_{1,{\rm E1}}=1$, and $\delta_{2,{\rm E1}}=\Theta(-\mu_2^*)+\Theta(\mu_2^*)\Theta(\mu_2^*-\mu)$, where $\mu^\ast_{1,2} $ are defined in~\eqref{eq:critical mu for AdSi}. 
Expanding $s_\pm$ in the order of $\delta_\mu$ gives
\begin{equation}
\label{eq:expansion about delta for spn}
    s_\pm=\frac{2(T^2\pm T^2)}{(T_{\rm max}^2-T^2)(T^2-T_{\rm min}^2)}+\frac{((T_0^2+T^2)\pm(T_0^2+T^2))\delta_\mu}{(T_{\rm max}^2-T^2)(T^2-T_{\rm min}^2)}\pm\frac{\delta_\mu^2}{4T^2}+\mathcal{O}(\delta_\mu^3),
\end{equation}
which means $s_-/s_+=\mathcal{O}(\delta_\mu^2)$.
We can plug \eqref{eq:expansion about delta for spn} into $F_{\rm num}$ and $F_{\rm den}$ in \eqref{eq:numerator and denominator of gamma equation} and expand them w.r.t. $\delta_\mu$ to get
\begin{subequations}
\label{eq:numerator and denominator of gamma equationn for gammaapp1}
\begin{equation}
\label{eq:numerator of gamma equation for gammaapp1}
    F_{\rm num}\approx\pi+\frac{\pi}{2l_1T}\delta_\mu-\frac{3\pi(T_0^2+T^2)}{16l_1T^3}\delta_\mu^2+\mathcal{O}(\delta_\mu^3),
\end{equation}
\begin{equation}
\label{eq:denominator of gamma equation for gammaapp1}
    F_{\rm den}=\pi-\left(\frac{1}{l_2T}+1\right)\cdot\frac{\pi}{2}\delta_\mu+\frac{3\pi}{16}\left(\frac{T_0^2+3T^2}{l_2T^3}+2\right)\delta_\mu^2+\mathcal{O}(\delta_\mu^3).
\end{equation}
\end{subequations}
Then we can simplify the equation for $\gamma$
\begin{equation}
\label{eq:simplified gamma equation for gammaapp1}
    \gamma=\frac{F_{\rm num}}{F_{\rm den}}\approx1+\frac{T_{\rm max}+T}{2T}\delta_\mu+\frac{(T_{\rm max}+T)(-3T_0^2+T(\frac{1}{l_2}+\frac{3}{l_1})-2T^2)}{16T^3}\delta_\mu^2+\mathcal{O}(\delta_\mu^3).
\end{equation}
It is worth mentioning that the right-hand side of \eqref{eq:simplified gamma equation for gammaapp1} is not symmetric by exchanging $l_1$ and $l_2$.
Defining $\gamma=\delta_\gamma+1$, we have
\begin{equation}
\label{eq:solution for Delta delta with condition}
\begin{split}
    \delta_\mu=\frac{2l_1l_2T}{l_1+l_2+l_1l_2T}\delta_\gamma+\frac{l_1^2l_2(-4l_2T-l_2^2T^2-4l_1T+l_1l_2T^2+2l_1l_2^2T^3+3l_2(l_1+l_2)(l_1^{-2}-l_2^{-2}))}{2(l_1+l_2+l_1l_2T)^3}\delta_\gamma^2+\mathcal{O}(\delta_\gamma^3)
\end{split}
\end{equation}

Now, we can calculate the on-shell action with \eqref{eq:total action final results}.
Comparing \eqref{eq:constraint for Mi with Li} and \eqref{eq:numerator and denominator of gamma equation}, we know that
\begin{equation}
    F_{\rm num} = \sqrt{-M_1}L_1 .
\end{equation}
Therefore, with $L_1+L_2=L$, the onshell action is
\begin{equation}
\label{eq:total action final results for gammaapp1}
\begin{split}
    I_{\rm tot}&=\frac{M_1 l_1 L_1+M_2 l_2 L_2}{2T_{\rm DCFT}}=\frac{M_1 l_1 L_1}{2T_{\rm DCFT}}\left(1+\frac{\mu l_2}{\gamma l_1}\right)=-\frac{1}{2T_{\rm DCFT}}\left(1+\frac{\mu l_2}{\gamma l_1}\right)(1+\gamma^{-1})\frac{l_1}{L}\cdot F_{\rm num}^2\\
    &\approx\frac{-1}{2T_{\rm DCFT}L}\left[2(l_1+l_2)\pi^2+(l_2-l_1)\pi^2\delta_\gamma+\frac{l_1(l_1-l_2+l_1l_2T)}{l_1+l_2+l_1l_2T}\pi^2\delta_\gamma^2\right],
\end{split}
\end{equation}
where the leading term is consistent with \eqref{eq:total action final results with L1L2}. 
This leads to Eq.~\eqref{eq:cusp_dimension_big} in the main text.

If $l_1=l_2=l$, the onshell action can be simplified as  
\begin{equation}
\label{eq:total action final results for gammaapp1 with l1l2}
    I_{\rm tot}\approx-\frac{1}{2T_{\rm DCFT}L}\left(4l\pi^2+\frac{l^2T}{2+lT}\pi^2\delta_\gamma^2\right).
\end{equation}
The lowest order of the correction is quadratic, because there is a $\mathbb{Z}_2$ symmetry of exchanging two AdS spacetime for $l_1=l_2$ and $L_1=L_2$.
Moreover, if $T=0$, the second order correction also vanishes.

Now, we discuss small cusp angles $\phi \approx 0$ or $\phi \approx 2\pi$. 
We consider the expansion around $\gamma=\frac{L_1}{L_2}\rightarrow0$ (or $\gamma=\frac{L_1}{L_2}\rightarrow\infty$).  

For $\gamma=\frac{L_1}{L_2}\rightarrow\infty$, as discussed in Sec.~\ref{sec:Bubble phase condition}, we have $\mu\rightarrow\infty$.
Then from \eqref{eq:simplified numerator with large mu} and \eqref{eq:F_den_infty_B21}, we arrive at
\begin{equation}
\label{eq:num and den with large mu}
    F_{\rm num}|_{\mu \rightarrow \infty} \approx 2\pi, \quad 
    F_{\rm den}|_{\mu \rightarrow \infty} \approx \frac{B(T,l_2,l_1)}{\sqrt{\mu}}, \quad  
    \gamma=\frac{F_{\rm num}}{F_{\rm den}}\Big|_{\mu \rightarrow \infty}\approx\frac{2\pi}{B(T,l_2,l_1)}\sqrt{\mu}+\mathcal{O}(1).
\end{equation}
Consequently, the onshell action reads
\begin{equation}
\label{eq:total action final results with large gamma}
\begin{split}
    I_{\rm tot}&=\frac{M_1 l_1 L_1+M_2 l_2 L_2}{2T_{\rm DCFT}}=-\frac{1}{2T_{\rm DCFT}L}\left(l_1+\frac{\mu l_2}{\gamma}\right)(1+\gamma^{-1})\cdot F_{\rm num}^2|_{\mu \rightarrow \infty}
    =-\frac{1}{2T_{\rm DCFT}L}\left(l_2\cdot B(T,l_2,l_1)^2\gamma+\mathcal{O}(1)\right).
\end{split}
\end{equation}

For $\gamma = \frac{L_1}{L_2}\rightarrow0$, as discussed in Sec.~\ref{sec:Bubble phase condition}, there are two possibilities.
If it is the bubble-solution phase with $B(T,l_1,l_2)<0$, then there is a solution $\mu_0>0$ such that $\gamma(\mu_0)=0$.
Similar to \eqref{eq:total action final results with large gamma}, the onshell action can be written as
\begin{equation}
\label{eq:total action final results with small gamma}
\begin{split}
    I_{\rm tot}=\frac{M_1 l_1 L_1+M_2 l_2 L_2}{2T_{\rm DCFT}}=-\frac{1}{2T_{\rm DCFT}L}\left(l_1\gamma+l_2\mu\right)(1+\gamma)\cdot F_{\rm den}^2|_{\mu = \mu_0}.
\end{split}
\end{equation}
From \eqref{eq:denominator of gamma equation}, we conclude that $F_{\rm den}^2(\mu_0)$ is also a constant.
On the other hand, if it is a no-bubble-solution phase, we have 
\begin{equation}
\label{eq:num and den with small mu}
    F_{\rm num}|_{\mu \rightarrow 0} \approx B(T,l_1,l_2), \quad 
    F_{\rm den}|_{\mu \rightarrow 0} \approx \frac{2\pi}{\sqrt{\mu}}, \quad
    \gamma=\frac{F_{\rm num}}{F_{\rm den}}\Big|_{\mu \rightarrow 0}\approx\frac{B(T,l_1,l_2)}{2\pi}\sqrt{\mu}.
\end{equation}
The onshell action becomes
\begin{equation}
\label{eq:total action final results with large gamma no bubble solution}
\begin{split}
    I_{\rm tot}
    &=-\frac{1}{2T_{\rm DCFT}L}\left(\frac{l_1\cdot B(T,l_1,l_2)^2}{\gamma}+\mathcal{O}(1)\right).
\end{split}
\end{equation}

The onshell action~\eqref{eq:total action final results with large gamma no bubble solution} and~\eqref{eq:total action final results with large gamma} leads to the cusp anomalous dimension in Eq.~\eqref{eq:cusp_dimension_small} in the main text.

\subsection{AdS geometry with defect on brane}
\label{sec:AdS geometry with defect on brane}

In this section, we consider a nontrivial corner with $\theta_0\neq\pi$ on the brane.
In the following, we focus on $l_1=l_2=l$ and $T_a=T_b=T$ with a general $\theta_0$.

\subsubsection{AdS geometry with defect on brane with general tension}
\label{sec:AdS geometry with defect on brane with general T}

Following the general procedure, we first get the intersecting point from~\eqref{eq:coefficient of equation with s} with $l_1 = l_2 = l$ and $T_a = T_b = T$ (results not shown here). 
Secondly, we simplify the variables $\iota_A$ and $\iota_B$ in~\eqref{eq:equations for theta with s}. 
From~\eqref{eq:equations for theta a with s}, for the expression in the sign function to be real, we require $l^2T^2+2\cos{\theta_0}<2$ for $0<lT<\sqrt{2}$ and $\cos{\theta_0}<2/(l^2T^2)-1$ for $\sqrt{2}<lT<2$. 
The range for $\theta_0$ in the region $\sqrt{2}<lT<2$ is larger than that in $0<lT<\sqrt{2}$.
Later, we will see that a real solution actually requires $l^2T^2+2\cos{\theta_0}<2$ for all $0<lT<2$. 
Using this condition, we have $\iota_A=-\iota_B={\rm sgn}(\theta_0-\pi){\rm sgn}(l^2T^2-1)$ for $\mu\rightarrow0$.

After a tedious calculation, under the limit $\mu\rightarrow0$, \eqref{eq:Ip1 with elliptic integrals} can be simplified as
\begin{equation}
\label{eq: I1p for mu to 0 with defect on brane}
\begin{split}
    I_1'|_{\mu \rightarrow 0}=&2(1-{\rm sgn}(\theta_0-\pi){\rm sgn}(l^2T^2-1))\left(\frac{-K(\frac{lT-2}{lT+2})-(l^2T^2-1)\cdot\Pi(-lT(lT-2),\frac{lT-2}{lT+2})}{\sqrt{lT(lT+2)}}\right)\\
    +&2{\rm sgn}(\theta_0-\pi){\rm sgn}(l^2T^2-1)\left(\frac{-K(u_0,\frac{lT-2}{lT+2})-(l^2T^2-1)\cdot\Pi(u_0,-lT(lT-2),\frac{lT-2}{lT+2})}{\sqrt{lT(lT+2)}}\right),
\end{split}
\end{equation}
where $u_0=\sqrt{\frac{l^2T^2+2\cos{\theta_0}-2}{(lT-2)(\sqrt{(l^2T^2+l^2T^2\cos{\theta_0}-2)(\cos{\theta_0}-1)}-lT\cos{\theta_0})}}$.
Here we can see that for $u_0$ to be real, the range of parameters is given by $l^2T^2+2\cos{\theta_0}<2$ for $0<lT<2$.
Therefore, we have 
\begin{equation}
    \tilde B(T,l,\theta_0) \equiv F_{\rm num}|_{\mu \rightarrow 0}=I_1'|_{\mu \rightarrow 0}+\pi(1-{\rm sgn}(I_1'|_{\mu \rightarrow 0})),
\end{equation}
where $I_1'|_{\mu \rightarrow 0}$ is given in \eqref{eq: I1p for mu to 0 with defect on brane}. 
Next, we check the value of $I_2'|_{\mu \rightarrow 0}$ in the range of $0<lT<2$ and $-1<\cos{\theta_0}<1-l^2T^2/2$ is finite and negative.
It in turn indicates that
\begin{equation}
    F_{\rm den}|_{\mu \rightarrow 0} = I_2'|_{\mu \rightarrow 0}+\frac{\pi}{\sqrt \mu}(1-{\rm sgn}(I_2'|_{\mu \rightarrow 0})) \rightarrow \frac{2\pi}{\sqrt{\mu}} .
\end{equation}
We have the behavior of $\gamma(\mu)$ near $\mu \rightarrow 0$
\begin{equation}
    \gamma(\mu \rightarrow 0) = \frac{F_{\rm num}}{F_{\rm den}}\Big|_{\mu \rightarrow 0} = \frac{\sqrt\mu}{2\pi} \cdot\tilde B(T,l,\theta_0).
\end{equation}
Similar to the discussion in Sec.~\ref{sec:Bubble phase condition}, the sign of $\tilde B(T,l,\theta_0)$ determines a phase transition.
Namely, $\tilde B(T,l,\theta_0) > 0$ is the no-bubble-solution phase, while $\tilde B(T,l,\theta_0) < 0$ is the bubble-solution phase. 
We plot two examples with $l_1=l_2=1, T=1.25$ and $\theta_0=\pi/2,3\pi/4$ in Fig.~\ref{fig:two example of gamma mu} (b).
It shows that for a small corner angle $\theta_0=\pi/2$, there is a finite solution $\mu_0>0$ for $\gamma\rightarrow0$, which corresponds to a bubble solution.
While, with a large corner angle $\theta_0=3\pi/4$, we have $\mu\rightarrow0$ for $\gamma\rightarrow0$, which corresponds to the no-bubble-solution phase.
Notice the opposite signs at the limit $\mu \rightarrow 0$. 

In the same manner, we can also get the asymptotic behavior of the onshell action with $\gamma\rightarrow0$ for the no-bubble-solution phase
\begin{equation}
\label{eq:total action final results with large gamma no bubble solution qith defect on brane general solution}
\begin{split}
    I_{\rm tot}&=-\frac{1}{2T_{\rm DCFT}L}\left(\frac{l}{\gamma} \cdot \tilde B(T,l,\theta_0)^2 +\mathcal{O}(1)\right).
\end{split}
\end{equation}
To further simplify the result, 
we can prove that for $0<lT<2$, $I_1'|_{\mu \rightarrow 0}<0$. 
Therefore, in the range $\theta_0<\pi$ and $lT>1$, $\Tilde{B}(T,l,\theta_0)$ reads 
\begin{equation}
\label{eq:tB more general}
    \tilde{B}(T,l,\theta_0)=2\pi-2\ \Xi_1(T,l,l)+\ \Xi_{u_0}(T,l,l),
\end{equation}
where $\Xi_{u_0}(T,l_1,l_2)$ is defined in \eqref{eq:Xi special function}, and $u,v$ are the same as \eqref{eq:expansion of parameters in the elliptic integrals in I1p with small mu} with $l_1=l_2=l$.
This gives Eq.~\eqref{eq:cusp_dimension_corner} in the main text.

\subsubsection{Special case of AdS geometry with a corner on branes}
\label{sec:AdS geometry with defect on brane with T 1}

In this section, we focus on a special case $lT=1$.
Proceeding as before, the range of $\theta_0$ in this special case is  $-1<\cos{\theta_0}<1/2$.
This range leads to $\iota_A=-\iota_B={\rm sgn}(\theta_0-\pi)$.
Actually, it is what we expect because of the $\mathbb{Z}_2$ symmetry in this case, which maps one brane to another.
Then, because we only care about the solution for $\gamma\rightarrow0$, we expand $I_1'$ in \eqref{eq:equation Ip with elliptic integrals} around $\mu\rightarrow0$. 
Most of the derivation is straightforward, but there is an expansion of the incomplete elliptic integral which is a little complicated.
For $\Pi(y_0=\sin{\pi/2-a\mu},u=1-b\mu^2,v=m)$, we can expand it around $\mu\rightarrow0$
\begin{equation}
\label{eq:third kind incomplete elliptic integral expasion}
\begin{split}
    &\Pi(\sin{(\pi/2-a\mu)},1-b\mu^2,m)=\int_0^{1-a^2\mu^2/2}\frac{{\rm d}t}{(1-(1-b\mu^2)t^2)\sqrt{(1-t^2)(1-mt^2)}}\\
    =&\Pi(1,1-b\mu^2,m)-\int_0^1\frac{\frac{a^2\mu^2}{2}{\rm d}x}{(1-(1-b\mu^2)(1-\frac{a^2\mu^2}{2}x)^2)\sqrt{(1-(1-\frac{a^2\mu^2}{2}x)^2)(1-m(1-\frac{a^2\mu^2}{2}x)^2)}}\\
    =&\Pi(1,1-b\mu^2,m)-\frac{\arctan{\sqrt{\frac{a^2}{b}}}}{\sqrt{b(1-m)}}\cdot\frac{1}{\mu}+\mathcal{O}(1).
\end{split}
\end{equation}
Then, using the expansion above, we can simplify the numerator in \eqref{eq:equation for mu}, and find that
\begin{equation}
\label{eq:num with small mu for special case T 1}
    \Tilde{B}(1/l,l,\theta_0) \approx \theta_0- \theta_0^c, \qquad \theta_0^c = \frac{2}{\sqrt{3}}\cdot K\left(-\frac{1}{3}\right). 
\end{equation}

We can calculate the denominator in \eqref{eq:equation for mu} and prove it is positive in the limit of $\mu \rightarrow 0$.  
The derivation of the expansion of the denominator is tedious, and there are also two expansions of equations are useful:
\begin{equation}
\label{eq:third kind incomplete elliptic integral expasion difference}
\begin{split}
    \Pi(\mu,-\frac{1}{3})-\Pi(\mu,-\frac{1}{3}+\frac{2\mu}{3})=&\int_0^{1}\frac{{\rm d}t}{(1-\mu t^2)\sqrt{(1-t^2)}}\left(\frac{1}{\sqrt{1+\frac{1}{3}t^2}}-\frac{1}{\sqrt{1+\frac{1-2\mu}{3}t^2}}\right)\\
    =&\left[\frac{3}{4}\cdot E\left(-\frac{1}{3}\right)-K\left(-\frac{1}{3}\right)\right]\cdot\mu+\mathcal{O}(\mu^2),
\end{split}
\end{equation}
and similar to \eqref{eq:third kind incomplete elliptic integral expasion}, we also have
\begin{equation}
\label{eq:third kind incomplete elliptic integral expasion for den}
\begin{split}
    &\Pi\left(\sin{(\pi/2-a\mu)},\mu,\frac{2\mu-1}{3}\right)=\int_0^{1-a^2\mu^2/2}\frac{{\rm d}t}{(1-\mu)t^2)\sqrt{(1-t^2)(1-\frac{2\mu-1}{3}t^2)}}\\
    =&\Pi(1,\mu,\frac{2\mu-1}{3})-\int_0^1\frac{\frac{a^2\mu^2}{2}{\rm d}x}{(1-\mu(1-\frac{a^2\mu^2}{2}x)^2)\sqrt{(1-(1-\frac{a^2\mu^2}{2}x)^2)(1-\frac{2\mu-1}{3}(1-\frac{a^2\mu^2}{2}x)^2)}}\\
    =&\Pi(1,\mu,\frac{2\mu-1}{3})-\frac{\sqrt{3}a}{2}\cdot\mu+\mathcal{O}(\mu^2).
\end{split}
\end{equation}
With these expansions, we find $I_2'|_{\mu \rightarrow 0}=-\frac{1}{\sqrt{3}}(6E(-1/3)+4K(-1/3))<0$ for $\mu\rightarrow0$, which leads to $F_{\rm den}|_{\mu\rightarrow 0}\approx2\pi/\sqrt{\mu} $.

Therefore, if $\theta_0>\theta_0^c$, the positive numerator leads to the no-bubble-solution phase with a divergent action which is proportional to $\gamma^{-1}$.
While, for $\theta_0<\theta_0^c$, there is a bubble phase with a finite action for $\gamma\rightarrow0$.
And remember that here we focus on $\cos{\theta_0}\in(-1,1/2)$ with $\theta_0\in(\frac{\pi}{3},\frac{5\pi}{3})$.

\subsection{Ising model with defects}
\label{sec:Ising model with defects}

In this section, we focus on the Ising CFT with periodic boundary condition and two defects. 
To be concrete, consider the following Hamiltonian
\begin{equation}
\label{eq:Ising CFT with defects}
H=-\frac{1}{2}\sum'_r [\sigma^z(r)+\sigma^x(r)\sigma^x(r+1)] - \frac{1}{2}  \kappa_a \sigma^x(L_1+L_2)\sigma^x(1) - \frac12 \kappa_b \sigma^x(L_1)\sigma^x(L_1+1),
\end{equation}
where $\sigma^\mu(r) $ denotes the Pauli matrix at site $r$ and the first summation omits the terms at the two defect bonds with strength $\kappa_\alpha$.

The Hamiltonian~\eqref{eq:Ising CFT with defects} can be exactly diagonalized~\cite{henkel1989ising}, and for the two defects case, we define $\phi=2\pi\alpha=2\pi p/q$, with coprime integers $p,q$, to denote the location of two defects.
With two defects strength $\kappa_a$ and $\kappa_b$, the eigenvalues of \eqref{eq:Ising CFT with defects} are related to the solution of the algebraic equation 
\begin{equation}
\label{eq:equation of Hamiltonian}
    (1-\kappa_a^2)(1-\kappa_b^2)\cos{[x(1-2\alpha)]}+(1+\kappa_a^2)(1+\kappa_b^2)\cos{x}+4\kappa_a\kappa_b(1-2Q)=0,
\end{equation}
where $Q=0$ $(1)$ corresponds to the odd (even) sector of \eqref{eq:Ising CFT with defects} after Jordan-Wigner transformation. 
With the solution for $x$, we can calculate the energy spectrum, particularly the lowest eigenvalue.
It is easy to check that~\eqref{eq:equation of Hamiltonian} has a periodicity $2\pi q$.
In general, we can define $z={\rm e}^{{\rm i}x/q}$, then~\eqref{eq:equation of Hamiltonian} becomes a polynomial of $z$ of order $2q$:
\begin{equation}
\label{eq:equation of complex form}
    \frac{1}{2}(1-\kappa_a^2)(1-\kappa_b^2)(z^{q-2p}+z^{-q+2p})+\frac{1}{2}(1+\kappa_a^2)(1+\kappa_b^2)(z^{q}+z^{-q})+4\kappa_a\kappa_b(1-2Q)=0,
\end{equation}
Therefore, \eqref{eq:equation of complex form} has $2q$ roots, and if $z_0$ is a root, $1/z_0$ is also a root.
With the roots $x_i\in[0,\Tilde{\gamma}(q)\cdot\pi]$ and $i=1,2,...,\Tilde{\gamma}(q)$, the universal $O(\frac{1}{L})$ part of the lowest eigenvalue is
\begin{subequations}
\label{eq:constant part for two defects case}
\begin{equation}
        E=\frac{2\pi \Tilde{\gamma}(q)}{L}\sum_{i=1}^{\Tilde{\gamma}(q)}\left[-\frac{1}{12}\left(\frac{1}{2}-6\Tilde{\Delta}_i^2\right)\right],
\end{equation}
\begin{equation}
        \Tilde{\Delta}_i=\frac{1}{2}-\frac{1}{2\pi \Tilde{\gamma}(q)}x_i,
\end{equation}
\end{subequations}
where $\Tilde{\gamma}(q)=q\ (q/2)$ for the odd (even) $q$.

In the following, we consider defect changing operator for distinct $\kappa_{a,b}$ at $\phi = \pi$ as well as the cusp anomalous dimension for two identical defects $\kappa_a=\kappa_b=\kappa$ at strong and weak defect limit.

\subsubsection{Defect changing operator in different sectors}

When $\phi = \pi$, $\alpha=1/2$, it is straightforward to get the solution of~\eqref{eq:equation of Hamiltonian}, and consequently,
\begin{equation}
    \tilde \Delta_Q(\kappa_a, \kappa_b) =  \frac1\pi \left| \arctan \kappa_a - (-1)^Q \arctan \kappa_b \right|.
\end{equation}
Note that there is only one solution we need to consider. 
Then the scaling dimension of the defect changing operator is given by
\begin{equation}
    \Delta_{ab}^{11}(\pi) = \pi \left[ \tilde \Delta_Q(\kappa_a, \kappa_b)^2 - \frac12 (\tilde \Delta_0(\kappa_a, \kappa_a)^2 + \tilde \Delta_0(\kappa_b, \kappa_b)^2) \right] = \frac{1}{\pi } \left( \arctan \kappa_a - (-1)^Q \arctan \kappa_b \right)^2.
\end{equation}
In the $Q=0$ sector, $\Delta_{ab}^{11}(\pi)$ vanishes when $\kappa_a = \kappa_b$ as one would expect.
However, in the $Q=1$ sector, when $\kappa_a = \kappa_b = \kappa$, $\Delta_{ab}^{11}(\pi) = \frac{4}{\pi } \arctan^2 \kappa  $ still nontrivially depends on $\kappa$. 
Note that this happens in the holographic model when $\theta_0 \ne \pi$.
Hence, the effect of a nontrivial corner term is similar to the different sector $Q=0,1$ here.

\subsubsection{Weak defect limit}
\label{sec:Weak defect limit}

We discuss the weak defect limit with expansion around $\kappa\approx1$.
To proceed, we consider $\alpha=p/q=p/(2n+1)$ with $p$ being an even integer and an integer $n$.
Actually, this parametrization does not guarantee $p,q$ are coprime, but we will find that the result below is still valid.
$q=2n+1$ gives $\Tilde{\gamma}(q)=q$.
Taking $\kappa=1$,~\eqref{eq:equation of Hamiltonian} gives $x=(2m+1)\pi$ for $m=0,1,...,n$.
Once we include a perturbation $\kappa = 1 - \delta\kappa$, two degenerate roots at $x=(2m+1)\pi$ for $0\leq m<n$ will split and give rise to $2n$ roots; while the correction to the solution $x=(2n+1)\pi$ leads to another root which is less than $(2n+1)\pi$ (the other one that is greater than $(2n+1)\pi$ is neglected).
To the second order in $\delta \kappa$, we set $x=(2m+1)\pi+a \cdot\delta\kappa+b\cdot\delta\kappa^2$ and plug it into~\eqref{eq:equation of Hamiltonian}.
The vanishing of the coefficient in $\delta\kappa$ and $\delta\kappa^2$ then leads to the equations of $a$ and $b$ as follows,
\begin{subequations}
\label{eq:equations for a and b}
\begin{equation}
        a^2-2+2\cos[(2m+1)\pi(1-2\alpha)]=0,
\end{equation}
\begin{equation}
        (1-a^2+ab)+\cos[2\pi(\alpha-m+2m\alpha)]+a(2\alpha-1)\sin[2\pi(\alpha-m+2m\alpha)]=0.
\end{equation}
\end{subequations}
The corresponding solutions are
\begin{subequations}
\label{eq:solutions for a and b}
\begin{equation}
        a=\pm2\cos{[\pi(\alpha-m+2m\alpha)]},
\end{equation}
\begin{equation}
        b=\pm\cos{[\pi(\alpha-m+2m\alpha)]}-(2\alpha-1)\sin{[2\pi(\alpha-m+2m\alpha)]}.
\end{equation}
\end{subequations}
With these corrections, the energy \eqref{eq:constant part for two defects case} becomes  
\begin{equation}
\begin{split}
\label{eq:simplify subleading contribution for small kappa}
        E&=-\frac{2\pi (2n+1)}{24L}\left\{\sum_{m=0}^{n-1}\left[2-12\left(2\left(\frac{1}{2}-\frac{(2m+1)\pi}{2\pi(2n+1)}\right)^2+2\left(\frac{a_m}{2\pi(2n+1)}\right)^2\delta\kappa^2\right.\right.\right.\\
        &\left.\left.\left.+2\left(\frac{1}{2}-\frac{(2m+1)\pi}{2\pi(2n+1)}\right)\frac{-b_m-b_m'}{2\pi(2n+1)}\delta\kappa^2\right)\right]+1-12\left(\frac{2\delta\kappa}{2\pi(2n+1)}\right)^2\right\}+O(\delta\kappa^3).
\end{split}
\end{equation}
where $a_m, b_m$ and $a_m', b_m'$ denote the $+$ and $-$ solution in~\eqref{eq:solutions for a and b}.
Performing the summation over $m$, we arrive at the final result
\begin{equation}
\label{eq:total contribution for small kappa}
        E=-\frac{\pi}{12L}\cdot\left[1+\frac{6}{\pi^2}\left(\frac{(1-2\alpha)\pi}{\sin{2\alpha\pi}}-1\right)\delta\kappa^2\right]+O(\delta\kappa^3),
\end{equation}
which leads to Eq.~\eqref{meq:ising_cusp} in the main text.

\subsubsection{Strong defect limit}
\label{sec:Strong defect limit}

We can also consider another limit $\kappa=\delta\kappa\ll1$.
In this limit, \eqref{eq:equation of Hamiltonian} becomes 
\begin{equation}
\label{eq:equation for near zero defect}
    (1+\delta\kappa^2)^2\cos{[x(1-2\alpha)]}+(1+\delta\kappa^2)^2\cos{x}+4\delta\kappa^2=0.
\end{equation}
Similarly, we consider $\alpha=p/q=p/(2n+1)$ with integer $n$.
If $\kappa=0$, then the zeroth order solution is $x=(\frac{1}{2}+m_1)\frac{\pi}{1-\alpha},(\frac{1}{2}+m_2)\frac{\pi}{\alpha}$ with $0\leq\frac{1}{2}+m_1\leq(1-\alpha)q=q-p$ and $0\leq\frac{1}{2}+m_2\leq\alpha q=p$.
$m_1=0,1,...,q-p-1$ and $m_2=0,1,...,p-1$ constitute the $q$ solutions.
To the second order in $\delta \kappa$, it is straightforward to obtain the correction for the $m_1$ type solution $x=\frac{(2m_1+1)\pi}{2(1-\alpha)}+a\delta\kappa+b\delta\kappa^2$ with $a=0$ and $b=\frac{2}{(1-\alpha)\tan{\frac{2m_1+1}{4(1-\alpha)}\pi}}$, and the correction to the $m_2$ type solution $x=\frac{(2m_2+1)\pi}{2\alpha}+a\delta\kappa+b\delta\kappa^2$ with $a=0$ and $b=\frac{2}{\alpha\tan{\frac{2m_2+1}{4\alpha}\pi}}$.

Next we need to substitute the corrected solutions above in \eqref{eq:constant part for two defects case}, and performing the summation.
The summation cannot be done analytically, but we obtain the zeroth order result:
\begin{equation}
\begin{split}
\label{eq:zeroth order constant part for small kappa limit}
        E_0&=-\frac{2\pi (2n+1)}{24L}\left\{\sum_{m_1=0}^{q-p-1}\left[1-12\left(\frac{1}{2}-\frac{\frac{2m_1+1}{2(1-\alpha)}\pi}{2\pi(2n+1)}\right)^2\right]+\sum_{m_2=0}^{p-1}\left[1-12\left(\frac{1}{2}-\frac{\frac{2m_2+1}{2\alpha}\pi}{2\pi(2n+1)}\right)^2\right]\right\}\\
        &=-\frac{\pi}{12N}\frac{1}{4\alpha(1-\alpha)},
\end{split}
\end{equation}
where we see the same type of divergence at $\alpha\rightarrow0$ ($\alpha\rightarrow1$).

\newpage
\section*{Appendix}

\subsection{Special functions and arc equations}
\label{sec:Arc equations and special functions}

\subsubsection{Special functions}
\label{sec:Special functions}

In this section, we summarize the elliptic integrals, and discuss some properties of them. 
For the complete elliptic integrals of the first, second and third kind, we have
\begin{subequations}
\label{eq:complete elliptic integrals}
\begin{equation}
\label{eq:first kind complete elliptic integral}
    K(v)=\int_0^1\frac{{\rm d}t}{\sqrt{(1-t^2)(1-vt^2)}},
\end{equation}
\begin{equation}
\label{eq:second kind complete elliptic integral}
    E(v)=\int_0^1\frac{\sqrt{1-vt^2}{\rm d}t}{\sqrt{1-t^2}},
\end{equation}
\begin{equation}
\label{eq:third kind complete elliptic integral}
    \Pi(u,v)=\int_0^1\frac{{\rm d}t}{(1-ut^2)\sqrt{(1-t^2)(1-vt^2)}}.
\end{equation}
\end{subequations}
For the incomplete elliptic integrals of the first, second and third kind, we have
\begin{subequations}
\label{eq:incomplete elliptic integrals}
\begin{equation}
\label{eq:first kind incomplete elliptic integral}
    K(y_0,v)=\int_0^{y_0}\frac{{\rm d}t}{\sqrt{(1-t^2)(1-vt^2)}},
\end{equation}
\begin{equation}
\label{eq:second kind incomplete elliptic integral}
    E(y_0,v)=\int_0^{y_0}\frac{\sqrt{1-vt^2}{\rm d}t}{\sqrt{1-t^2}},
\end{equation}
\begin{equation}
\label{eq:third kind incomplete elliptic integral}
    \Pi(y_0,u,v)=\int_0^{y_0}\frac{{\rm d}t}{(1-ut^2)\sqrt{(1-t^2)(1-vt^2)}},
\end{equation}
\end{subequations}
where $K(v)=K(1,v)$, $E(v)=EK(1,v)$ and $\Pi(u,v)=\Pi(1,u,v)$.

With the definition above, in the following we list some properties of the complete elliptic integrals, which are used in the main text.
For $K(v)$, changing integral variable $t=\sqrt{1-s^2}$, it is easy to show that
\begin{equation}
\label{eq:property of K}
    K(v)=\frac{1}{\sqrt{1-v}}K\left(\frac{v}{v-1}\right).
\end{equation}
Similarly, for $\Pi(u,v)$, with $t=\sqrt{1-s^2}$, we have
\begin{equation}
\label{eq:property of Pi}
    \Pi(u,v)=\frac{1}{(1-u)\sqrt{1-v}}\Pi\left(\frac{u}{u-1},\frac{v}{v-1}\right).
\end{equation}
Besides, for the special case $v=0$, we have
\begin{equation}
\label{eq:v 0 for K and Pi}
    K(y_0,0)=\arcsin{y_0},\qquad \Pi(y_0,u,0)=\frac{\arctan{\left(y_0\sqrt{\frac{1-u}{1-y_0^2}}\right)}}{\sqrt{1-u}}.
\end{equation}
Finally, as mentioned in Ref.~\cite{bachas2021phases}, with $\delta\ll1$, the elliptic integrals have the property that
\begin{equation}
    K\left(-\frac{a}{\delta}\right)\approx\Pi\left(u,-\frac{a}{\delta}\right)\approx-\frac{\ln{\delta}\cdot\sqrt{\delta}}{2\sqrt{a}}+\mathcal{O}(\sqrt{\delta}).
\end{equation}

\subsubsection{Arc equations with special functions}
\label{sec:Arc equations with special functions}

In this section, we show the detailed derivation of \eqref{eq:equation Ip with elliptic integrals}.
Here, we derive \eqref{eq:Ip1 with elliptic integrals} explicitly, and \eqref{eq:Ip2 with elliptic integrals} can be obtained similarly.
For $I_1'$, with $s=-\sigma/M_1$ and \eqref{eq:constraint for Mi with Li}, we have 
\begin{equation}
\label{eq:rescaled Ip1 with M1}
    I_1'=\sqrt{-M_1}\cdot I_1=\int_{s_+}^{+\infty}(\dot{x}_1^a+\dot{x}_1^b)(-M_1)^{-3/2}{\rm d}s-\int_{s_+}^{s_0}(\iota_A\ \dot{x}_1^a-\iota_B\ \dot{x}_1^b)(-M_1)^{-3/2}{\rm d}s,
\end{equation}
where
\begin{equation}
\label{eq:rescaled x1d with different T}
    (-M_1)^{-3/2} \dot{x}_1^\alpha=-l_1\left.\frac{\mu-1+s\left(T_\alpha^2+T_0^2\right)}{2(-l_1^2+s)\sqrt{sA(s-s_+)(s-s_-)}}\right|_\alpha,
\end{equation}
in which we take $T=T_\alpha$ in the expression of $s_0$ and $s_\pm$.
Then define $y^2=s_+/s$, we have
\begin{equation}
\label{eq:transform part of Ip1 to elliptic integrals}
\begin{split}
    &\int_{s_+}^{s_0}\dot{x}_1^\alpha(-M_1)^{-3/2}{\rm d}s=\int_{1}^{\sqrt{s_+/s_0}}l_1 \frac{s_+(T_\alpha^2+T_0^2)+(\mu-1)y^2}{(s_+-l_1^2 y^2)\sqrt{As_+}\sqrt{(1-y^2)(1-\frac{s_-}{s_+}y^2)}}{\rm d}y\\
    =&\frac{l_1}{\sqrt{As_+}}\int_{1}^{\sqrt{s_+/s_0}}\left(\frac{-(\mu-1)l_1^{-2}}{\sqrt{(1-y^2)(1-\frac{s_-}{s_+}y^2)}}+\frac{(\mu-1)l_1^{-2}s_+ +s_+(T_\alpha^2+T_0^2)}{(s_+-l_1^2 y^2)\sqrt{(1-y^2)(1-\frac{s_-}{s_+}y^2)}}\right){\rm d}y\\
    =&\frac{l_1}{\sqrt{As_+}}\left\{\frac{\mu-1}{l_1^{2}}\left[K\left(\frac{s_-}{s_+}\right)-K\left(\sqrt{\frac{s_+}{s_0}},\frac{s_-}{s_+}\right)\right]-(T_\alpha^2+T_0^2+\frac{\mu-1}{l_1^{2}})\left[\Pi\left(\frac{l_1^2}{s_+},\frac{s_-}{s_+}\right)-\Pi\left(\sqrt{\frac{s_+}{s_0}},\frac{l_1^2}{s_+},\frac{s_-}{s_+}\right)\right]\right\}.
\end{split}
\end{equation}
Plugging \eqref{eq:transform part of Ip1 to elliptic integrals} into \eqref{eq:rescaled Ip1 with M1} and taking $s_0\rightarrow+\infty$ for the first integral in \eqref{eq:rescaled Ip1 with M1}, with $K(0,v)=\Pi(0,u,v)=0$, we can obtain \eqref{eq:Ip1 with elliptic integrals}.

\subsection{On-shell action with a corner}
\label{sec:Review of the reference bachas2021phases}

In this section, we derive the on-shell action, following the method in Ref.~\cite{bachas2021phases}. 
We work in Euclidean signature. 
The continuity of induced metric on the interface brane $(0, r=r(\sigma),x=x(\sigma))$ leads ${\rm d}s^2=f(\sigma){\rm d}\tau^2+g(\sigma){\rm d}\sigma^2$ with $f(\sigma)$ and $g(\sigma)$ given in \eqref{eq:continuity condition for metrics}.
For \eqref{eq:simplified eom b}, with $e^\mu_\tau=(1,0,0)$ and $e^\mu_x=(0,\dot{r},\dot{x})$, the normal direction $n$ of the brane for $AdS_1$ reads $n_\mu=\Tilde{A}(0,-\dot{x},\dot{r})$ with $\Tilde{A}>0$ and $\Tilde{A}^2=\frac{l^2}{\dot{x}^2(r^2-Ml^2)+\dot{r}^2l^2/r^2}$.
The extrinsic curvature is then
\begin{equation}
\label{eq:extrinsic curvature}
    K_{\tau\tau}=-\Gamma^\rho_{\tau\tau}n_\rho=-(Mr-\frac{r^3}{l^2})A\cdot(-\dot{x})=-\dot{x}\frac{r^2}{l}\frac{f}{\sqrt{fg}}.
\end{equation}


Now we derive the onshell action \eqref{eq:total action}. 
For AdS spacetime with radius $l_i$, we have $R_i=-6/l_i^2$ and $\sqrt{g_i}{\rm d}^3x=l_i r_i{\rm d}r_i{\rm d}x_i{\rm d}\tau$.  
The bulk term $I_{\rm EH}$ is
\begin{equation}
\label{eq:simplified EH action}
    I_{\rm EH}=\sum_{i=1,2}\frac{2}{l_i^2}\int_{M_i}\sqrt{g_i}=\sum_{i=1,2}\int_{M_i}\frac{2r_i}{l_i}{\rm d}r_i{\rm d}x_i{\rm d}\tau.
\end{equation}
Defining the vector $\mathbf{n}'=(0,r^2,0)$, by the Stoke's theorem, we have $\int_\Omega(\nabla\cdot \mathbf{n}'){\rm d}V=\int_{\partial\Omega}\mathbf{n}'\cdot{\rm d}\mathbf{S}$.
It leads $\int_\Omega2r{\rm d}r{\rm d}x{\rm d}\tau=\int_{\partial\Omega}r^2 \hat{\mathbf{n}}'\cdot{\rm d}\mathbf{S}$, where $\hat{\mathbf{n}}'$ is the normalized vector of $\mathbf{n}'$ with unit length.
Hence, the bulk term can be expressed as the integral on their surface that
\begin{equation}
\label{eq:simplified EH action on surface}
    I_{\rm EH}=\sum_{i=1,2}\frac{1}{l_i}\left(\int_{B_i}r_i^2\hat{\mathbf{n}}_i'\cdot{\rm d}\mathbf{S}_i+\int_{W}r_i^2\hat{\mathbf{n}}_i'\cdot{\rm d}\mathbf{S}_i\right)=\sum_{i=1,2}\frac{1}{l_i}\left(\int_{B_i}r_i^2{\rm d}x_i{\rm d}\tau+\int_{W}r_i^2\hat{\mathbf{n}}_i'\cdot{\rm d}\mathbf{S}_i\right),
\end{equation}
where $W = W_a \cup W_b$. 
Using the parameter $\sigma$, the second term in \eqref{eq:simplified EH action on surface} can be evaluated to be 
\begin{equation}
\label{eq:simplified EH action on surface second term}
    \sum_{i=1,2}\frac{1}{l_i}\int_{W}r_i^2\hat{\mathbf{n}}_i'\cdot{\rm d}\mathbf{S}_i=-\sum_{i=1,2}\frac{1}{l_i}\int_{W}r_i^2 {\rm d}\tau{\rm d}x_i=-\int_{W}\left(\frac{r_1^2 \dot{x}_1}{l_1}+\frac{r_2^2 \dot{x}_2}{l_2}\right) {\rm d}\tau{\rm d}\sigma = \sum_\alpha T_\alpha \int_{W_\alpha}\sqrt{fg}\ {\rm d}\tau{\rm d}\sigma.
\end{equation}
where in the last step we used \eqref{eq:simplified eom b}. 
Besides, with the induced metric on the brane ${\rm d}s^2=f(\sigma){\rm d}\tau^2+g(\sigma){\rm d}\sigma^2$, the $I_T$ term is 
\begin{equation}
\label{eq:simplified tension term}
    I_T=T\int_{W}\sqrt{fg}\ {\rm d}\sigma{\rm d}\tau.
\end{equation}
The extrinsic curvature term $I_{\rm surface}$ can be written as
\begin{equation}
\label{eq:simplified surface term}
\begin{split}
    I_{\rm surface}=& -\sum_{i=1,2}\int_{B_i}\sqrt{\hat{g}_i}\ K_i- \int_{W}\sqrt{\hat{g}}(K_1+K_2) = -\sum_{i=1,2}\int_{B_i}\sqrt{\hat{g}_i}\ K_i -2T\int_{W}\sqrt{fg}\ {\rm d}\sigma{\rm d}\tau
\end{split}
\end{equation}
where we used $K_1+K_2=2T$ for the extrinsic curvature on the brane 
and $\int_{W}\sqrt{\hat{g}}=\int_{W}\sqrt{fg}\ {\rm d}\sigma{\rm d}\tau$.
Because the codimension-2 integral on the corner in the action will not contribute to the onshell result, the total action is
\begin{equation}
\label{eq:simplified total action}
\begin{split}
    I_{\rm tot} &=\sum_{i=1,2} \left[ \frac{1}{l_i}\int_{B_i}r_i^2{\rm d}x_i{\rm d}\tau-\int_{B_i}\sqrt{\hat{g}_i}\ K_i+\frac{1}{l_i}\int_{B_i}\sqrt{\hat{g}_i} \right].
\end{split}
\end{equation}

In the following, we calculate the integral in \eqref{eq:simplified total action} explicitly.
For the first term in \eqref{eq:simplified total action}, with the time direction length $T_{\rm DCFT}^{-1}$ and CFT coordinate $r_{i,\infty}$, we have 
\begin{equation}
\label{eq:first term in Itot evaluation}
    \frac{1}{l_i}\int_{B_i}r_i^2{\rm d}x_i{\rm d}\tau=\frac{r_{i,\infty}^2 T_{\rm DCFT}^{-1}L_i}{l_i}.
\end{equation}
For the second term in \eqref{eq:simplified total action}, the induced metric on $B$ is ${\rm d}s^2=(r_\infty^2-Ml^2){\rm d}\tau^2+r_\infty^2{\rm d}x^2$.
With ${e'}_\tau^\mu=(1,0,0)$ and ${e'}_x^\mu=(0,0,1)$, ${n'}_\mu=A'(0,1,0)$ with $A'=\sqrt{\frac{l^2}{r_\infty^2-Ml^2}}$, the extrinsic curvature becomes 
\begin{equation}
\label{eq:CFT extrinsic curvature}
    K_{\tau\tau}=-\Gamma^\rho_{\tau\tau}n_\rho=-\Gamma^r_{\tau\tau}\sqrt{\frac{l^2}{r_\infty^2-Ml^2}}=\frac{r_\infty}{l}\sqrt{r_\infty^2-Ml^2},~
    K_{xx}=-\Gamma^\rho_{xx}n_\rho=-\Gamma^r_{xx}\sqrt{\frac{l^2}{r_\infty^2-Ml^2}}=\frac{r_\infty}{l}\sqrt{r_\infty^2-Ml^2},
\end{equation}
where $\Gamma^r_{\tau\tau}=\Gamma^r_{xx}=Mr_\infty-r_\infty^3/l^2$.
Then, we arrive at
\begin{equation}
\label{eq:extrinsic curvature on CFT}
    K=g^{\mu\nu}K_{\mu\nu}=K_{\tau\tau}\cdot(r_\infty^2-Ml^2)^{-1}+K_{xx}\cdot r_\infty^{-2}=\frac{\frac{2r_\infty}{l}-\frac{Ml}{r_\infty}}{\sqrt{r_\infty^2-Ml^2}}.
\end{equation}
Now, the second term in \eqref{eq:simplified total action} is
\begin{equation}
\label{eq:evaluation of K integral on B}
    \int_{B_i}\sqrt{\hat{g}_i}\ K_i=\int_{B_i}\sqrt{(r_{i,\infty}^2-M_i l_i^2)r_{i,\infty}^2}\cdot\frac{\frac{2r_{i,\infty}}{l_i}-\frac{M_i l_i}{r_{i,\infty}}}{\sqrt{r_{i,\infty}^2-M_i l_i^2}}{\rm d}x_i{\rm d}\tau=\frac{(2r_{i,\infty}^2-M_il_i^2) T_{\rm DCFT}^{-1}L_i}{l_i}.
\end{equation}
Similarly, the last term in \eqref{eq:simplified total action} becomes
\begin{equation}
\label{eq:evaluation of fifth integral in Itot}
    \frac{1}{l_i}\int_{B_i}\sqrt{\hat{g}_i}=\frac{1}{l_i}\int_{B_i}\sqrt{(r_{i,\infty}^2-M_i l_i^2)r_{i,\infty}^2}{\rm d}x_i{\rm d}\tau=\frac{\sqrt{(r_{i,\infty}^2-M_i l_i^2)r_{i,\infty}^2} T_{\rm DCFT}^{-1}L_i}{l_i}.
\end{equation}
Finally, combining \eqref{eq:first term in Itot evaluation}, \eqref{eq:evaluation of K integral on B} and \eqref{eq:evaluation of fifth integral in Itot}, the onshell action reads 
\begin{equation}
\label{eq:total action final results}
\begin{split}
    I_{\rm tot}&=\sum_{i=1,2}\left(\frac{1}{l_i}\int_{B_i}r_i^2{\rm d}x_i{\rm d}\tau-\int_{B_i}\sqrt{\hat{g}_i}\ K_i+\frac{1}{l_i}\int_{B_i}\sqrt{\hat{g}_i}\right)\\
    &=\sum_{i=1,2}\left(r_{i,\infty}^2-(2r_{i,\infty}^2-M_il_i^2)+\sqrt{(r_{i,\infty}^2-M_i l_i^2)r_{i,\infty}^2}\right)\frac{ T_{\rm DCFT}^{-1}L_i}{l_i}\xrightarrow{r_{i,\infty}\rightarrow\infty}\frac{M_1 l_1 L_1+M_2 l_2 L_2}{2T_{\rm DCFT}}.
\end{split}
\end{equation}
Due to the cancellation of the integral on the interface brane with the help of the Stoke's theorem, it is the same as the result in Ref.~\cite{bachas2021phases},

\subsection{Sweeping transition and non-self-intersecting condition without corners}
\label{sec:Sweeping transition and non-self-intersecting condition}

In this section, we will review and generalize the discussion of the sweeping transition in Ref.~\cite{bachas2021phases}, and also discuss the monotonicity of brane solutions.
We will also consider a few special cases. 

\begin{figure}
    \centering
    \subfigure[]{\includegraphics[width=0.33\textwidth]{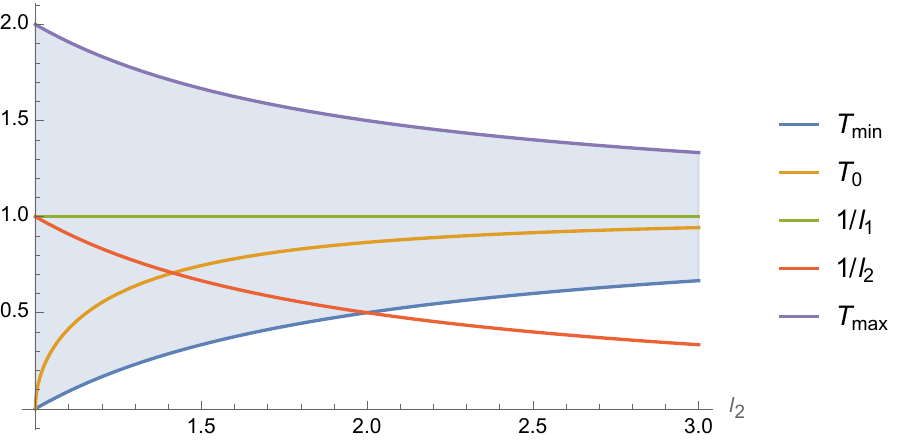}}\qquad\quad
    \subfigure[]{\includegraphics[width=0.33\textwidth]{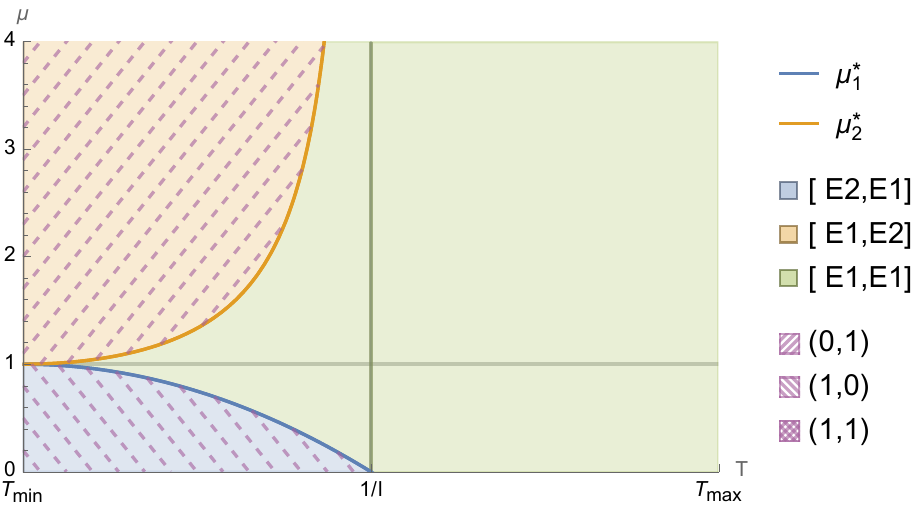}}\\
    \subfigure[]{\includegraphics[width=0.32\textwidth]{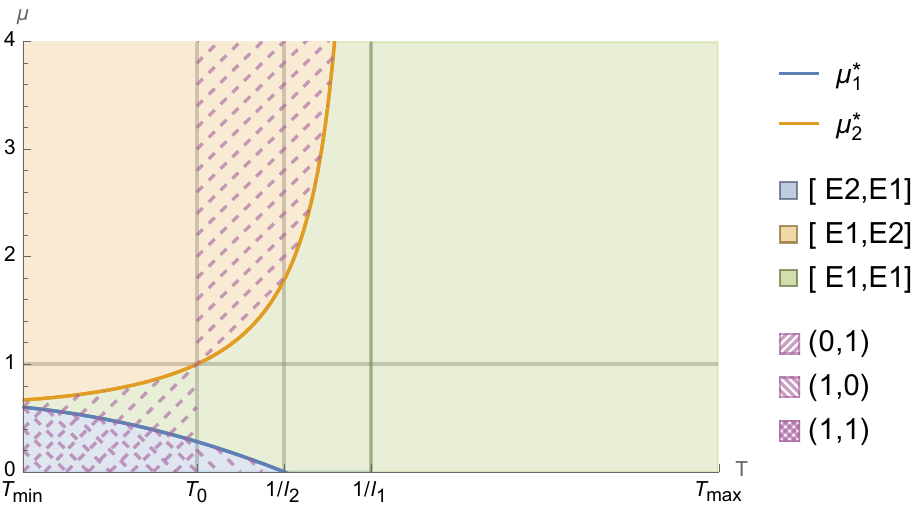}}
    \subfigure[]{\includegraphics[width=0.32\textwidth]{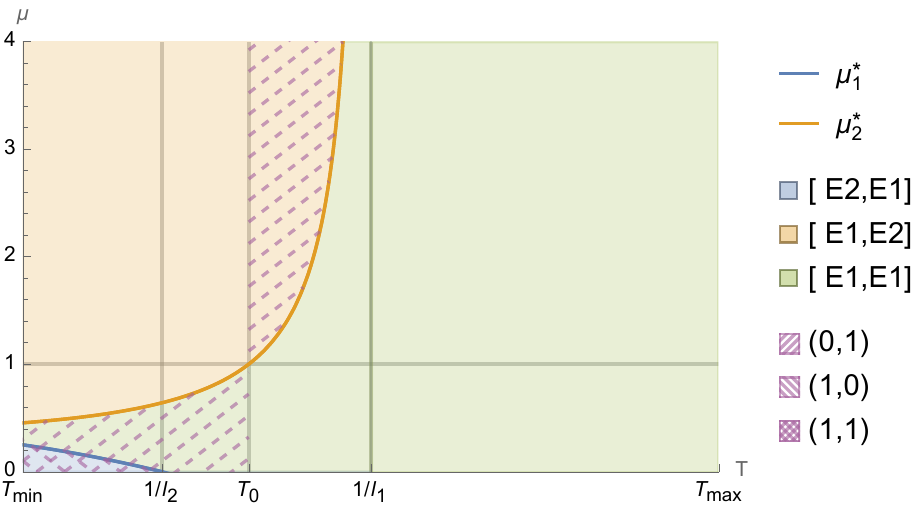}}
    \subfigure[]{\includegraphics[width=0.32\textwidth]{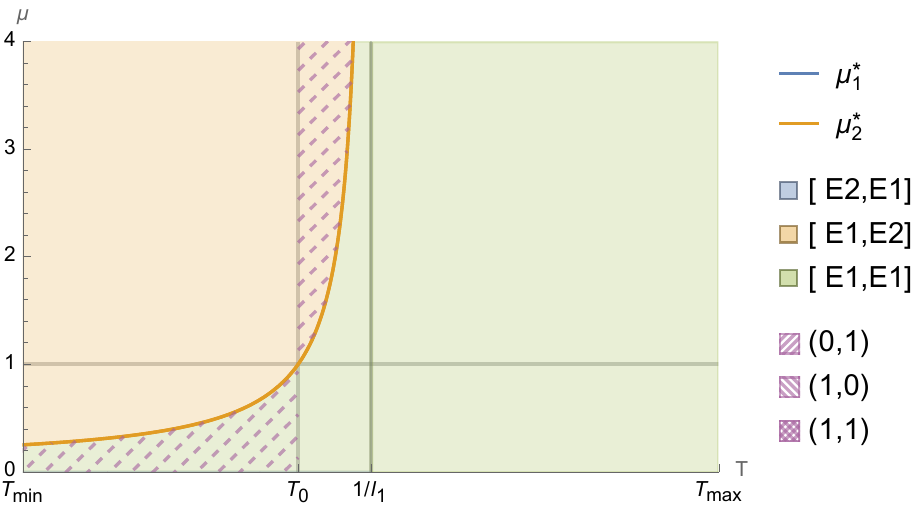}}
    \caption{(a) The values of special tensions with $l_1=1$, where the blue region corresponds to the valid tension $T_{\rm min}<T<T_{\rm max}$. 
    There are three regions of $l_2$, which have different orders of $T_0$, $T_{\rm min}$ and $l_2^{-1}$. 
    The crossing points are $l_2=\sqrt{2}l_1$ and $l_2=2l_1$. 
    Different phase diagrams for sweeping transition for (b) $l_2=l_1$, (c) $l_1<l_2<\sqrt{2}l_1$, (d) $\sqrt{2}l_1<l_2<2l_2$ and (e) $l_2>2l_1$. 
    The blue, orange and green region corresponds to different phases. 
    ${\rm [Ei,Ej]}$ means $AdS_{1,2}$ is in ${\rm Ei,Ej}$ phases with $i,j=1,2$. 
    The shadowed regions with $(a_1,a_2)$ indicate the monotonicity of branes.
    $a_i=0,1$ corresponds to the monotonic or non-monotonic brane in $AdS_i$, respectively. 
    Monotonic branes in both AdS spaces are depicted without shadow lines.}
    \label{fig:Different phase diagram for sweeping transition}
\end{figure}

Firstly, we review the sweeping transition between E1 and E2 phase in Ref.~\cite{bachas2021phases}.
For the single brane geometry with brane coordinate $(r=r(\sigma),x=x(\sigma))$ on a constant time slice, the critical point is given by that the center $r=0$ is located on the interface brane.
With the reflection symmetry, it means that the critical point corresponds to $r(\sigma_+)=\sqrt{\sigma_++Ml^2}=0$, in which $\sigma_+$ is a solution of $A\sigma^2+2B\sigma+C=0$ with coefficients \eqref{eq:definition of ABC}.
Therefore, using $\sigma_+=\sigma_+^c \equiv -M_i l_i^2$, we can get the critical $\mu_i^\ast$
\begin{equation}
\label{eq:critical mu for AdSi}
    \mu_1^*=\frac{l_1^2}{l_2^2}-T^2l_1^2=\frac{1-T^2l_2^2}{l_2^2/l_1^2},\qquad
    \mu_2^*=\frac{1}{l_2^2/l_1^2-T^2l_2^2}=\frac{l_1^2/l_2^2}{1-T^2 l_1^2}.
\end{equation}
Remember that we assume $l_1<l_2$.
Now we can determine different phases.
With $\sigma_+=(-B+\sqrt{B^2-AC})/A$, for different limits we have
\begin{equation}
\label{eq:asympototic expansion of sigma}
    \sigma_+\approx
    \begin{cases}
        \frac{M_2}{-2T/l_1+(T^2+T_0^2)},&M_2\gg M_1,\\
        \frac{M_1}{-2T/l_2+(T^2-T_0^2)},&M_2\ll M_1.
    \end{cases}
\end{equation}
Then from \eqref{eq:explicit expression of x}, for $\dot{x}_1$ with $\mu=\frac{M_2}{M_1}\gg1$, we have
\begin{equation}
\label{eq:dot x1 with M2ggM1}
\begin{split}
    {\rm sgn}(\dot{x}_1(\sigma_+))=&{\rm sgn}\left[-l_1\frac{M_1-M_2+\sigma_+\left(T^2+T_0^2\right)}{2(l_1^2 M_1+\sigma_+)}\right]\approx{\rm sgn}\left[\frac{M_2}{2\sigma_+}-\frac{T^2+T_0^2}{2}\right]
    =-{\rm sgn}\left[\frac{(T-T_{\rm min})(T_{\rm max}-T)}{2}\right]=-1,
\end{split}
\end{equation}
which corresponds to the E1 phase for $AdS_1$. 
Hence, when $\mu > \mu_1^\ast$ ($\mu < \mu_1^\ast$), $AdS_1$ is in E1 (E2) phase. 
Similarly, from \eqref{eq:explicit expression of x}, with $\mu=\frac{M_2}{M_1}\ll1$, we have
\begin{equation}
\label{eq:dot x2 with M2llM1}
\begin{split}
    {\rm sgn}(\dot{x}_2(\sigma_+))=&{\rm sgn}\left[-l_2\frac{M_2-M_1+\sigma_+\left(T^2-T_0^2\right)}{2(l_2^2 M_2+\sigma_+)}\right]\approx{\rm sgn}\left[\frac{M_1}{2\sigma_+}-\frac{T^2-T_0^2}{2}\right]
    =-{\rm sgn}\left[\frac{(T+T_{\rm min})(T_{\rm max}-T)}{2}\right]=-1,
\end{split}
\end{equation}
which means $AdS_2$ is in E1 phase. 
Hence, when $\mu < \mu_1^\ast$ ($\mu > \mu_1^\ast$), $AdS_2$ is in E1 (E2) phase. 
Therefore, with the discussion above, we express $\delta_{i,{\rm E1}}$ explicitly that
\begin{equation}
\label{eq:explicit expression of deltaiE1}
    \delta_{1,{\rm E1}}=\Theta(\mu-\mu_1^*),\qquad \delta_{2,{\rm E1}}=\Theta(-\mu_2^*)+\Theta(\mu_2^*)\theta(\mu_2^*-\mu),
\end{equation}
where $\Theta(x)$ is the step function that is nonzero only for $x>0$.
Because $\mu_1^\ast$ changes signs at $T= l_2^{-1}$, i.e.,
\begin{equation}
    \mu_1^\ast({T<l_2^{-1}}) > 0, \quad \mu_1^\ast({T>l_2^{-1}}) < 0,
\end{equation}
it means that when $T > l_2^{-1}$, $AdS_1$ is always in the E1 phase.
A plot of the phase diagram is shown in Fig.~\ref{fig:Different phase diagram for sweeping transition} (c,d,e). 
Notice that the difference between (c,d) and (e) comes from whether $T_{\rm min} < l_2^{-1} $, i.e., if $T_{\rm min} < l_2^{-1}$, there is a transition for $AdS_1$ as shown in (c,d), otherwise $T_{\rm min} > l_2^{-1}$, there are no transitions for $AdS_1$ as shown in (e). 
On the other hand, the difference between (c) and (d) is due to the
possibility of monotonicity of the gradient $\dot x_i$ as we will detail next.

In general, we should start \eqref{eq:explicit expression of x}, as $\dot{x}_i = 0$ gives a non-monotonic brane solution. 
It is easy to show the denominator of the right-hand side in \eqref{eq:explicit expression of x} is non-negative. 
On the other hand, the numerator a linear function in $\sigma$, so there is only one transition point where the brane will change the direction.
Setting the numerator of the right-hand side of \eqref{eq:explicit expression of x} to be zero, we get the zero point, $\sigma_{1,2}$,
\begin{equation}
\label{eq:zeros of dotxi}
    \sigma_1=\frac{M_2-M_1}{T^2+T_0^2},\qquad\sigma_2=\frac{M_1-M_2}{T^2-T_0^2}.
\end{equation}
Because the range of the brane parameter is $\sigma \ge \sigma_+$, in order for the zero point to be accessible for a brane solution, it requires 
\begin{equation} \label{eq:non-monotonic_condition}
0<\sigma_+/\sigma_i<1.
\end{equation}
It is easy to check that $\sigma_i=\sigma_+$ leads $\mu=\mu_i^*$, and the following non-monotonicity condition,
\begin{equation}
    \begin{split}
    0 < \sigma_+/\sigma_1 < 1 & \Rightarrow 0 < \mu < \mu_1^\ast, \\
    0 < \sigma_+/\sigma_2 < 1 & \Rightarrow 
    \begin{cases}  0 < \mu < \mu_2^\ast & \quad T < T_0 \\   \mu > \mu_2^\ast & \quad T_0 < T < l_1^{-1}
    \end{cases}.
    \end{split}
\end{equation}
Note that the condition for $AdS_2$ depends on $T_0$.
This is the reason for the difference between Fig.~\ref{fig:Different phase diagram for sweeping transition} (c) and (d). 
More explicitly, in Fig.~\ref{fig:Different phase diagram for sweeping transition} (c,d,e), we use the label $(a_1,a_2)$ with $a_i=0$ ($a_i=1$) to show if the corresponding phase has (non-)monotonic branes in $AdS_i$.
For $T_{\rm min} > l_2^{-1}$, there is no non-monotonic brane for $AdS_1$, but there exist non-monotonic branes for $AdS_2$ as shown in figure~\ref{fig:Different phase diagram for sweeping transition} (e).
For $T_{\rm min} < l_2^{-1}$, there exist non-monotonic branes for both $AdS_1$ and $AdS_2$ as shown in Fig.~\ref{fig:Different phase diagram for sweeping transition} (c) and (d).
Fig.~\ref{fig:Different phase diagram for sweeping transition} (a) summarizes the relation among the allowed range of tensions, $T_0$, and $l_2^{-1}$.

\begin{figure}
    \centering
    \includegraphics[width=0.5\textwidth]{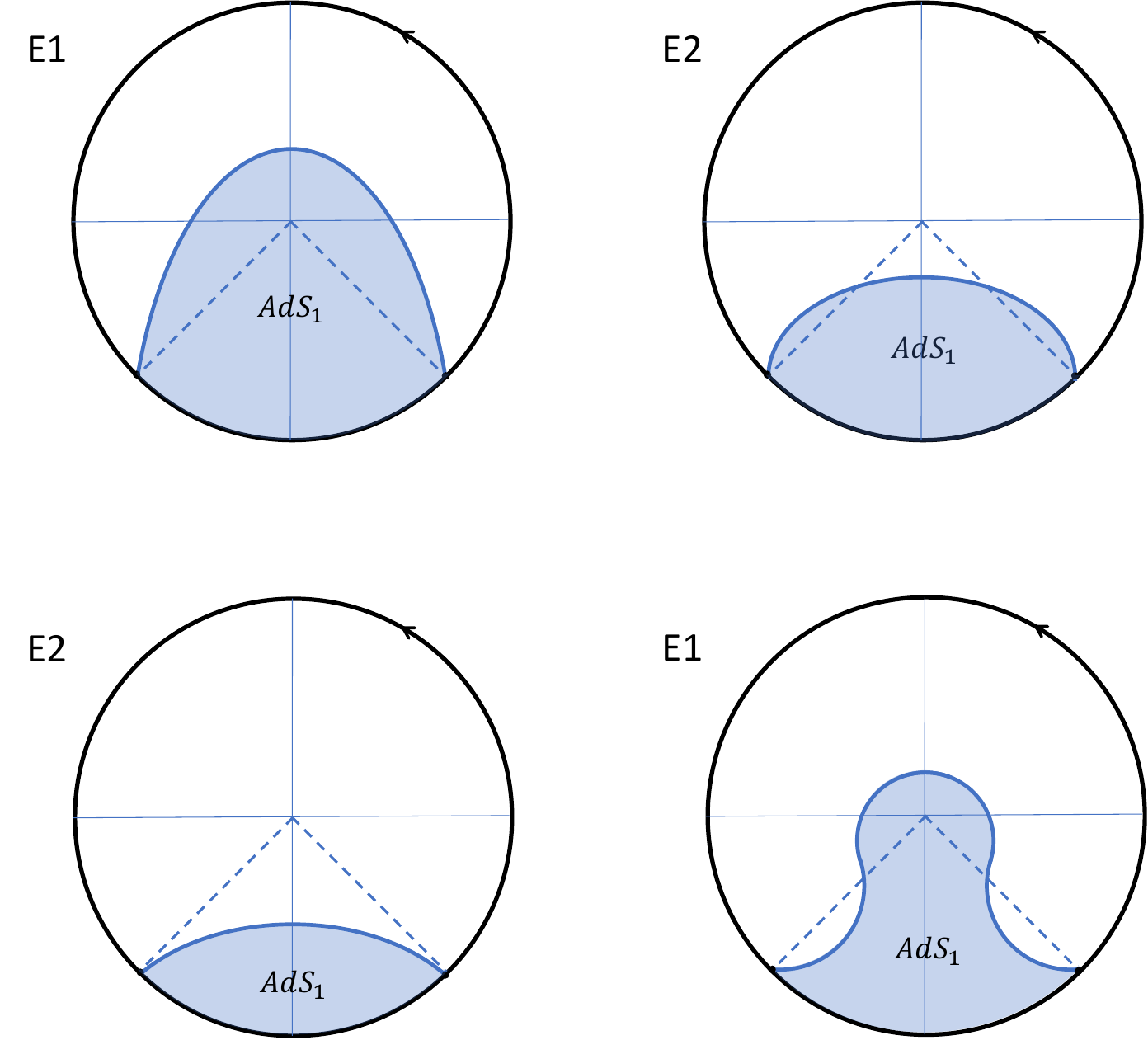}
    \caption{Different configurations of branes:  
    Left upper panel: E1 phase with monotonic brane. 
    Left lower panel: E2 phase with monotonic brane. 
    Right upper panel: E2 phase with non-monotonic brane. 
    Right lower panel: E1 phase with non-monotonic brane.}
    \label{fig:Different phase transition diagram}
\end{figure}

The phase diagrams in Fig.~\ref{fig:Different phase diagram for sweeping transition} seem complicated because the phases E1 and E2 are not directly related to the monotonicity of branes.
However, we can understand it by combining the E1 and E2 phases and the angle between the brane and the asymptotic boundary.
With the metric \eqref{eq:metric for two AdS}, near the asymptotic boundary $r\rightarrow\infty$, we can ignore $M_i l_i^2$.
Then, using the coordinate transformation $r_i=l_i/z_i$, we obtain the Poincare metric
\begin{equation}
\label{eq:Poincare metric}
    {\rm d}s^2=\frac{l_i^2(-{\rm d}t^2+{\rm d}z_i^2+{\rm d}x_i^2)}{z_i^2}.
\end{equation}
Then, with the result in Ref.~\cite{Anous_2022,Sun_2023}, we can solve the brane equation $x_i=\tan{\psi_i}z_i$ with 
\begin{equation}
\label{eq:ange of brane equation with Poincare metric}
    \sin{\psi_1}=\frac{l_1}{2T}\left(T^2+\frac{1}{l_1^2}-\frac{1}{l_2^2}\right),\qquad \sin{\psi_2}=-\frac{l_1}{2T}\left(T^2+\frac{1}{l_2^2}-\frac{1}{l_1^2}\right).
\end{equation}
Here we assume the asymptotic boundary of $AdS_1$ is located within $x_1 < 0$, while the asymptotic boundary of $AdS_2$ is within $x_2 > 0$ and the defect A is located at $x_i = 0$. 
The positive direction of the angle $\psi_i$ is defined as pointing outsides the $AdS_i$ bulk.
Now we can consider the brane equations.
Since $l_1<l_2$, \eqref{eq:ange of brane equation with Poincare metric} indicates $\sin{\psi_1}>0$ and $0<\psi_1<\frac{\pi}{2}$.
Therefore, we expect that the E1 phase, which includes the center, will be equivalent to the monotonic brane for $AdS_1$, which is shown in Fig.~\ref{fig:Different phase transition diagram} left upper panel.
While, the E2 phase, which excludes the center, will be equivalent to the non-monotonic brane for $AdS_1$, which is shown in Fig.~\ref{fig:Different phase transition diagram} right upper panel.
A phase transition between E1 and E2 shown in figure~\ref{fig:Different phase transition diagram} (a) also leads to a transition of the monotonicity of branes.
However, for $AdS_2$ the situation is more complicated.
If $T>T_0$ with $-\frac{\pi}{2}<\psi_2<0$, it is similar to $AdS_1$ that the E1 (E2) phase is equivalent to (non-)monotonic branes.
However, if $T_{\rm min}<T<T_0$ with $0<\psi_2<\frac\pi2$, it is opposite to $AdS_1$ that the E2 (E1) phase is equivalent to (non-)monotonic branes, which is shown in Fig.~\ref{fig:Different phase transition diagram} lower panels.
Therefore, we also have the phase transition for the monotonicity of branes within the same phase E1 or E2. 


Monotonicity is related to self-intersecting branes,
because a self-intersecting brane requires the non-monotonicity from the discussion above. 
For example, as the E1 phase shown in Fig.~\ref{fig:Different phase transition diagram} right lower panel, the points $\dot{x}(\sigma)=0$ may touch each other and make the brane self-intersecting. 
Note that the self-intersecting solution may not be physical~\cite{PhysRevD.106.066020}, so it is important to address it.
Numerically, we have checked some non-monotonic examples, and we expect that all of them will not give self-intersecting branes.

In the following, we discuss some special cases.
(i) $L_1=L_2=L/2$, i.e., $\gamma=1$.
In this case, as mentioned in Ref.~\cite{bachas2021phases}, we always have the solution $\mu=1$.
And with \eqref{eq:critical mu for AdSi}, we have $\mu_2^*(T_0)=1$.
Therefore, as shown in Fig.~\ref{fig:Different phase transition diagram} (c), (d) and (e), at $\mu = 1$, by increasing the tension $T$, the AdS geometry changes from [E1,E2] to [E1,E1], at the critical point $T=T_0$.
Moreover, there are no non-monotonic branes in the solution.
This result is used in \eqref{eq:simplified L12 with L1L2}. 
(ii) $l_1=l_2=l$.
In this case, because $\mu_1^*=1-l^2T^2$ and $\mu_2^*=1/(1-l^2T^2)$, we have $\mu_1^*(T_{\rm min})=\mu_2^*(T_{\rm min})=1$ with $T_{\rm min}=0$. 
The phase diagram is Fig.~\ref{fig:Different phase diagram for sweeping transition} (b).
And similar to the discussion above, we find that in this phase diagram, if and only if the AdS spacetime is in E2 phase, it has non-monotonic branes. 
Moreover, if we also have $L_1=L_2=L/2$, then there is only one phase [E1,E1] with monotonic branes.

\subsection{Brane solution for the symmetric case}
\label{Brane equation for symmetric case}


In this section, we derive the brane solution for $L_1=L_2=L/2$without corner terms. 
In this case, $\gamma = 1$, and have $\mu=1$.
Then we can solve the quadratic equation of $\sigma$ with coefficients \eqref{eq:definition of ABC}.
Letting $M_1=M_2=M$, we have $C=0$, $\sigma_-=0$ and $\sigma_+=-\frac{2B}{A}=-4T^2M/A$.
The differential equations \eqref{eq:explicit expression of x} can be simplified to
\begin{subequations}
\label{eq:explicit expression of x Z2 symm}
\begin{equation}
\label{eq:explicit expression of x1 Z2 symm}
    \dot{x}_1=-l_1\frac{T^2+T_0^2}{2(l_1^2 M+\sigma)\sqrt{A(\sigma-\sigma_+)}}=\frac{-l_1(T^2+T_0^2)}{2(l_1^2 M+\sigma)\sqrt{A\sigma+4T^2M}},
\end{equation}
\begin{equation}
\label{eq:explicit expression of x2 Z2 symm}
    \dot{x}_2=-l_2\frac{T^2-T_0^2}{2(l_2^2 M+\sigma)\sqrt{A(\sigma-\sigma_+)}}=\frac{-l_2(T^2-T_0^2)}{2(l_2^2 M+\sigma)\sqrt{A\sigma+4T^2M}}.
\end{equation}
\end{subequations}
In the following, we use the integral
\begin{equation}
\label{eq:integral ab}
    \int_{-b}^{\Tilde{\sigma}}\frac{{\rm d}\sigma}{(\sigma+a)\sqrt{\sigma+b}}=\frac{2}{\sqrt{a-b}}\arctan{\sqrt{\frac{b+\Tilde{\sigma}}{a-b}}},
\end{equation}
where $-b>-a$ and $\Tilde{\sigma}>-b>0>a$.
With $4T^2-Al_1^2=l_1^2(T^2+T_0^2)^2>0$, and $4T^2-Al_2^2=l_2^2(T^2-T_0^2)^2>0$, we can apply the integral above to get the brane equations that
\begin{equation}
\label{eq:explicit brane equation for x1}
\begin{split}
    x_1(\Tilde{\sigma})=&\int_{\sigma_+}^{\Tilde{\sigma}}\dot{x}_1{\rm d}\sigma=\frac{-l_1(T^2+T_0^2)}{2\sqrt{A}}\int_{-\frac{4T^2M}{A}}^{\Tilde{\sigma}}\frac{{\rm d}\sigma}{(\sigma+l_1^2 M)\sqrt{\sigma+4T^2M/A}}\\
    =&\frac{-l_1(T^2+T_0^2)}{\sqrt{l_1^2MA-4T^2M}}\arctan{\sqrt{\frac{\Tilde{\sigma}+4T^2M/A}{l_1^2M-4T^2M/A}}}=\frac{-1}{\sqrt{-M}}\arctan{\sqrt{\frac{\Tilde{\sigma}+4T^2M/A}{l_1^2M-4T^2M/A}}}.
\end{split}
\end{equation}
as well as 
\begin{equation}
\label{eq:explicit brane equation for x2}
\begin{split}
    x_2(\Tilde{\sigma})=&\int_{\sigma_+}^{\Tilde{\sigma}}\dot{x}_2{\rm d}\sigma=\frac{-l_2(T^2-T_0^2)}{2\sqrt{A}}\int_{-\frac{4T^2M}{A}}^{\Tilde{\sigma}}\frac{{\rm d}\sigma}{(\sigma+l_2^2 M)\sqrt{\sigma+4T^2M/A}}\\
    =&\frac{-l_2(T^2-T_0^2)}{\sqrt{l_2^2MA-4T^2M}}\arctan{\sqrt{\frac{\Tilde{\sigma}+4T^2M/A}{l_2^2M-4T^2M/A}}}=\frac{-{\rm sgn}(T^2-T_0^2)}{\sqrt{-M}}\arctan{\sqrt{\frac{\Tilde{\sigma}+4T^2M/A}{l_2^2M-4T^2M/A}}}.
\end{split}
\end{equation}

\begin{figure}
    \centering
    \subfigure[]{\includegraphics[width=0.35\textwidth]{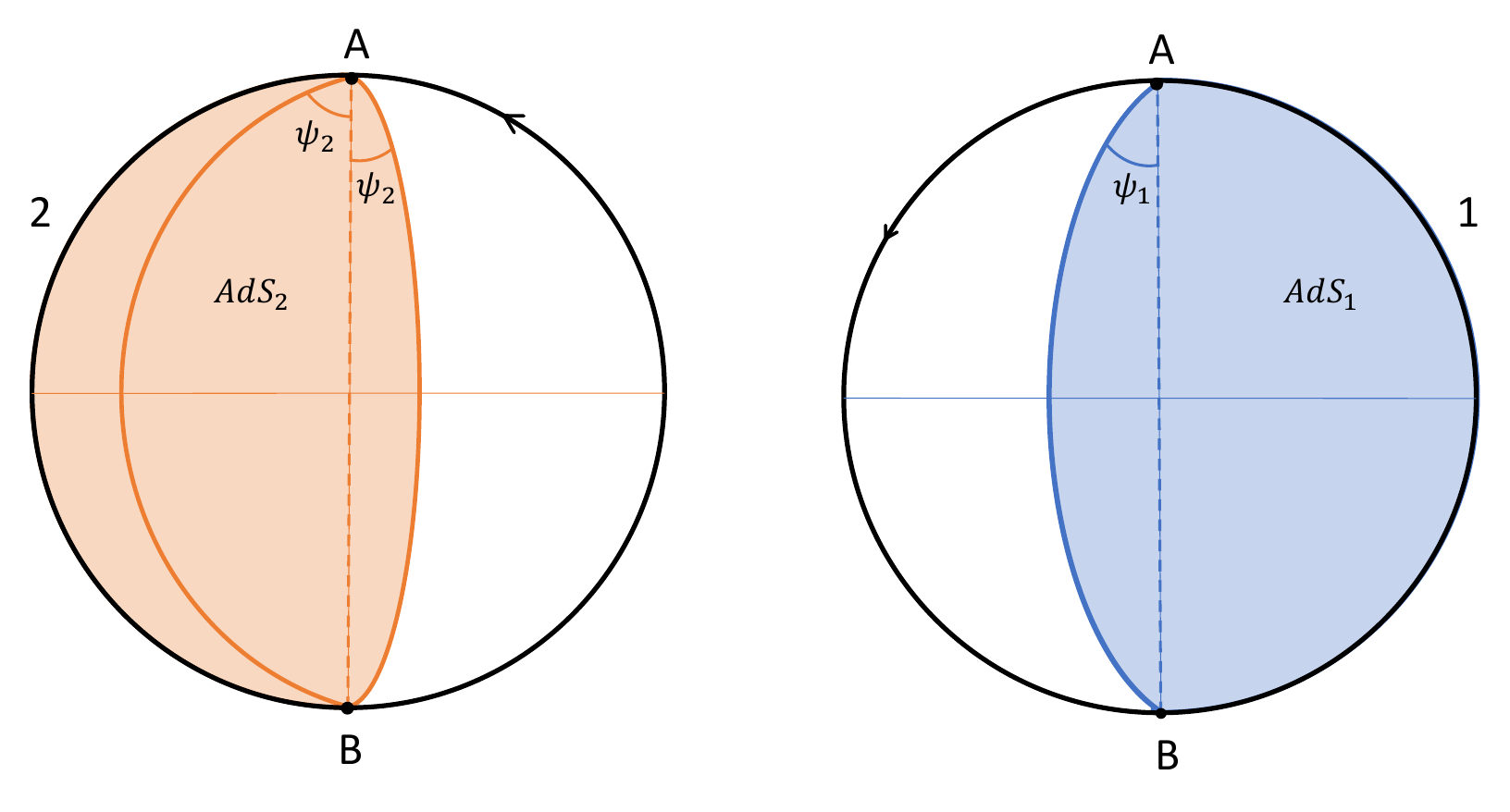}}
    \caption{Brane trajectory in the equipartition case $L_1=L_2$. 
    Note that $l_1 < l_2$. 
    $AdS_1$ always has $\psi_1>0$, which is shown in the right panel. 
    While $\psi_2$ can have different signs for $AdS_2$.}
    \label{fig:Brane diagram}
\end{figure}

As discussed in Appendix~\ref{sec:Sweeping transition and non-self-intersecting condition}, the brane is always monotonic.
Therefore, the monotonically decreasing function $x_1(\Tilde{\sigma})$ in \eqref{eq:explicit brane equation for x1} corresponds to $0<\psi_1 < \frac\pi2$.  
While for $x_2(\Tilde{\sigma})$, it relies on the sign of $T^2-T_0^2$, and it is consistent with $\sin{\psi_2}$ because $\sin{\psi_2}= - \frac{l_2}{2T}(T^2-T_0^2)$.
We show the diagram of the geometry in Fig.~\ref{fig:Brane diagram}.
Moreover, with \eqref{eq:explicit brane equation for x1} and \eqref{eq:explicit brane equation for x2}, we have $x_1(+\infty)-x_1(\sigma_+)=-\frac{\pi}{2}\cdot\frac{1}{\sqrt{-M}}$ and $x_2(+\infty)-x_1(\sigma_+)=-{\rm sgn}(T^2-T_0^2)\frac{\pi}{2}\cdot\frac{1}{\sqrt{-M}}$.
Because of $L=2\pi/\sqrt{-M}$, the interface brane connects antipodal points at the asymptotic boundary.
This is the same as the result of end-of-the-world brane~\cite{miyaji2022holographic}.


\end{document}